\documentclass[preprint]{aastex}

\usepackage{color}

\newcommand{\ar}{[\ion{Ar}{2}]}
\newcommand{\si}{[\ion{Si}{2}]}
\newcommand{\sthree}{[\ion{S}{3}]}
\newcommand{\sfour}{[\ion{S}{4}]}

\newcommand{\ofour}{[\ion{O}{4}]}
\newcommand{\neon}{[\ion{Ne}{2}]}
\newcommand{\fetwo}{[\ion{Fe}{2}]}
\newcommand{\fe}{Fe-K}
\newcommand{\xsi}{\ion{Si}{13}}
\newcommand{\casa}{Cas~A}
\newcommand{\kps}{km~s$^{-1}$}

\begin{document}

\title{The Three-Dimensional Structure of Cassiopeia A}

\author{Tracey DeLaney\altaffilmark{1}, Lawrence Rudnick\altaffilmark{2},
M. D. Stage\altaffilmark{3}, J. D. Smith\altaffilmark{4},
Karl Isensee\altaffilmark{2}, Jeonghee Rho\altaffilmark{5},
Glenn E. Allen\altaffilmark{6}, Haley Gomez\altaffilmark{7},
Takashi Kozasa\altaffilmark{8}, William T. Reach\altaffilmark{5},
J. E. Davis\altaffilmark{6}, and J. C. Houck\altaffilmark{6}}
\altaffiltext{1}{Department of Physics and Engineering, West Virginia Wesleyan College, 59 College Avenue, Buckhannon, WV 26201; delaney\_t@wvwc.edu}
\altaffiltext{2}{Astronomy Department, University of Minnesota, Minneapolis,
MN 55455; larry@astro.umn.edu, isensee@astro.umn.edu}
\altaffiltext{3}{Mount Holyoke College, Department of Astronomy, 206 Kendade,
50 College Street, South Hadley, MA 01075; mstage@mtholyoke.edu}
\altaffiltext{4}{Ritter Astrophysical Observatory, University of Toledo,
Toledo, OH 43606 ; jd.smith@utoledo.edu}
\altaffiltext{5}{\emph{Spitzer} Science Center, California Institute of
Technology, MS 220-6, Pasadena, CA 91125; rho@ipac.caltech.edu,
reach@ipac.caltech.edu}
\altaffiltext{6}{Massachusetts Institute of Technology, Kavli Institute for
Astrophysics and Space Research, 77 Massachusetts Avenue, NE80, Cambridge, MA
02139; gea@space.mit.edu, houck@space.mit.edu, davis@space.mit.edu}
\altaffiltext{7}{School of Physics and Astronomy, Cardiff University, Queens
Building, The Parade, Cardiff CF24 3AA, UK; haley.morgan@astro.cf.ac.uk}
\altaffiltext{8}{Department of Cosmosciences, Graduate School of Science,
Hokkaido University, Sapporo 060-0810, Japan; kozasa@mail.sci.hokudai.ac.jp}

\begin{abstract}

We used the \emph{Spitzer} Space Telescope's Infrared Spectrograph to
map nearly the entire extent of Cassiopeia A between 5-40~\micron.  Using infrared and \emph{Chandra} X-ray Doppler velocity measurements, along with the locations of optical ejecta beyond the forward shock, we constructed a 3-D model of the remnant.  The structure of \casa\ can be characterized into a spherical component, a tilted thick disk, and multiple ejecta jets/pistons and optical fast-moving knots all populating the thick disk plane.  The Bright Ring in \casa\ identifies the intersection between the thick plane/pistons and a
roughly spherical reverse shock.  The ejecta pistons indicate a radial
velocity gradient in the explosion. Some ejecta pistons are bipolar with
oppositely-directed flows about the expansion center while some ejecta pistons
show no such symmetry. Some ejecta pistons appear to maintain the integrity
of the nuclear burning layers while others appear to have punched through the
outer layers.  The ejecta pistons indicate a radial velocity gradient in the
explosion.  In 3-D, the Fe jet in the southeast occupies a ``hole'' in the
Si-group emission and does not represent ``overturning'', as previously
thought.  Although interaction with the circumstellar medium affects the detailed appearance of the remnant and may affect the visibility of the southeast Fe jet, the bulk of the symmetries and asymmetries in \casa\ are intrinsic to the explosion.

\end{abstract}


\section{Introduction}
\label{sec:intro}

The three dimensional structure of core-collapse supernova explosions is of
considerable interest, with implications for the explosion mechanisms
\citep[e.g.][]{awm03,wbp07}, the role of rotation \citep[e.g.][]{mbg07},
pulsar ``natal kicks'' \citep[e.g.][]{spj04} and for the subsequent
interaction of ejecta with the circumstellar medium/pre-supernova wind
\citep{svg08}.  Asymmetries in the explosion and subsequent fallback have
implications for mixing and the subsequent progression of explosive
nucleosynthesis \citep{jwh08}.  ``Jet-induced'' scenarios may play a role
in forming supernova remnants such as Cassiopeia~A (\casa) \citep{wmc08},
and at their extreme, jet-dominated explosions may be responsible for
gamma-ray bursts \citep[e.g.][]{zwh04,mkm05}.  There is plentiful evidence
for asymmetric supernova explosions from structural and spectral data
\citep{wwh02} and from spectropolarimetry \citep[e.g.][]{whh01, tkm08}.

\casa\ provides a unique opportunity to study the three-dimensional supernova
explosion structure because of the high velocities of its clumpy ejecta, seen
both through Doppler and proper motions.  It is the result of a core-collapse
supernova approximately 330 years ago \citep{fhm06} and is bright across the
electromagnetic spectrum.  The brightest emission in \casa\ is concentrated
onto the 200$\arcsec$ diameter Bright Ring where ejecta from the explosion
are illuminated after crossing through and being compressed and heated by the
reverse shock \citep{mfc04,pf07}.  In a few locations, the position of the
reverse shock itself has been identified just inside of the Bright Ring from
the rapid turn-on of optical ejecta \citep{mfc04}.

While most of the observed ejecta at all wavebands are concentrated on the Bright Ring, optical ejecta are also identified at and beyond the forward shock. The forward shock is visible as a thin ring of X-ray filaments approximately 300$\arcsec$ in diameter \citep{gkr01}.  The canonical explanation for the far outer knots is that they are dense optical ejecta
\citep[$n_e\approx$4000--10,000 cm$^{-3}$;][]{fg96} which cool
quickly \citep[$t_c\approx$20--30 years;][]{kv76} after passing through the
reverse shock.  They are reheated after they overtake the forward shock and
encounter slow-moving ambient material which drives a slow shock into the
supersonic dense ejecta and produces optical emission.  In addition to the
largely symmetric-appearing Bright Ring and forward shock, jets of Si- and
S-rich ejecta extend out to large distances in the northeast and southwest
with optical ejecta identified at a distance of $\approx200\arcsec$ beyond
the forward shock radius \citep{hlb04}.  The jets may denote an axis of
symmetry of the supernova explosion \citep[e.g.][]{fhm06}.  The optical knots
are rapidly expanding with proper motions of 1000s of \kps\ on the Bright
Ring and up to 14,000 \kps\ for the optical ejecta at the tip of the jets
\citep{kv76,fhm06}.  To the southeast, X-ray images show Fe-rich
ejecta at a greater radius than Si-rich ejecta which has been interpreted as
an overturning of ejecta layers during the supernova explosion \citep{hrb00}.

Large-scale Doppler mappings have been carried out at optical wavelengths
primarily using S and O emission lines \citep{lmu95, rhf95}.  These
studies showed that the ejecta defining the Bright Ring outline a shell.
Assuming that the shell is spherical, the Doppler velocities of ejecta knots
can be converted into equivalent distances from the expansion center
along the line of sight.  The bulk of the ejecta are found to lie nearly in
the plane of the sky and they are organized into distinct velocity structures
such as the two complete rings - one blue shifted and one red shifted - that
form the northern part of the Bright Ring.  

A subsequent census of outlying
optical ejecta showed that the northeast Jet is oriented within 6$\degr$ of
the plane of the sky with an opening angle of $\approx$25$\degr$ \citep{fg96}.
The remainder of the outer optical knots are mostly located within
$\pm20\degr$ of the plane of the sky and form a giant ring at the location
of the forward shock \citep{fes01} although there are gaps in this
distribution to the north and south \citep{fhm06}.  None of the optical
searches, either imaging or spectral, reveal fast-moving ejecta knots
projected near the center of the remnant or a population of very-fast-moving
outer optical knots projected onto the Bright Ring.  One interpretation
offered by \citet{fes01} for this ``plane-of-the-sky'' effect was that a
near tangent viewing angle is required to detect ejecta knots.  However,
other effects may play a role in the lack of projected outer ejecta knots.
For instance, imaging searches rely on large proper motions to detect outer
ejecta knots and the filters used for the images may not have been broad enough
to detect the highly red- or blue-shifted emission.  While the spectral
searches certainly had the wavelength range to detect fast-moving projected
ejecta knots, the outer optical knots tend to be \emph{very} faint unless
they are encountering dense ambient media \citep{fes01,fhm06}.  Given
the small amount of H$\alpha$ emission projected against \casa\ \citep{fes01},
any projected outer ejecta knots may be too faint to have been detected in
the spectral searches.  Therefore, despite the absence of a population of
projected ejecta near the center of Cas A, either with velocities typical of
outer ejecta or velocities typical of Bright Ring ejecta, the Bright Ring and
the outer optical knot ring were believed to be limb-brightened shells
\citep{lmu95, fes01}.

Spectra of the X-ray-emitting ejecta have also been mapped using a number of
telescopes.  The first mapping with the \emph{Einstein} X-ray Observatory
\citep{mcc83} and a subsequent mapping with \emph{ASCA} \citep{hgt94} were at
very low spatial resolution (1\arcmin-2\arcmin) and showed large-scale
asymmetries in Doppler structure. These asymmetries were confirmed at higher
spatial resolution with \emph{XMM-Newton} \citep[20\arcsec;][]{wbv02,wbv03}
and the first 50-ks ACIS observation with the \emph{Chandra} X-ray
Observatory \citep[4\arcsec;][]{hsp01}.  At their improved spatial
resolution, \citet{wbv02} found that the Si-rich ejecta forms the same
general set of ring structures as the optical emission but the two distinct
rings to the north are difficult to identify. The Fe-rich emission is
concentrated in three main areas to the north, southeast, and west, but
\citet{wbv03} determined that most of the Fe mass is concentrated within what
they describe as a bipolar, double cone that is oriented at -55$\degr$ from
north and 50$\degr$ out of the plane of the sky.  \citet{hrb00} suggest an
overturning in the explosion in the southeast, with the Fe emission extending
out to the forward shock.  \citet{wbv02} also found that in the north the
centroid of the Fe-rich ejecta appears behind the Si-rich ejecta perhaps
identifying another overturning region.  A higher spatial resolution
(1$\arcsec$) analysis using the \emph{Chandra} ACIS 1~Ms data set plus two
archival 50-ks data sets also shows the same optical ring structures in
Si-rich emission and shows the strong \fe\ regions as dynamically distinct
\citep{sah04,dha05}.  A more recent analysis using \emph{Chandra} High Energy
Transmission Grating (HETG) spectra of the Si-He$\alpha$ triplet showed that,
on 1$\arcsec$ spatial scales, the Si-rich ejecta has a great deal of
substructure that is reminiscent of the variations seen in the optical data
\citep{lds06}.

In this paper, we present a Doppler analysis of \casa's ejecta using infrared
data from the \emph{Spitzer} Space Telescope \citep{err06} and X-ray data
from the archival \emph{Chandra} ACIS 1~Ms observation plus two other 50-ks
observations \citep{sah06, dha05}.  For the first time, we combine
multiwavelength data into a full 3-dimensional reconstruction of \casa.
To this model, we add previously published X-ray results from
\citet{lds06} and the jets and outer optical knots from \citet{fes01} and
\citet{fg96}.  Based on this model, we draw conclusions about the multiple
kinematic components and several major asymmetries in \casa's explosion.

\section{\emph{Spitzer} Observations and Data Analysis}
\label{sec:obs}

\begin{figure}
\epsscale{1}
\begin{center}
\plottwo{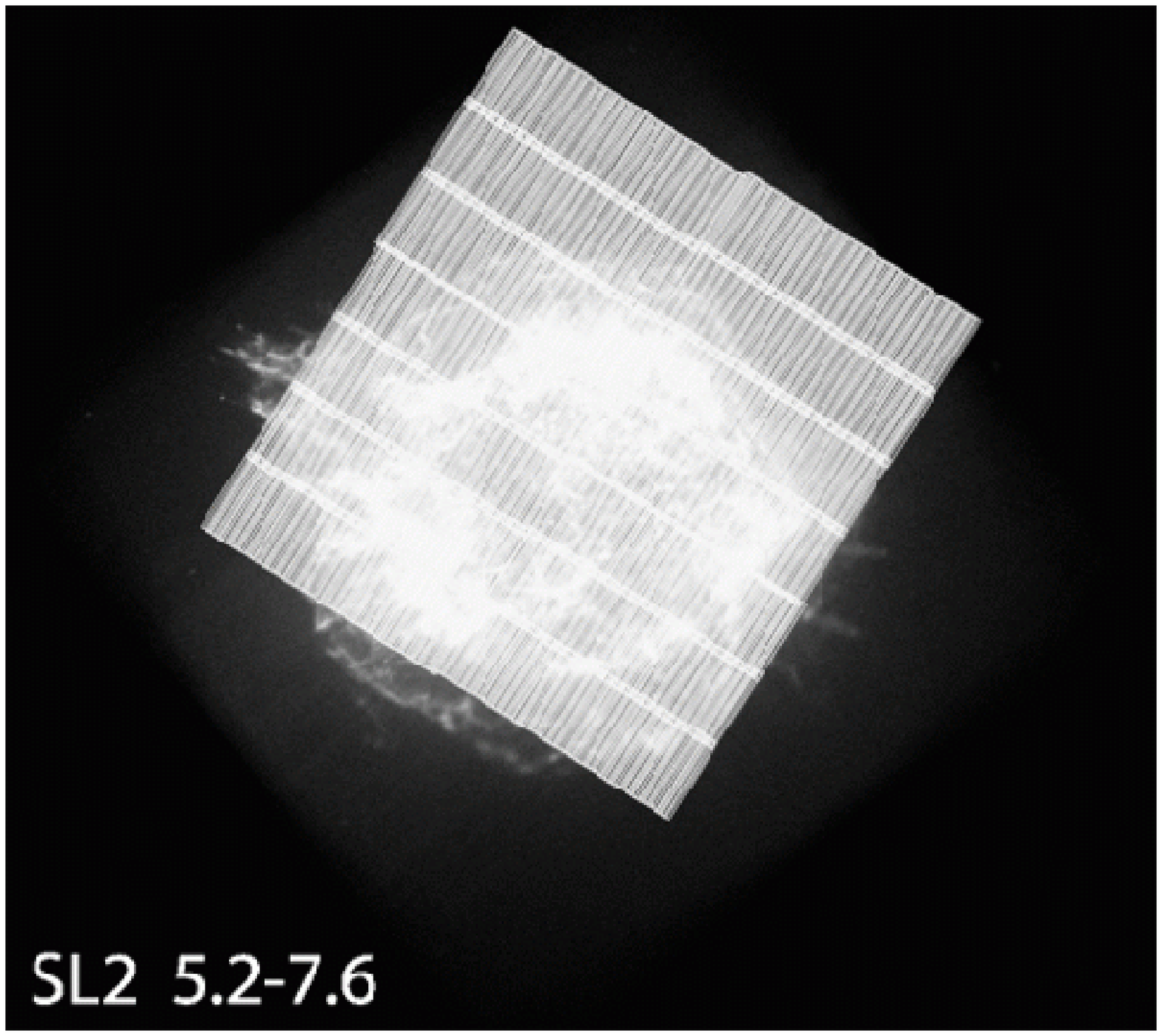}{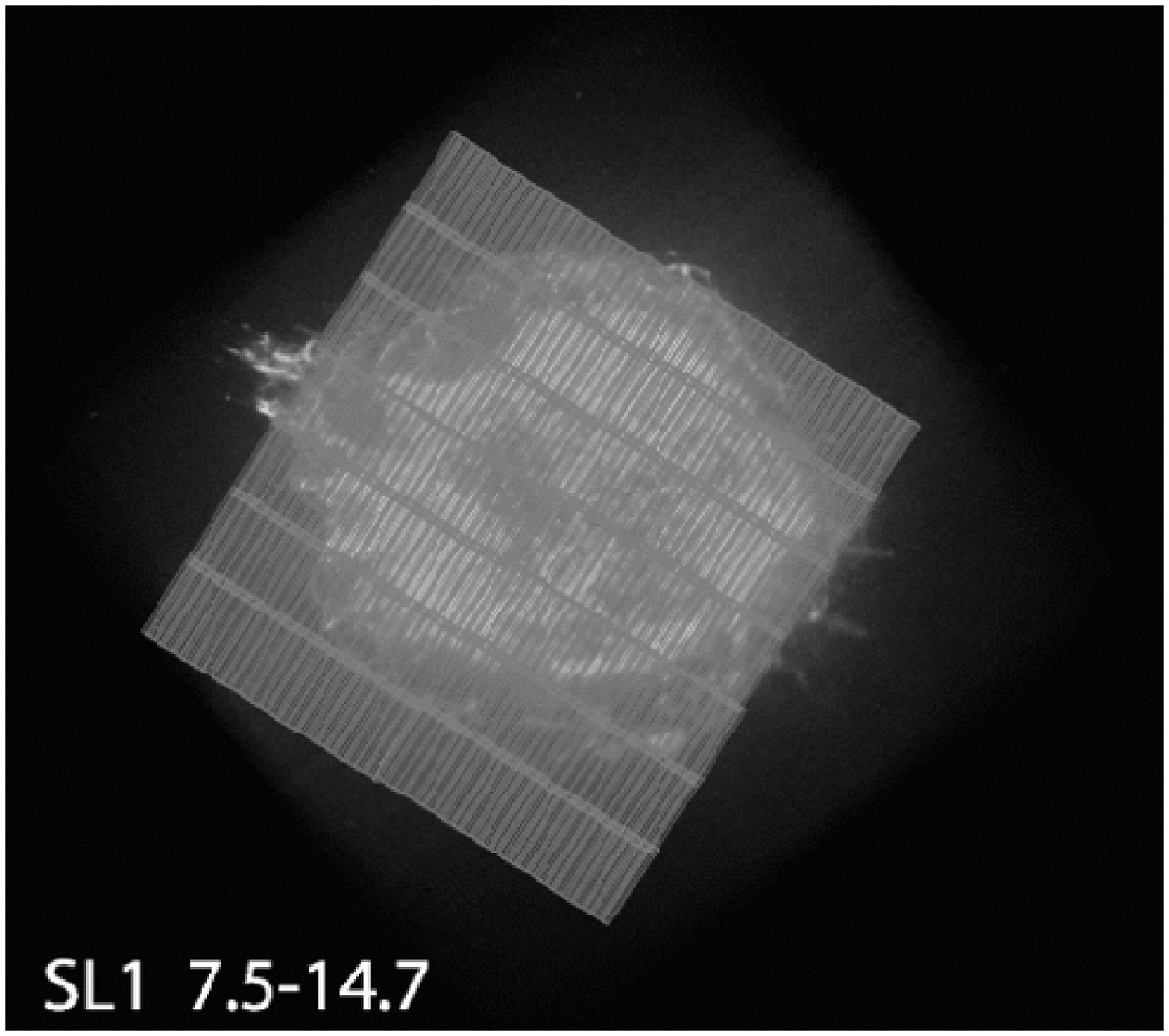}
\plottwo{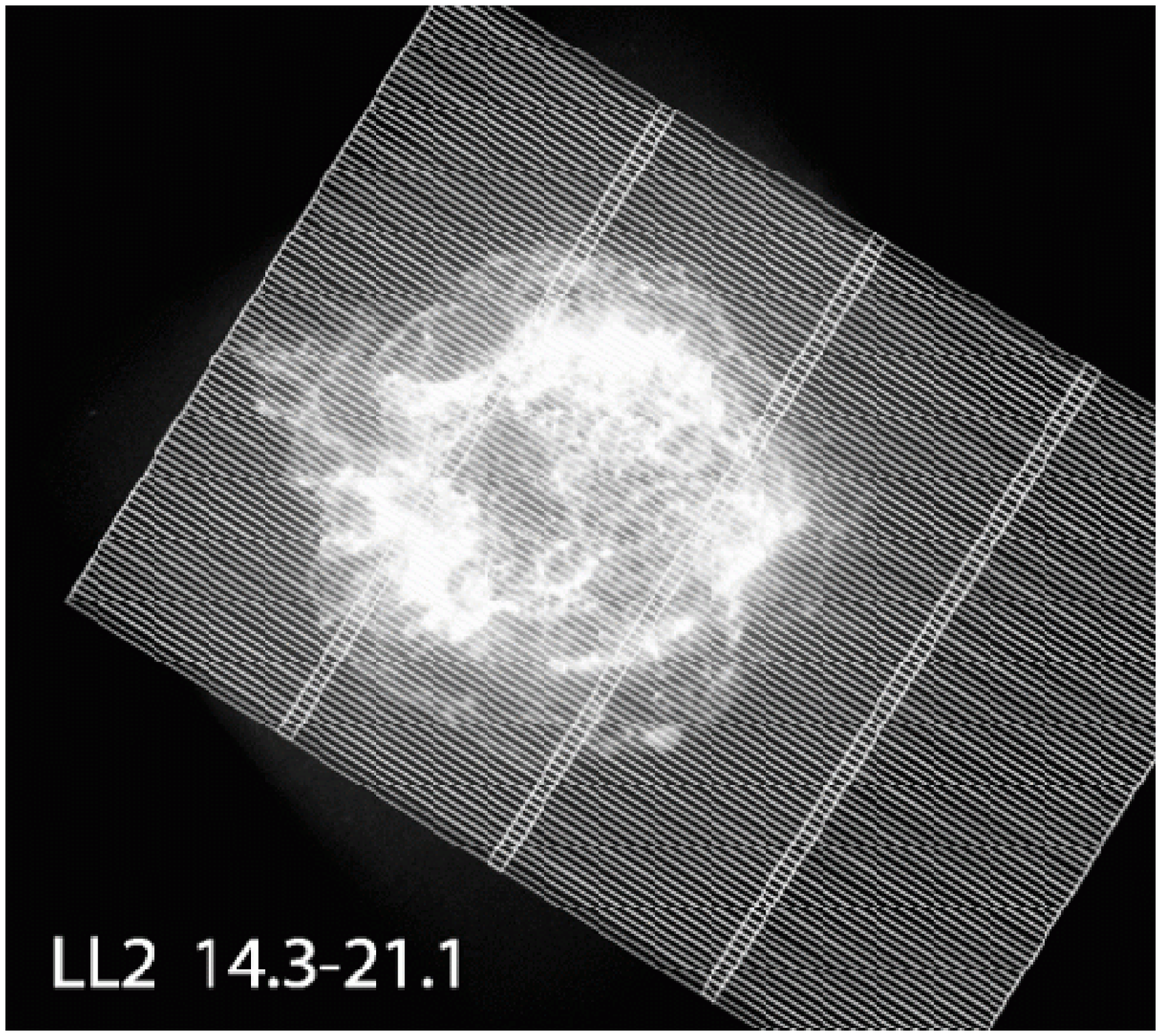}{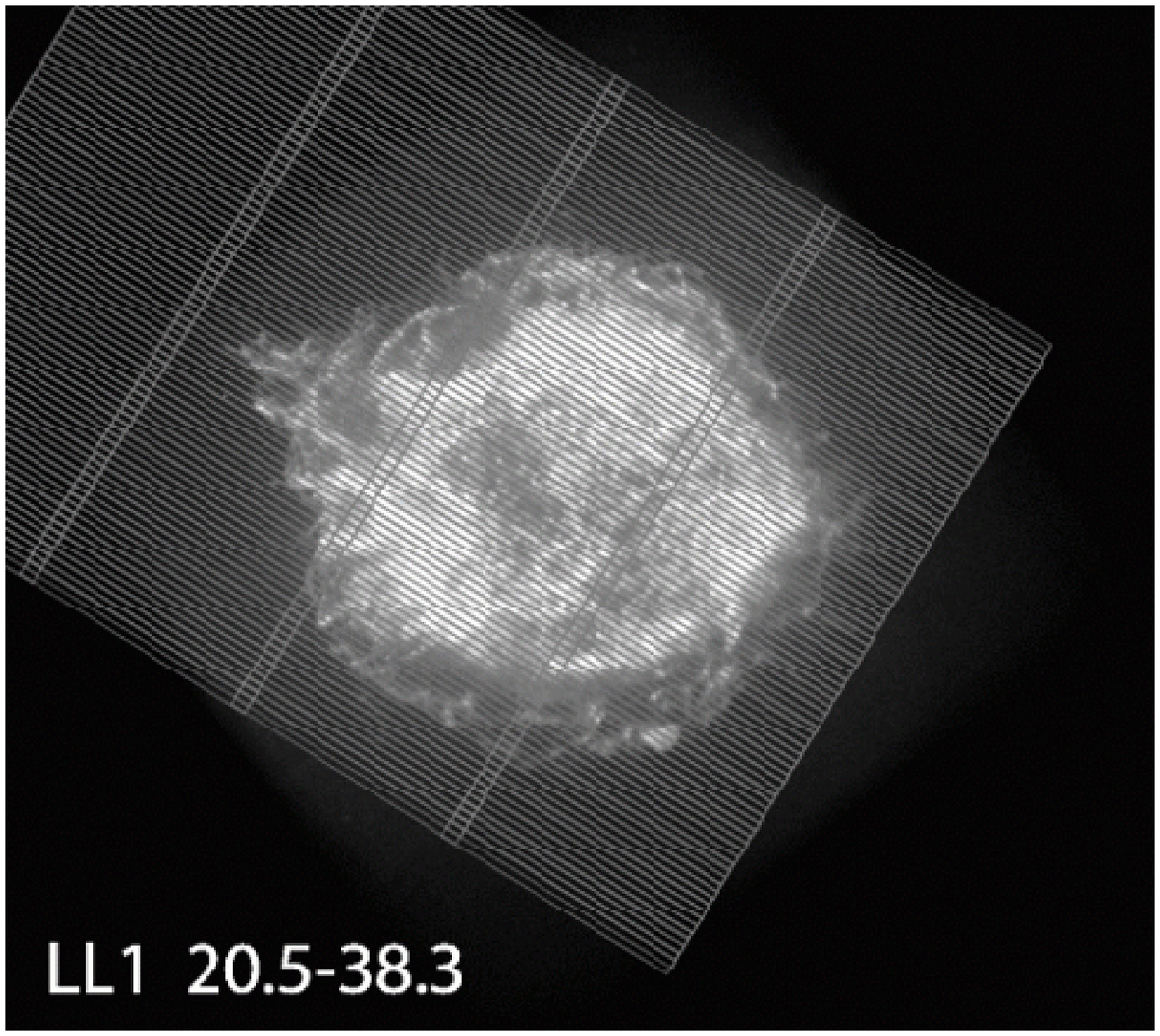}
\end{center}
\caption{The slit positions for the \emph{Spitzer} IRS mapping overlaid on
the \emph{Chandra} broadband X-ray image.  The wavelength ranges in microns
are indicated for each of the short-low (SL) and long-low (LL) orders.
\label{slits}}
\end{figure}

The \emph{Spitzer} Infrared Spectrograph (IRS) was used on 2005 January 13 to
spectrally map nearly the full extent of \casa\ with portions of the
outer structures missing from some slits as shown in Figure~\ref{slits}.
Low-resolution spectra (resolving power of $\sim$60-128) were taken between
5-15~\micron\ (short-low module, SL) and 15-38~\micron\ (long-low module, LL)
with each module including two orders of wavelength.  The long-wavelength
(15-38~\micron) spectra were taken in a single large map with 4$\times$91
pointings, using a single 6 s ramp at each position.  To achieve the spatial
coverage with the short-wavelength (5-15~\micron) slit, a set of four
quadrant maps were made, two with 4$\times$87 pointings and two with
3$\times$87 pointings, using a 6 s ramp at each position.  The mapped area
ranged from 6\farcm3$\times$5\farcm9 (SL) to 11\farcm0$\times$7\farcm8 (LL),
with offsets between the maps produced in each of the two orders in each
module of 3\farcm2 (LL) and 1\farcm3 (SL), along the slit direction.  The
effective overlap coverage of all modules and orders is
4\farcm9$\times$5\farcm8.  The data used here were processed with the S12
version of the IRS pipeline, using the CUBISM package \citep{sad07} to
reconstruct the spectra at each slit position, subtract the sky background,
and create 3-dimensional data cubes as described in \citet{smith09}.  The
statistical errors at each position in the data cube are calculated using
standard error propagation of the BCD-level uncertainty estimates produced
by the IRS pipeline.

\subsection{The Bright Ring and Diffuse Interior Emission}
\label{sec:emlin}

\begin{figure}
\epsscale{1}
\plotone{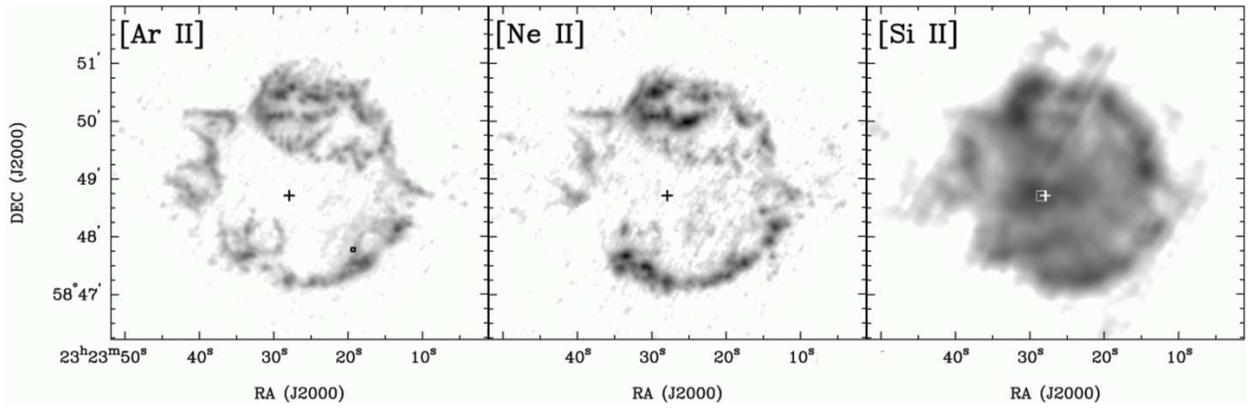}
\caption{Left: 6.99~\micron\ \ar\ image and center: 12.81~\micron\ \neon\
image showing the Bright Ring. Right: 34.8~\micron\ \si\ image showing the
Diffuse Interior Emission.  For reference, the location of the
central compact object is denoted by the crosses.  The \ar\ spectrum in
Figure~\ref{gaussfit} was extracted from the small box region on the
southwest Bright Ring.  The \si\ spectrum in Figure~\ref{gaussfit} was
extracted from the central box near the CCO.
\label{lineimages}}
\end{figure}

There are a number of bright infrared ionic emission lines in \casa\ from
elements such as  Ar, Ne, Si, S, Fe, and the 26~\micron\ blend of Fe and O.
For the most part, images made in these emission lines bear a close
resemblance to optical images that primarily contain S and O emission
\citep{err06}.  In order to characterize the Bright Ring, we have chosen the
6.99~\micron\ \ar\ line and the 12.81~\micron\ \neon\ line.  The \ar\ line is
relatively bright everywhere, even in Ne and O bright regions, and it is
relatively isolated from other emission lines except for very weak
6.63~\micron\ [\ion{Ni}{2}] and possibly 6.72~\micron\ \fetwo, although the
latter transition is a$^4$F$_{9/2}-$a$^6$D$_{7/2}$ which we expect to be
negligible even given the \fetwo\ detected in \casa\ at other infrared
wavelengths \citep{smith09,gf01,rrr03}.  The \neon\ line was used in order to
calculate the \neon/\ar\ ratio, which varies considerably from place-to-place
on the Bright Ring and may indicate inhomogeneous mixing between the
oxygen-burning and carbon-burning layers of the progenitor star
\citep{err06,smith09}.  The disadvantage of the \ar\ and \neon\ lines is that
the SL mapping did not extend out to the full extent of the Jet and
Counterjet. The \ar\ and \neon\ images are shown in Figure~\ref{lineimages}.

In addition to the Bright Ring, there is Diffuse Interior Emission seen in
the 10.51~\micron\ \sfour\ line, the 18.71~\micron\ \sthree\ line,
the 33.48~\micron\ line of \sthree, and the 34.82~\micron\ line of \si\
\citep{rkr08,smith09}.  Bright interior emission is seen at 26~\micron\ as
well.  Since we detect no emission from any of the other infrared Fe lines
(5.35~\micron, 17.9~\micron, etc) in the Diffuse Interior Emission, we
believe that the 26~\micron\ line is \ofour\ with little or no contribution
from \fetwo.  Support for this interpretation is provided by the
high-resolution \emph{Spitzer} spectra in \citet{isensee09}\footnote{Note
that the 26~\micron\ line is bright in the interior \emph{and} on the Bright
Ring.  We now know from the high-resolution \emph{Spitzer} spectra that in
the interior, the 26~\micron\ line is \ofour, however, on the Bright Ring,
there is emission from both \fetwo\ and \ofour\ (K. Isensee, private
communication).}.  The \si\ image, shown in Figure~\ref{lineimages}, shows
both emission associated with the
Bright Ring that is presumably shocked supernova ejecta, as well as emission
in the interior.  The Diffuse Interior Emission bears a striking resemblance
to the free-free absorption observed at 74 MHz \citep{kpd95}.  That absorption
is thought to occur due to cold, photoionized ejecta that have not yet crossed
the reverse shock \citep[e.g.][]{hf88}.  We discuss the physical conditions
in the unshocked ejecta more fully in \S\ref{sec:disk}.

This Diffuse Interior Emission is brightest and covers most of the remnant in
the \ofour\ and in the long wavelength \sthree\ and \si\ lines. We have
chosen to use both the 33.48~\micron\ \sthree\ line and the 34.82~\micron\
\si\ line to map the interior emission.  These lines are not entirely free of
contamination from nearby lines -- there is a small probability of
[\ion{Fe}{3}] at 33.04~\micron\ and a reasonable probability of \fetwo\ at
35.35~\micron.  However, these potentially contaminating lines lie on
opposite sides of the \sthree\ and \si\ lines, making it easier to detect
their presence.  The \ofour\ line is less suitable for mapping the remnant
because, although \ofour\ dominates in the interior, on the Bright Ring there
is the possibility of contamination from \fetwo\ emission.

\subsection{Preview of Velocity Structure: The First Moment Map}
\label{sec:velstruc}

\begin{figure}
\epsscale{0.7}
\plotone{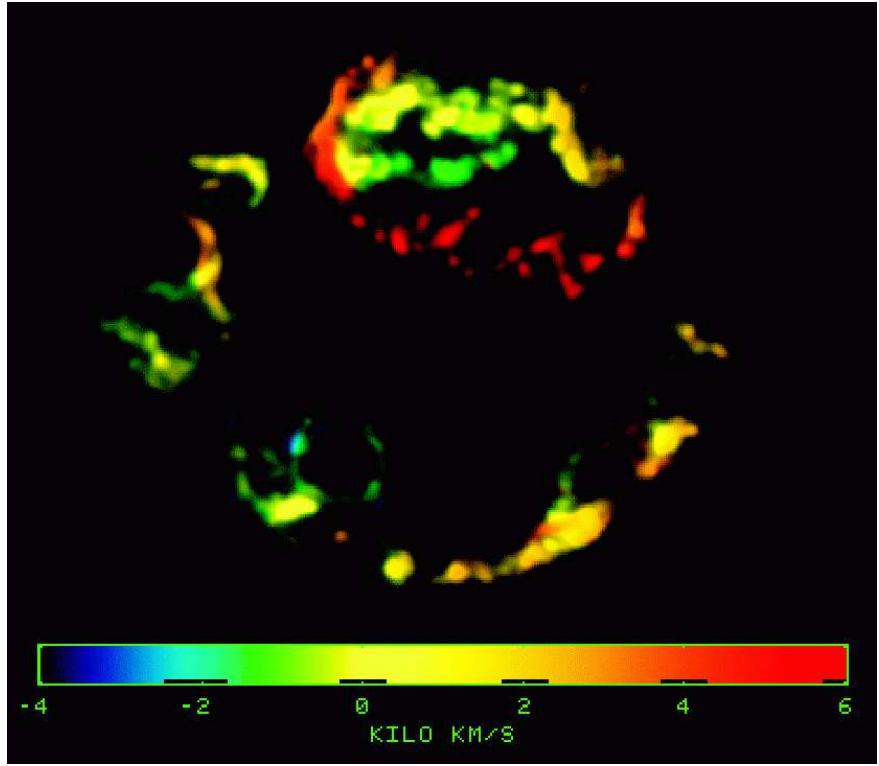}
\caption{1st moment map for \ar\ showing the major Doppler structures
that define the Bright Ring.  Note that in locations with more than one
Doppler component, the color represents a brightness-weighted average
velocity.
\label{ardopimage}}
\end{figure}

Figure~\ref{ardopimage} shows the 1st moment map of the \ar\ line, calculated
over the range -7100 to +10600~\kps.  This image presents a preview of
the Doppler structure of \casa\ to be discussed in more detail below.  The
color scale ranges from $\approx -4000$ to $+6000$~\kps\ from blue to red.
In approximately 1/3 of the locations, there is more than one Doppler
component along the line of sight, so the first moment represents only a
weighted average of those components.  The \ar\ 1st moment map is very
similar to the optical Doppler map of \citet{lmu95} which was reconstructed
from the fitting of spectra.  The double ring structures to the north -- one
red-shifted and one blue-shifted -- are clearly defined as are the
blue-shifted ``parentheses'' southeast of center and the distinct Doppler
structures at the base of the northeast jet.  The velocity range is
the same as found for the optical ejecta \citep{lmu95,rhf95} meaning that the
infrared ejecta have experienced very little or no deceleration and are
effectively in free expansion as \citet{tfv01} note for the optical ejecta.
We do not construct a similar 1st moment map for \si\ because $\approx$3/4 of
the lines of sight contain more than one Doppler component.

\subsection{Gaussian Fitting}
\label{sec:irgfit}

In order to determine velocities for the structures in the remnant,
we fit Gaussian profiles to each selected emission line as described below.
The data cubes were first spatially binned by 2 pixels in the x and y
directions to improve signal-to-noise, resulting in effective resolutions
of 3$\farcs$7 for the \ar\ emission and 10$\farcs$2 for the \si\ emission.
Rather than concentrate on individual knots or clumps of ejecta, we
systematically fit every spatial pixel in our binned cubes that had an
emission line detection at greater than 3$\sigma$.  The spectral resolution
is approximately the same for all of our selected emission lines
($R\approx100$).  The corresponding velocity resolution varies from
2500--3000 \kps.

The Gaussian fitting was carried out using the Interactive Data Language (IDL)
and the routines
MPFIT\footnote{\url{http://cow.physics.wisc.edu/$_\textrm{\~{}}$craigm/idl/fitting.html}}
for the Gaussian component(s) and POLY-FIT to model the local continuum as a
straight line.  We began with an automated routine in which only 1 or 2
Gaussian components were allowed per emission line with no constraint on their
positions with respect to zero velocity -- i.e. they could both be negative,
both positive, or one negative and one positive.  The \sthree\ and \si\ lines
were jointly fit because they were close enough in wavelength that highly
red-shifted S could blend with highly blue-shifted Si.  The automated process
only allowed fits that met the following criteria: 1) Gaussian
full-width-half-max within 500~\kps\ of the spectral resolution of the
line being fit; 2) a signal-to-noise ratio greater than 3; 3) statistical
velocity errors less than 1000~\kps; 4) velocity values less than
$\pm 20,000$~\kps; and 5) a reduced $\chi^2$ value less than 70 for
\sthree\ and \si\ and less than 25 for \ar\ and \neon.  The automated fits
were then examined by hand for accuracy and some fitting was redone manually
to reject ``false positives'' and recover ``false negatives.''

\begin{figure}
\epsscale{0.6}
\plotone{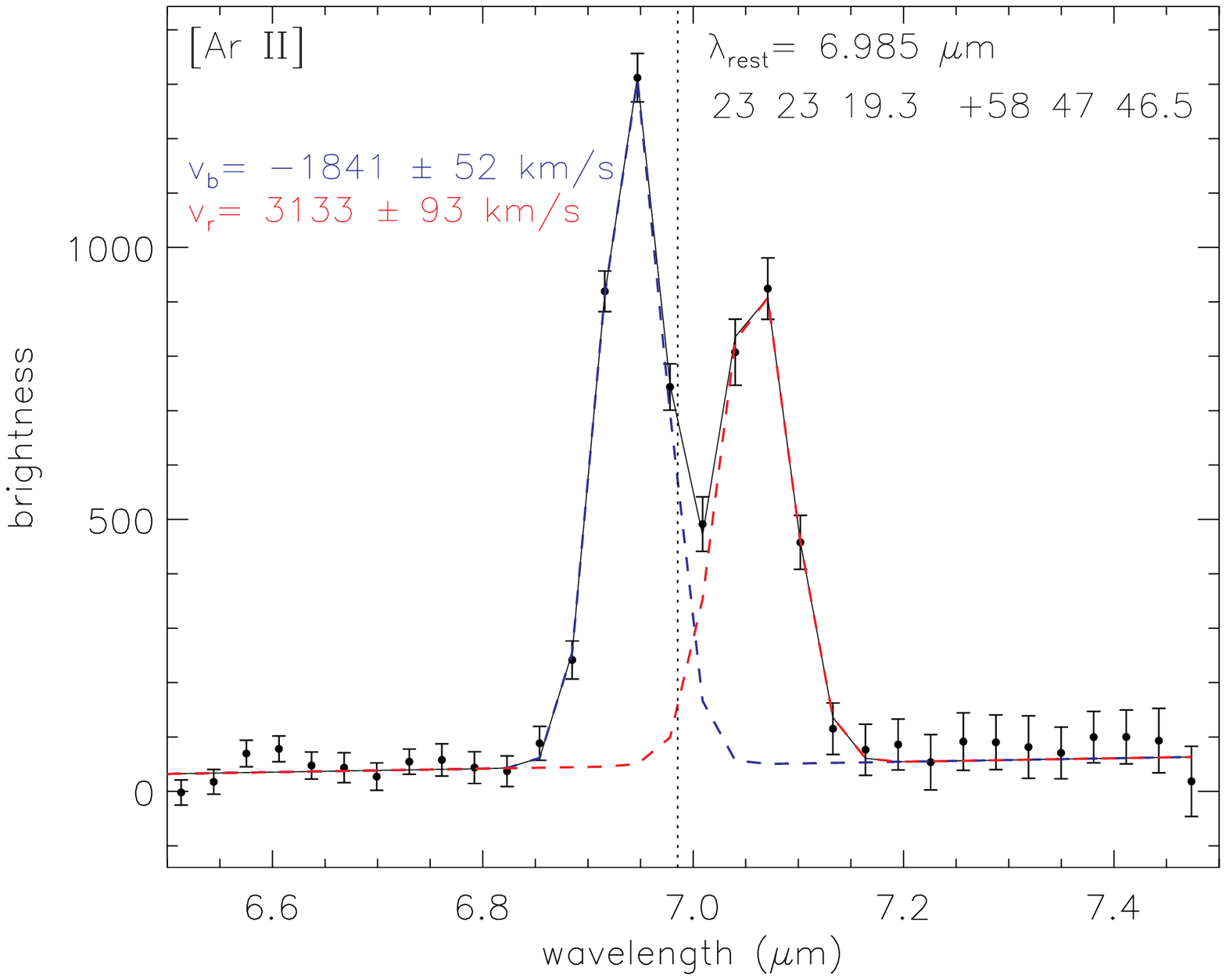}
\plotone{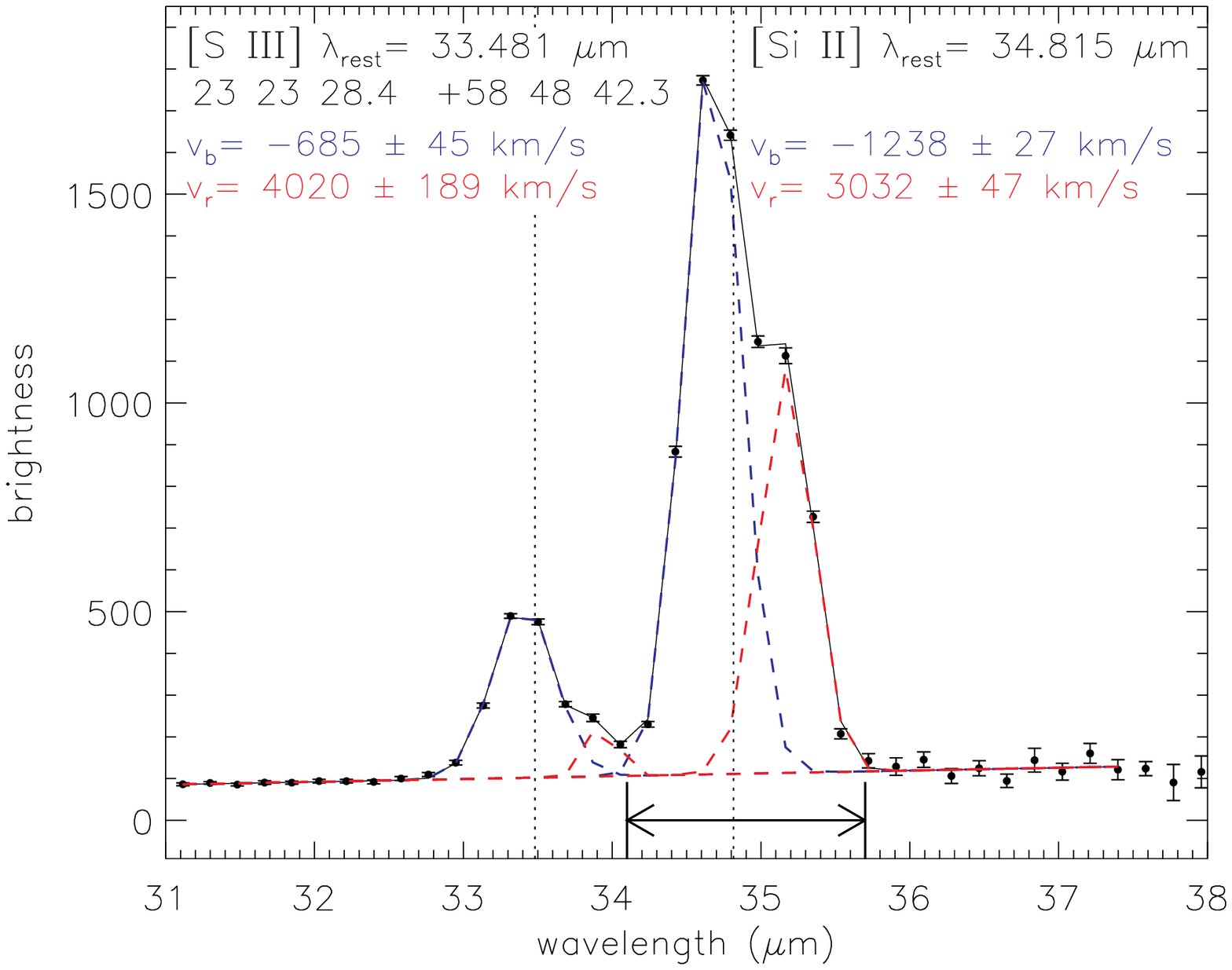}
\caption{Top: Gaussians plus continuum fit (solid line) to the \ar\ emission
in a region on the Bright Ring (see Figure~\ref{lineimages}) with 2
line-of-sight Doppler components (dashed red and blue lines).  Bottom: Joint
Gaussians plus continuum fit to the \sthree\ and \si\ emission in a region
near the center of the remnant (see Figure~\ref{lineimages}), each with 2
line-of-sight Doppler components.  The velocities of the blue ($v_b$) and
red ($v_r$) components are indicated with 1-$\sigma$ statistical errors.
The rest wavelengths of the \ar, \sthree, and \si\ lines are indicated by the
vertical dotted lines.  The wavelength range plotted in Figure~\ref{gausssi}
is indicated at the bottom of the plot.  \label{gaussfit}}
\end{figure}

The top panel of Figure~\ref{gaussfit} shows a typical 2-Gaussian fit to
the \ar\ emission at a location on the Bright Ring with two major velocity
structures superposed along the line-of-sight.  Typical formal velocity
errors from the Gaussian fitting averaged about 100~\kps, however the
actual velocity uncertainties, determined by experiments with different
binning and Gaussian width constraints, are closer to 200~\kps.  For
the weaker \neon\ line, the actual velocity uncertainties were about
400~\kps.  The \ar\ and \neon\ fits were compared for each location
and where their velocities agreed within 1$\sigma$, the ratio between the
fitted Gaussian heights was computed to find the regions where \neon\
is relatively strong.

The bottom panel of Figure~\ref{gaussfit} shows a typical 2-Gaussian fit to
the \sthree\ and \si\ emission in a region near the center.  Actual velocity
uncertainties for the \si\ emission average about 300~\kps\ and for the much
weaker \sthree\ line, the actual velocity uncertainties are about 700~\kps.
The fitted velocities for the corresponding \sthree\ and \si\ components
generally agree within the velocity errors and there is no systematic pattern
in the velocity differences that would indicate strong contamination by
either the 33.04~\micron\ [\ion{Fe}{3}] line or the 35.35~\micron\ \fetwo\
line.

\begin{figure}
\epsscale{1}
\plotone{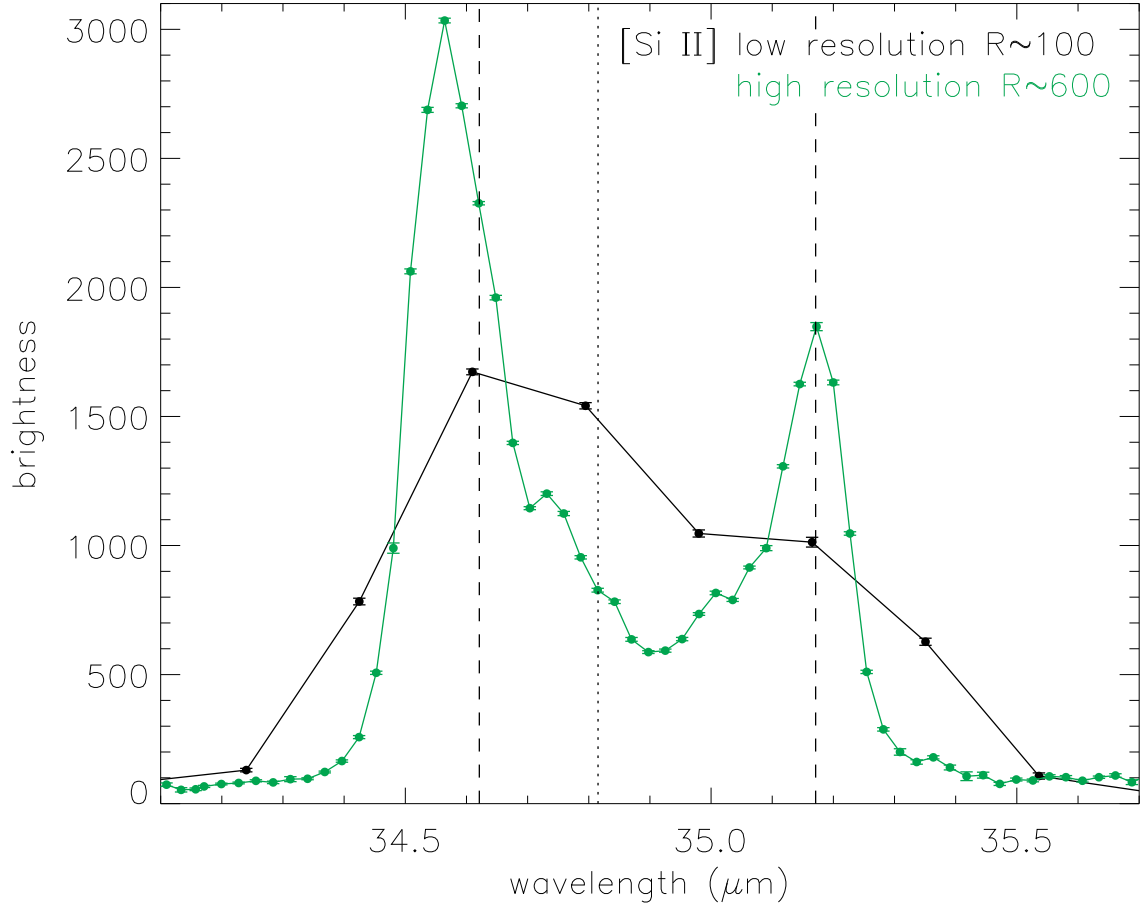}
\caption{High-resolution \si\ spectrum (green) from same region
as the right panel of Figure~\ref{gaussfit} showing two primary Doppler
components with smaller velocity structures extending to the center of the
remnant.  The low-resolution spectrum is plotted in black for comparison and
the low-resolution line centroids are indicated by the vertical dashed lines.
The vertical dotted line denotes the rest wavelength of \si.  \label{gausssi}}
\end{figure}

All of the final accepted fits are well described by either 1 or 2
line-of-sight Gaussian components.  However, we know from the optical images
of \casa\ that there is a great deal of structure on small scales with typical
knot sizes between 0$\farcs$2 and 0$\farcs$4 \citep{fmc01}.  Our
high-resolution infrared spectra taken with \emph{Spitzer's} Short-High and
Long-High IRS modules show that the spatial substructure exhibits velocity
substructure as well.  \citet{isensee09} show that small regions of the
Bright Ring and Diffuse Interior Emission have many velocity components with
total Doppler ranges of as much as 2000~\kps.  At angular resolutions
of 4$\arcsec$--10$\arcsec$ and spectral resolutions from 2500~\kps\ to
3000~\kps\ we are insensitive to the small-scale substructure.  However, based
on optical Doppler maps \citep{lmu95} and our own Doppler map in
Figure~\ref{ardopimage}, major structures are separated by many thousands of
\kps, so  even at low spatial and spectral resolution we are mapping the
gross dynamic behavior in \casa.  To illustrate this point, we show in
Figure~\ref{gausssi} the high-resolution ($R=600$, velocity resolution =
500 \kps) spectrum of the \si\ emission displayed in the right panel of
Figure~\ref{gaussfit}.  The high-resolution spectrum is, indeed, dominated
by two major velocity components, but there is also weaker emission extending
into the center of \casa.  The vertical dashed lines indicate the Gaussian
centers of the low-resolution fits and demonstrate the systematic
uncertainties of averaging over multiple velocity components.  A more
detailed comparison between the low- and high-resolution spectra near the
center of \casa\ is presented in \citet{isensee09}.

\section{\emph{Chandra} Data Analysis}
\label{sec:xrayanal}

A spatially resolved spectral analysis was carried out using the nine
individual ACIS pointings from the million-second observation of \casa\ in
2004 \citep{hlb04} and two additional 50-ks ACIS observations from 2000
\citep{hhp00} and 2002 \citep{dr03}.  The initial analysis of the
\emph{Chandra} ACIS data is described fully in \citet{dha05} and
\citet{sah06}.  Briefly, standard processing and filtering were
applied to the eleven individual pointings and the data were re-projected
onto a common sky-coordinate plane, maintaining separate files for each
pointing.  A grid was created with 1$\arcsec$ spacing, approximately
450$\arcsec$ by 450$\arcsec$, centered on \casa.  Regions in this grid were
adaptively sized to contain at least 10,000 counts and ranged from
$1\arcsec \times 1\arcsec$ to $7\arcsec \times 7\arcsec$.  The regions were
allowed to overlap so that fits to regions larger than
$1\arcsec \times 1\arcsec$ are not independent.

One spectral model for each sky region was jointly fitted to the set of eleven
spectra using the Interactive Spectral Interpretation System
\citep[ISIS;][]{hd00}.  Each spectrum was associated with its own effective
exposure, response functions, and the background spectrum from the same
pointing.  The background was not subtracted from the source spectrum, but
was instead added to the model.  The background is dominated by photons from
\casa\ in the wings of the point-spread function of the telescope and in the
CCD readout streak.  The background for every source spectrum was drawn from
the same representative sky region.

The model consisted of fifteen Gaussian components for the He-like K$\alpha$
lines of O, Ne, Mg, Si, S, Ar, Ca, Ti, and Fe, the He-like K$\beta$ lines of
Si and S, the H-like K$\alpha$ line of Si, and L lines for multiple Fe ions.
The model included interstellar absorption column density and used a
bremsstrahlung continuum.  We chose to use Gaussian+continuum fits rather
than non-equilibrium ionization (NEI) models for several reasons.  First, our
primary purpose was to determine the strength and position (shift) of the
line emission, not to model the emission history.  Second, the available
NEI models, while allowing variable element abundances, use a single redshift
for the plasma and they do not account for the dynamical evolution of the
shocked gas.  While we could have followed the examples of \citet{lh03},
\citet{hl03}, and \citet{hl09} by tracing the shock evolution along with
the NEI evolution, we would still have had to model multiple plasmas along
the line-of-sight to decouple for instance Si-dominated ejecta from
overlapping and dynamically distinct Fe-dominated ejecta.  Multiple-component
plasma models often have essentially unconstrained line fits which can lead
to artifacts that plague automated fitting routines such as the linear
structures found in the analysis of \citet{ylc08}.  Using Gaussians
essentially unties the elements to allow for multiple Doppler shifts and
allows for model-independent Doppler shift determinations.  The joint
best-fit Gaussian component centers were recorded for each line in each sky
region and used to create FITS images of the fitted line centers for each
line.

Given that the Si He$\alpha$ line is the brightest emission line in 
the X-ray data, it is the natural choice for Doppler analysis.  However, we 
found significant variations between the velocities derived from the ACIS data
and the velocities derived from the 2001 HETG observation for the 17 regions
measured by \citet{lds06} at high spectral resolution.  As shown in Figure 
\ref{hetgacis}, there were large velocity variations 
($\approx1500$ km s$^{-1}$) between the data sets as well as both a 
systematic velocity scaling and offset.  The ACIS data are plotted using a 
rest wavelength of 6.648{\AA} corresponding to the \xsi\ He$\alpha$ 
resonance line that should dominate the spectrum (note that the forbidden line 
is at 6.740{\AA} so that the expected weighted average rest wavelength should 
be between the resonance and forbidden lines).  In order for the ACIS data to 
match the HETG data, a rest wavelength of 6.6169{\AA} is required and the 
ACIS velocities must be scaled by 1.67.  The velocity discrepancies between 
the HETG and ACIS data might possibly be the result of ACIS energy 
calibration issues near the \xsi\ line (G. Allen, private communication), 
although a follow-up analysis is required to confirm this hypothesis.  We 
will revisit the ACIS Si data in \S \ref{sec:vvsr} to further demonstrate the 
unsuitability of these data for our study.  We did not pursue the analysis of 
the S, Ar, or Ca emission lines from the Si group of elements because they 
would likely have the same velocity issues.  Since the X-ray Si-group emission 
traces out the same Bright Ring structures as the infrared \ar\ emission 
\citep{err06} and the gross Doppler structure of the X-ray Si emission is the 
same as for the optical emission \citep{hsp01,wbv02,sah04,dha05}, we feel that 
the Bright Ring is well-sampled without the ACIS Si (or Si-group) data.  

We therefore chose to focus on the \fe\ line
as a complement to the infrared lines.  It is distributed differently than
the Si-group and O/Ne emission \citep{hlb04} and is a separate dynamical
component from the other X-ray ejecta \citep{wbv02,wbv03}.  The \fe\ is
preferable to the three Fe-L lines from our Gaussian fitting because it is
not contaminated by nearby Ne or O lines nor any other lines in the Fe-L
forest.  The energy calibration near the 6.6 keV \fe\ line should also be
quite accurate due to the 5.9 keV and 6.4 keV Mn-K emission from the onboard
energy calibration sources mounted in the Science Instrument Module on
\emph{Chandra}.

The left panel of Figure~\ref{fekim} shows an image of the \fe\ emission
in \casa\ made using spectral tomography as described in \citet{drf04}.  The
spectral tomography technique is designed to separate overlapping spectral
structures and to spatially visualize the spectral components.  The technique
involves taking differences between images from two different energies with
a scale factor chosen to accentuate features of interest:
$M(f) \equiv M_{(6-7)} - f \times M_{(4-6)}$, where $f$ is the scale factor and $M(f)$ is the residual image at that scale factor.  The two input images to
the spectral tomography analysis were a 4-6 keV image, $M_{(4-6)}$, to model the continuum emission (both thermal and non-thermal), and a 6-7 keV image, $M_{(6-7)}$, that contains emission from the \fe\ line as well as continuum emission either from thermal or non-thermal processes.  We chose to scale the 4-6 keV image by 1/6 prior to subtraction based on the histogram of $\log(M_{(4-6)}/M_{(6-7)})$, which is Gaussian-shaped with a tail at high ratios.  The specific choice of scale factor, $f=1/6$, was made to isolate regions where \fe\ emission dominates (low values of $\log(M_{(4-6)}/M_{(6-7)})$) from other regions.  This difference is shown in the left panel of Figure~\ref{fekim} ($\equiv M(1/6)$) which has been smoothed to 3$\arcsec$ resolution.  Regions that are bright (have positive residuals) in Figure~\ref{fekim} represent ejecta with bright \fe\ emission.  Regions that
are dark (have negative residuals) in Figure~\ref{fekim} represent features
where the continuum is bright but there is no or weak \fe\ emission.  Note
that the negative residuals occur in regions with strong non-thermal emission as well as Si-rich (thermal) regions that are relatively Fe-poor \citep[e.g][]{drf04}. Thus, the tomography technique has separated \fe-rich regions from \fe-poor regions.  The \fe-rich emission is localized to three regions -- the southeast, the north, and the west as noted by e.g. \citet{hlb04}.  The right
panel of Figure~\ref{fekim} shows the fitted line centers for the bright \fe\
regions \citep{sah04}.  The velocity pattern matches the Doppler image of
\citet{wbv02} almost exactly with a red-shifted lobe of emission to the north
and a blue-shifted lobe of emission to the southeast.  The approximate
velocity range in the image is $\pm4000$~\kps\ without accounting for
ionization effects.

\citet{hl03} and \citet{lh03} show that the ionization age of the X-ray
plasma varies between $\approx 0.2-12\times10^{11}$ s cm$^{-3}$ from
place to place in \casa.  The centroid of the \fe\ line  shifts blueward
at higher ionization age because the ``line'' is actually a line complex
dominated by the He$\alpha$ forbidden, intercombination, and resonance
line triplet as well as the Ly$\alpha$ line.  At a higher ionization age, the
Ly$\alpha$ line becomes brighter relative to the He$\alpha$ triplet and so
the ``blended'' line shifts to higher energies simulating a blue shift.  Since
we only fit a single Gaussian to the \fe\ line complex, our centroids are
affected by ionization as well as Doppler shift.  We demonstrate the
importance of this ``ionization effect'' in Figure~\ref{ion} where we show
spectra between 6-9 keV (containing \fe\ and Ni-K lines) extracted from the
million-second dataset for regions A12 and A13 of \citet{hl03}.  These two
regions are identified in the left panel of Figure~\ref{fekim} and have
ionization ages of $8\times10^{11}$ and $1\times10^{11}$ s cm$^{-3}$,
respectively \citep{hl03}.  A wavelength shift of only 0.017{\AA}
(0.0635 keV, the separation between the \fe\ He$\alpha$ forbidden and
resonance lines) results in a velocity shift of 2746~\kps.  In
\S~\ref{sec:fedist} we correct for ionization effects in the southeast Fe
lobe by artificially collimating that structure.  We make no corrections for
ionization anywhere else.

\section{Reconstructing the 3-D Distribution}

The Doppler velocities and sky positions of the infrared and X-ray data can
be used to reconstruct their three-dimensional positions.  In order to convert
Doppler velocity to equivalent line-of-sight spatial position we must assume
that, prior to encountering the reverse shock, the ejecta were freely
expanding from a single point in time, so that distance~$\propto$~velocity.
This behavior is expected for a point explosion in space, however if
distance~$\not \propto$~velocity, then this project is not feasible because
there would be no way to map velocity into distance.  We must also choose a
geometry, and for this we adopt the simplest physical picture, a spherical
shell, consistent with the optical \citep{rhf95} and infrared data on the
Bright Ring.  We do not pursue more complex models involving ellipticity or
disjoint hemispheres because \citet{rhf95}, using better spatial and spectral
resolution, did not find compelling evidence that those sorts of models
better described their optical data.  In all cases, we measure projected
radius on the sky with respect to the optical ejecta expansion center derived
by \citet{tfv01}. This location is $\approx7\arcsec$ north of the central
compact object and we note that it is not at the geometrical center of the
Bright Ring. \citet{rhf95} considered several different centers in their
analysis and found that it had minimal impact on their results.

\subsection{Converting Doppler Velocity to Line-of-Sight Distance}
\label{sec:vvsr}

In order to convert Doppler velocity to line-of-sight distance independently
from the optical determinations, we plot Doppler velocity \emph{vs.}
projected radius on the sky as shown for the \ar\ (red) and \si\ (grey) data
in Figure~\ref{vvsr}.  The error boxes at the bottom of the figure indicate
the spatial resolution and the mean Doppler velocity errors for the two
populations.  The \ar\ data are concentrated onto a relatively thin,
semi-circular shell with about a 15\% scatter in thickness.  The Doppler
velocities range from $\approx$ -4000 to 6000~\kps.  The
velocity distribution is not centered at 0~\kps, but is offset towards
positive velocities.  The velocity range and offset and the detailed
structure of the \ar\ distribution are virtually the same as the optical data
in Figure~3 of \citet{rhf95}.

The \ar\ emitting regions can all be fit on the same spherical shell,
consistent with the scenario where ejecta become visible after being heated
by the reverse shock and later fade from view with observed lifetimes
$\sim$30 years \citep{vdb83}.  Since the shell is narrow,
only 1000-2000~\kps\ thick, we can relate Doppler velocity to spatial
distance by fitting a spherical expansion model.  In velocity \emph{vs.}
radius space, the model is a semicircle which we have chosen to parameterize
in terms of the center of the velocity distribution, $v_c$, the minimum
velocity at which the semicircle crosses the velocity axis, $v_m$, and a
scale factor, $S$, that relates the velocity axis to the spatial axis:
\begin{displaymath}(r_p/S)^2 + (v_D - v_c)^2 = (v_c - v_m)^2,\end{displaymath}
where $r_p$ is the observed projected radius and $v_D$ is the observed
Doppler velocity.  The scale factor, $S$, is used to directly convert Doppler
velocity into distance in or out of the plane of the sky.  Using a
least-squares fit to the data, we find $v_c=859\pm100$~\kps,
$v_m=-4077\pm200$~\kps, and $S=0\farcs022\pm0\farcs002$~per~\kps.  Our values
agree with the \citet{rhf95} values of $v_{c_{\mathrm{Reed}}}=770\pm40$~\kps,
$v_{m_{\mathrm{Reed}}}=-4520\pm100$~\kps, and
$S_{\mathrm{Reed}}=0\farcs019\pm0\farcs001$~per~\kps, although we do expect these
values to evolve over time as the remnant evolves and the reverse shock
encounters slower and slower ejecta.  Our best-fit model is shown as a red
semicircle in Figure~\ref{vvsr}.  We note that the adoption of a different
value of $v_c$ would not change the geometry of the remnant, but would result
in a translation of the entire remnant along the line of sight.

Also shown in Figure~\ref{vvsr} is a fiducial semi-circle representing the
location of the reverse shock in the velocity coordinates of the \ar\
ejecta.  The projected radius of the reverse shock is taken from
\citet{gkr01} and is plotted to be concentric to the spherical model that
describes the ejecta.  The reverse shock location was identified based on
X-ray Si emission and indicates the location where ejecta ``turn on'' after
being shock heated.  A spatial analysis of X-ray Si emission and infrared Si
and S emission determined that the infrared and X-ray ejecta ``turn on'' at
the same radius indicating that the timescales for shock heating up to
infrared or X-ray emitting temperatures are short \citep{smith09}.  One
might expect to see a progression from infrared/optical emission to X-ray
emission as ejecta ionize up to He/H-like states, and this may happen for
individual knots, but some ejecta knots ionize so quickly to X-ray emitting
temperatures that the inside edge of the Bright Ring as defined by
optical/infrared emission is the same as for the X-ray emission.  Although
ejecta knots continue moving downstream from the reverse shock, new ejecta
continuously encounter the reverse shock and turn on thus maintaining a
demarcation line at the reverse shock location \citep{mfc04,pf07}.

We identify the forward shock by a small line segment
at a radius of 153$\arcsec$.  Since the forward shock and the Bright Ring do
not share the same projected geometric center \citep{gkr01}, we make no
assertion as to any offset between these structures along the velocity axis.
Note that the velocities at which the reverse shock model crosses the
velocity axis do not represent the actual space velocity of that shock, which
is not constrained by the current analysis.  Rather, it represents the
velocity that an undecelerated ejecta knot would have at the spatial position
of that shock at the epoch of observation, assuming our distance-to-velocity
scale factor.

Some of the \si\ data are found on the same shell as the \ar, but most of the
\si\ is found interior to the fiducial reverse shock.  Since the \si\ do not
form a well-defined shell in Figure~\ref{vvsr} we cannot independently fit
for its scale factor between Doppler velocity and spatial distance.
Henceforth we will treat the \si\ as though it were in free expansion with
the \ar.  We justify our free expansion assumption and discuss the \si\
distribution more fully in \S\ref{sec:disk}.

In order to plot Doppler velocity \emph{vs.} projected radius for the X-ray
\fe\ data, we must first convert the fitted line centers to Doppler
velocities. This is not just a simple mathematical calculation because the
\fe\ ``line'' is composed of the unresolved He$\alpha$ forbidden,
intercombination, and resonance triplet of lines, the barely resolved
Ly$\alpha$ line, and, depending on the ionization age, lines from lower
ionization \fe\ species.  Because we modeled the \fe\ line complex with a
single Gaussian component, we must determine what average rest wavelength to
use for the velocity calculations and we must account for ionization
variations that will change the brightness ratios of these lines and mimic
changes in Doppler velocity.

One way to approach this problem is to assume that the spherical expansion
model derived from the infrared ejecta also applies to the X-ray ejecta.
However, we know from proper motion measurements that the X-ray ejecta, on
average, are decelerated, with an expansion rate of 0.208~\%~yr$^{-1}$
\citep{dr03}.  We can use the ratio between the X-ray expansion rate and the
free expansion rate \citep[0.304~\%~yr$^{-1}$;][]{tfv01} to scale the
spherical expansion model parameters into the decelerated velocity space
appropriate to the X-ray ejecta.  These scaled parameters are:
$S_x=0\farcs032$~per~\kps, $v_{c_x}=589$~\kps, and $v_{m_x}=-2806$~\kps.
Note that in the reference frame of the decelerated ejecta, the model
$v_{max}$ and $v_{min}$ simply scales from the free expansion values.  The
center of the velocity distribution, $v_{c_x}$, reflects the midpoint between
$v_{max}$ and $v_{min}$ in the decelerated space and is also a simple scaling
of the model.  Although the average expansion rate of the X-ray ejecta is
0.208~\%~yr$^{-1}$, the X-ray ejecta have experienced a range of
decelerations with a 1$\sigma$ variation of 0.08~\%~yr$^{-1}$ \citep{drf04}.
The resulting effective velocity error from using the wrong
distance-to-velocity scale factor is less than 1300~\kps\ in most cases, which
is $<20$\% of the total spread in X-ray velocities.  In the case of the \fe\
ejecta, the deceleration error is small ($<1/2$) compared to the ionization
error, but in the case of the \xsi\ ejecta, the deceleration error is the
dominant error.  

The distance-to-velocity scaling for the X-ray ejecta is
appropriate under the condition that the X-ray emitting material was in free 
expansion until encountering the reverse shock and the mass wall of swept-up 
ejecta immediately downstream.  The subsequent deceleration and heating of the 
ejecta leads to the X-ray emission.  The X-ray ejecta are then assumed to 
continue expanding at a slower, but constant, velocity.  In reality, the 
ejecta should continuously decelerate as they move downstream from the 
reverse shock.  However, the expansion rate of the X-ray ejecta is roughly 
constant with radius according to the proper motion data of \citet{drf04}.  
The errors introduced by the ``constant velocity'' assumption are dwarfed by 
the inherent variation in deceleration experienced by the ejecta at the 
reverse shock as discussed above.  This is not unexpected based on 
hydrodynamic models.  For example, in the snapshot in time shown in Figure~1 
of \citet{tm99}, the difference in ejecta velocity between the reverse 
shock and contact discontinuity is only a few percent.  This expected 
difference in velocity is much much smaller than the variation observed in the 
actual data \citep{drf04}.  Thus, the assumption of constant velocity motion 
after deceleration at the reverse shock is appropriate, to first order, for 
the X-ray ejecta.  The X-ray-emitting ejecta on the Bright Ring are therefore 
cospatial with the optically- and infrared-emitting ejecta, but are 
decelerated with respect to those ejecta and consequently require a different 
scaling factor to convert Doppler velocity to distance.  

Figure~\ref{vvsrxray} shows the Doppler velocity \emph{vs.} projected radius
for the \fe\ data with the scaled model plotted in red.  The \fe\ data are
plotted using a rest wavelength of 1.8615{\AA} which provides the best fit
(by eye) to the spherical model.  The \fe\ error box at the bottom of the
figure is an indication of the degree of uncertainty due to ionization
differences.  The fiducial reverse shock is indicated by the black semicircle
and has been mapped into the X-ray ejecta velocity space the same way as for
the infrared data in Figure~\ref{vvsr}, but now the velocities represent the
speed of decelerated ejecta at the reverse shock location.

The distribution of the \fe\ emission in Figure~\ref{vvsrxray} is very
interesting.  The northern, red-shifted emission appears to form a partial
spherical shell similar to the infrared \ar\ emission.  The thickness of
this shell is about 20-30$\arcsec$, which is nearly as thin as the infrared
\ar\ shell.  The blue-shifted \fe\ ejecta in the southeast of the remnant are
distributed quite differently.  Rather than a thin, shell-like structure, the
blue-shifted ejecta extend as a column, approximately 70$\arcsec$ long, and
at an angle of $\approx27\degr$ out of the plane of the sky.

We also show in Figure~\ref{vvsrxray} the X-ray HETG \xsi\ data from
\citet{lds06}.  With only 17 small regions, the \xsi\ data sparsely
populate the Bright Ring shell.  However, since the velocities of the \xsi\
data are determined through dispersed HETG spectra, they are completely
independent of the ACIS \fe\ data.  Therefore the HETG \xsi\ data provide
confirmation that the rest wavelength used to calculate the \fe\ Doppler
velocities is correct.

\begin{figure}
\epsscale{1}
\plotone{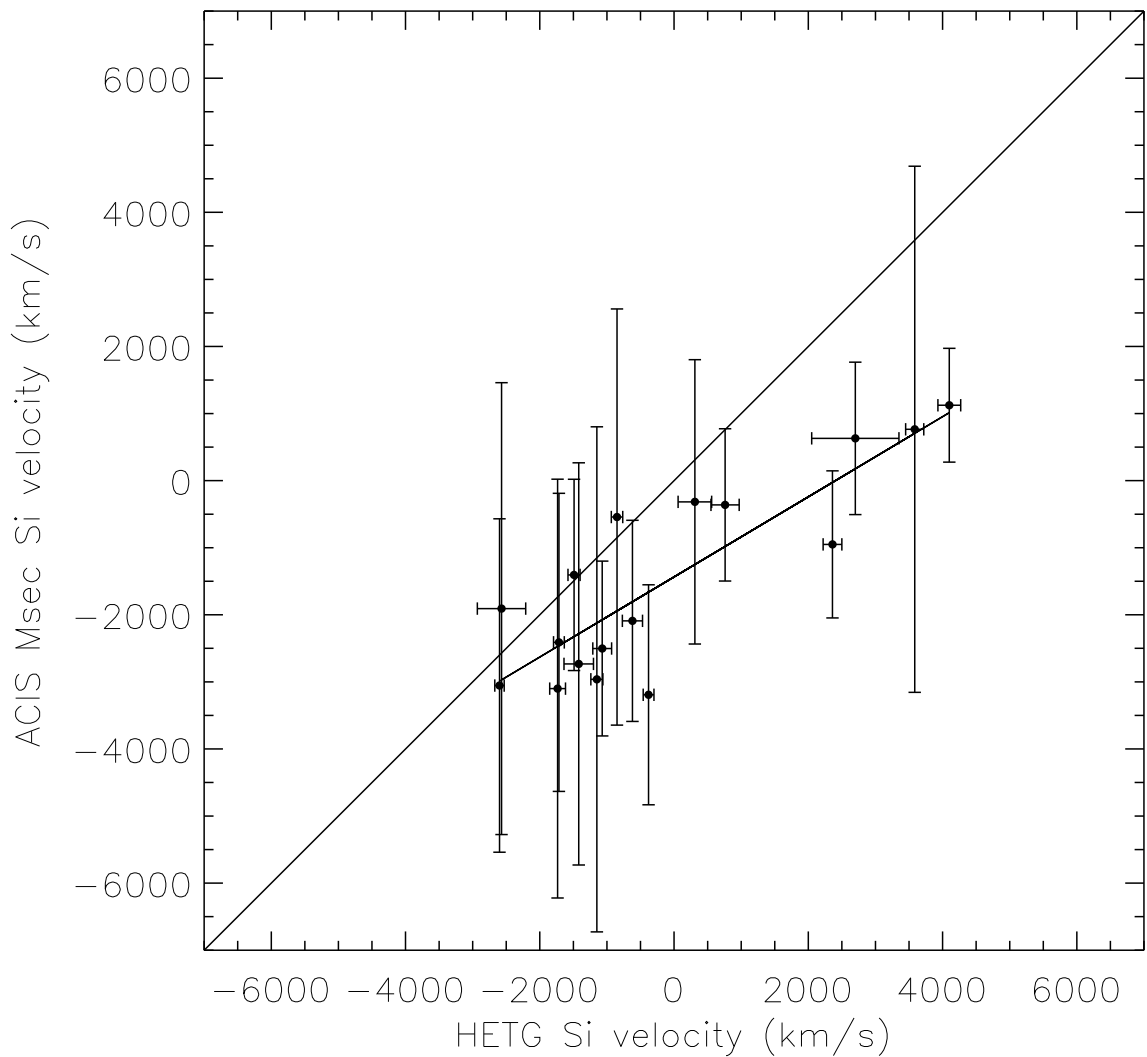}
\caption{HETG-derived Si Doppler velocity compared to ACIS-derived Si Doppler 
velocity for the sample of 17 ejecta knots measured by \citet{lds06}.  The 
best-fit straight line to the data is shown and indicates both a velocity 
scaling and offset between the two data sets.  The rest wavelength used to 
compute the ACIS velocities was 6.648{\AA}.
\label{hetgacis}}
\end{figure}

\begin{figure}
\epsscale{0.44}
\plotone{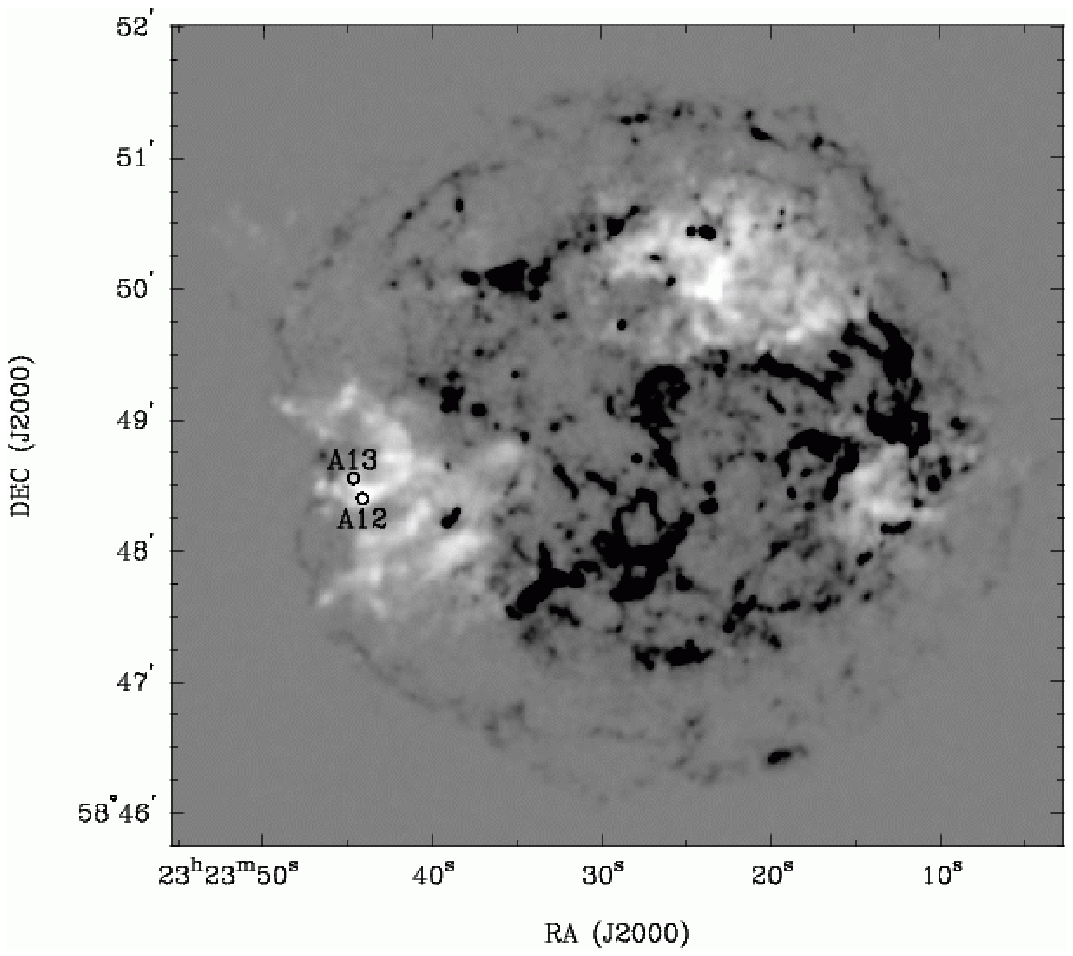}
\epsscale{0.5}
\plotone{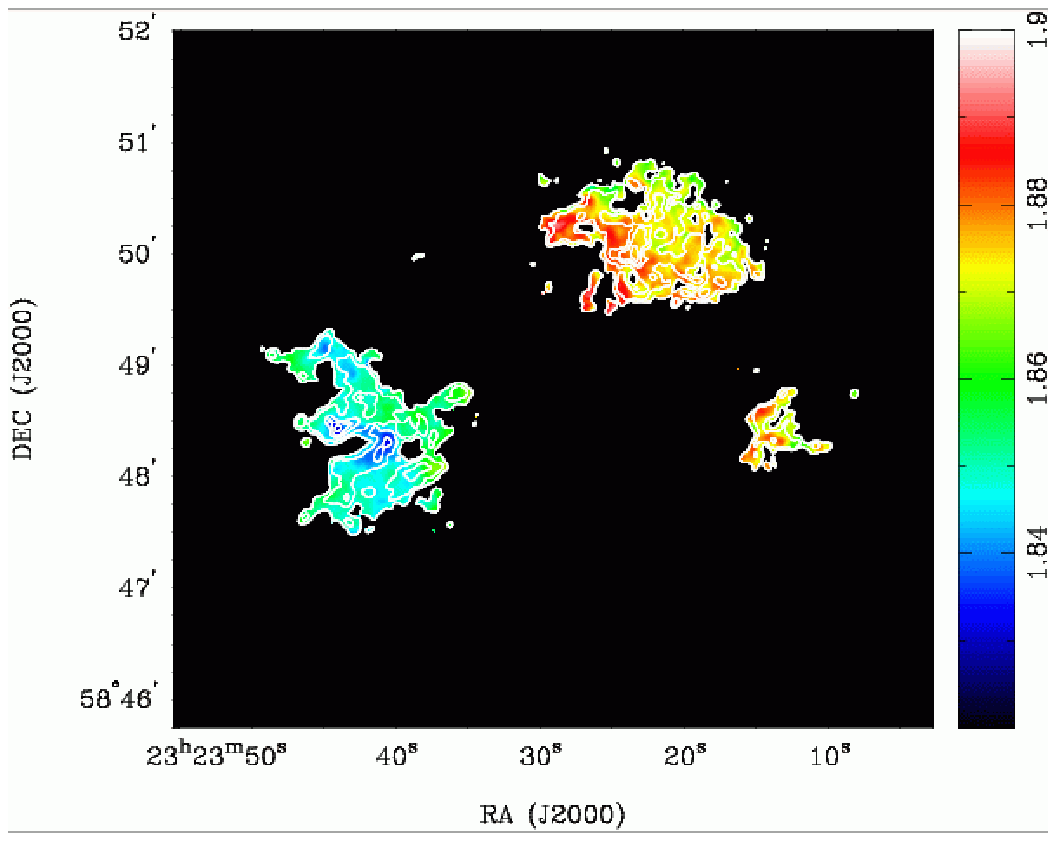}
\caption{Left: Spectral tomography image showing the \fe\ emission as positive
brightness.  The spectra from regions A12 and A13 of \citet{hl03} are shown
in Figure~\ref{ion}.  Right: Effective Doppler map showing the fitted
Gaussian line centers measured in {\AA}ngstroms.  The approximate velocity
range is $\pm4000$~\kps, however ionization effects are not accounted
for.  The contours represent brightness levels on the spectral tomography
image. \label{fekim}}
\end{figure}

\begin{figure}
\epsscale{0.7}
\plotone{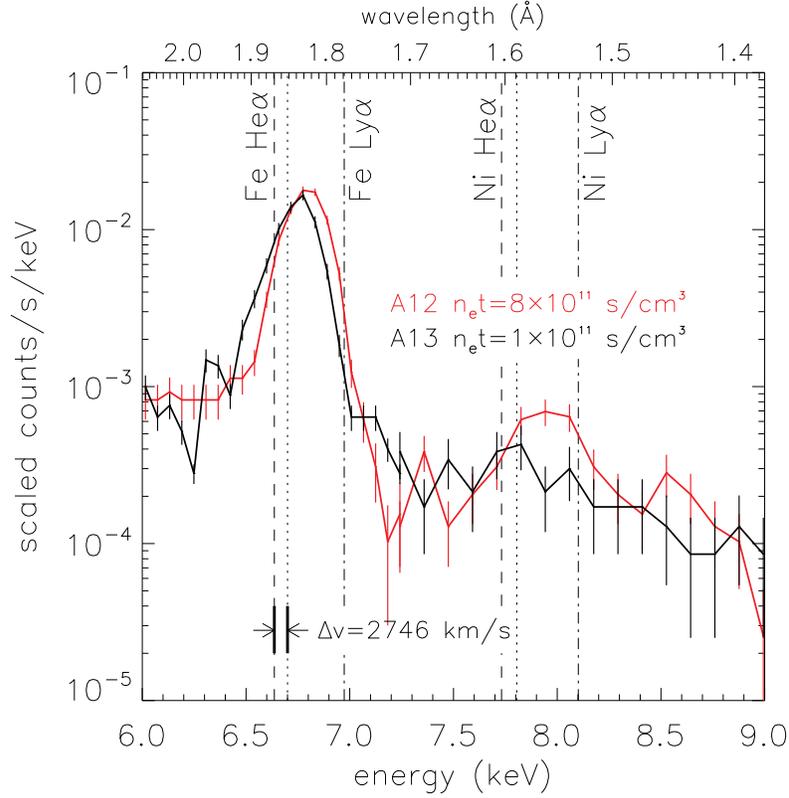}
\caption{X-ray spectra for regions A12 and A13 of \citet{hl03} (marked in
Figure~\ref{fekim}) showing the \fe\ and Ni-K emission lines.  The higher
ionization age of region A12 results in an apparent blueshift in the
spectrum.  The rest wavelengths of the dominant \fe\ and Ni-K He$\alpha$
forbidden (dashed line) and resonance (dotted line) lines and Ly$\alpha$
lines (dot-dashed line) are indicated.  At the bottom of the plot, we show
that a shift of 0.017{\AA} corresponds to an apparent Doppler shift of
2746~\kps.  \label{ion}}
\end{figure}

\begin{figure}
\epsscale{0.7}
\plotone{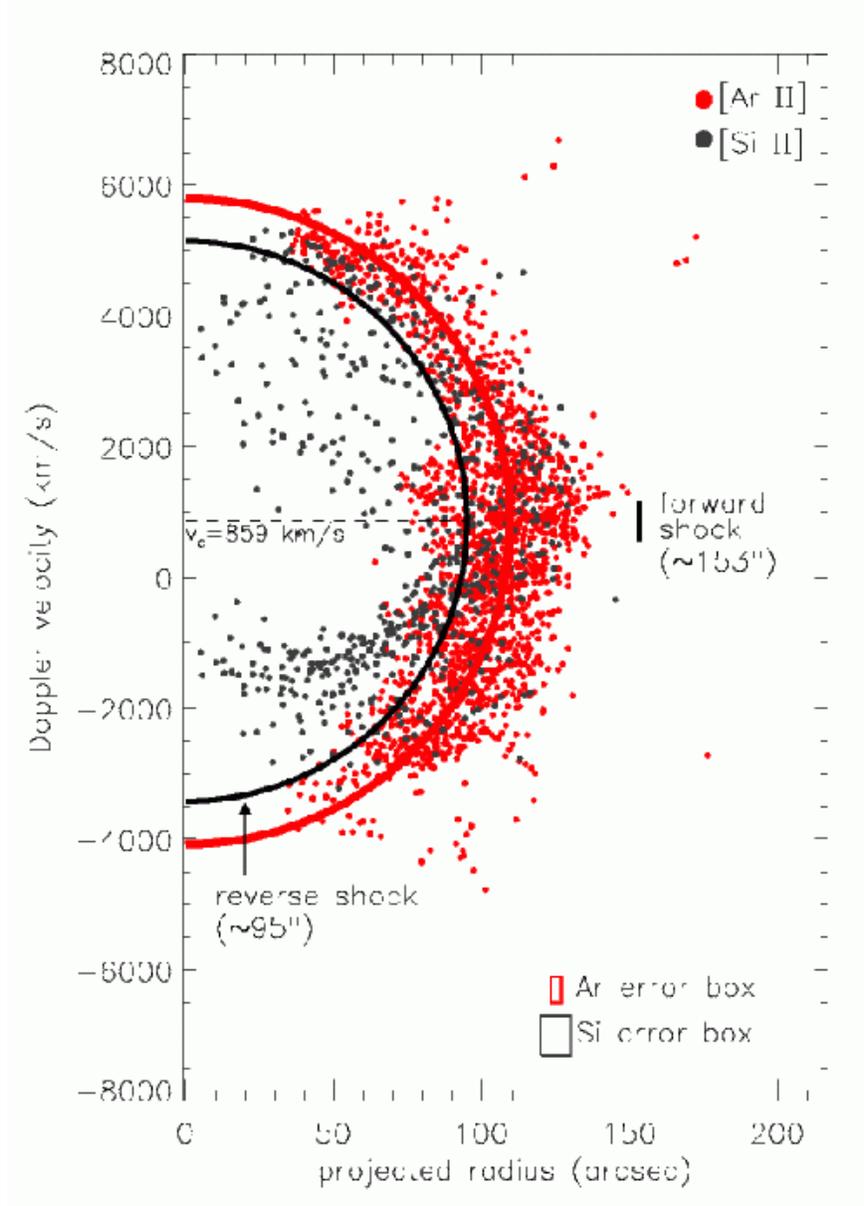}
\caption{Doppler velocity \emph{vs.} projected radius for \ar\ in
red and \si\ in grey.  The mean velocity errors and spatial
resolution are indicated by error boxes at the bottom of the figure.  The
red semicircle represents the best-fit spherical expansion model. The black
semicircle represents the reverse shock and the projected radius of the
forward shock is also indicated.  Note that the center of the spherical
expansion model is not at zero velocity.
\label{vvsr}}
\end{figure}

\begin{figure}
\epsscale{0.7}
\plotone{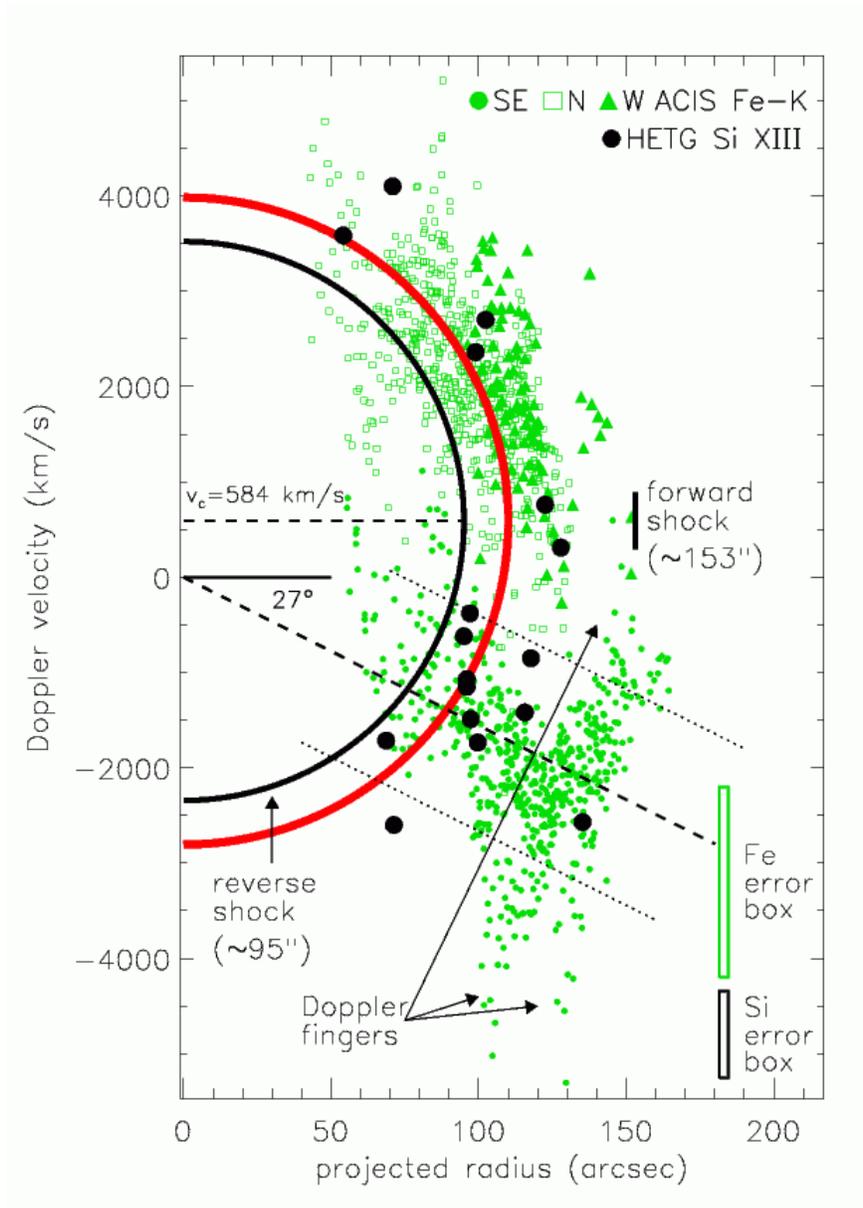}
\caption{Doppler velocity \emph{vs.} projected radius for \fe\ in
green and HETG \xsi\ in black.  Different symbols are used to represent the 
north (open square), west (filled triangle), and southeast (filled circle) \fe\
distributions.  The mean velocity errors and spatial
resolution are indicated by error boxes at the bottom of the figure.  The
red semicircle represents the best-fit spherical expansion model from
Figure~\ref{vvsr}, but scaled appropriately for the decelerated X-ray
ejecta.  The black semicircle represents the reverse shock in this
decelerated reference frame.  The forward shock projected radius is
indicated at 153$\arcsec$.  A few Doppler fingers that are due to ionization
effects are identified.  These correspond to fingers 1-3 in
Figure~\ref{feviews}.
\label{vvsrxray}}
\end{figure}

\begin{figure}
\epsscale{0.7}
\plotone{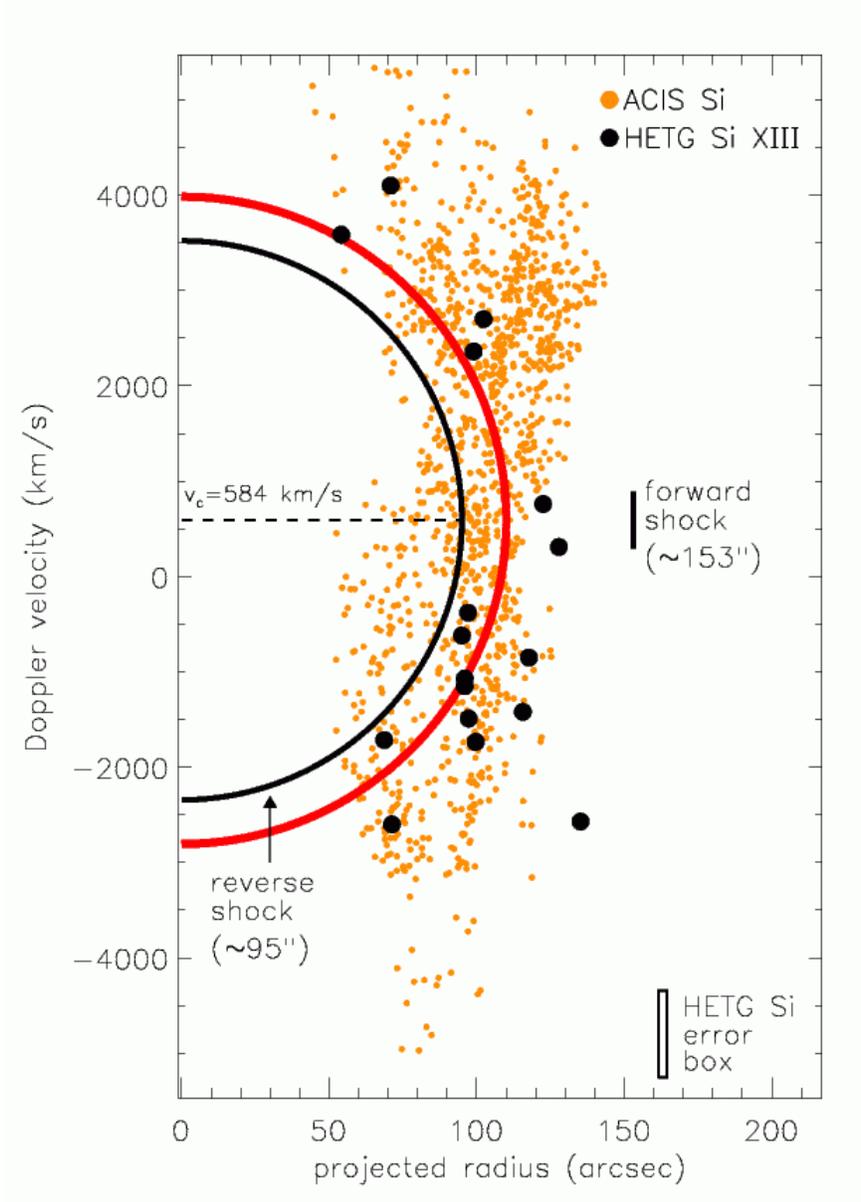}
\caption{Doppler velocity \emph{vs.} projected radius for ACIS Si in
orange and HETG \xsi\ in black.  The mean velocity error and spatial
resolution for the HETG data is indicated by the error box at the bottom of 
the figure.  The red semicircle represents the best-fit spherical expansion 
model from Figure~\ref{vvsr}, but scaled appropriately for the decelerated 
X-ray ejecta.  The black semicircle represents the reverse shock in this
decelerated reference frame.  The forward shock projected radius is
indicated at 153$\arcsec$.
\label{vvsracis}}
\end{figure}

In Figure~\ref{vvsracis} we show the Doppler velocity \emph{vs.} projected 
radius for the full X-ray ACIS Si data and the X-ray HETG \xsi\ data from 
\citet{lds06}.  The ACIS data are plotted using a rest wavelength of 
6.6169{\AA} that was derived from Figure~\ref{hetgacis} by forcing the 
best fit straight line to cross the diagonal line at 0 \kps.  In addition, 
the ACIS velocities have been multiplied by 1.67 to scale them to the HETG 
velocities based on Figure~\ref{hetgacis}.  Just like the ACIS \fe\ data, the 
ACIS Si data show extended Doppler fingers both towards negative velocities 
and towards positive velocities even beyond the top of the plotted area.  This 
is unexpected because the \xsi\ He$\alpha$ line complex was fit with a 
separate Gaussian component from the \ion{Si}{14} Ly$\alpha$ line and 
so ionization effects should not be an issue as they are for the ACIS \fe\ 
data.  The extensive amount of Doppler fingers essentially destroys the 
fine-scale Dopper structure of the Bright Ring.  Due to the many issues with 
the X-ray ACIS Si data, no more useful Doppler information about the Si-group 
ejecta can be derived from the ACIS Si data set than we already have from the 
infrared \ar.

Using the velocity-to-arcsecond scale factors derived for the freely-expanding
ejecta (infrared \ar\ and \si) and for the decelerated ejecta (X-ray \fe\ and
HETG \xsi), we plot their three-dimensional distributions as well as place the
fiducial reverse shock and a fiducial central compact object (CCO) into 3D
with the data.  The data have been smoothed with a 3-dimensional Gaussian to
``glue'' neighboring voxels together such that the symbol size is
$\approx15\arcsec$ for \si, $\approx10\arcsec$ for \xsi, and
$\approx6\arcsec$ for the other ejecta.  We did not encode brightness
information into these reconstructions.  We discuss the various 3D ejecta
distributions and model components below.

\subsection{Illuminating the Spherical Reverse Shock}

\begin{figure}
\epsscale{1}
\plottwo{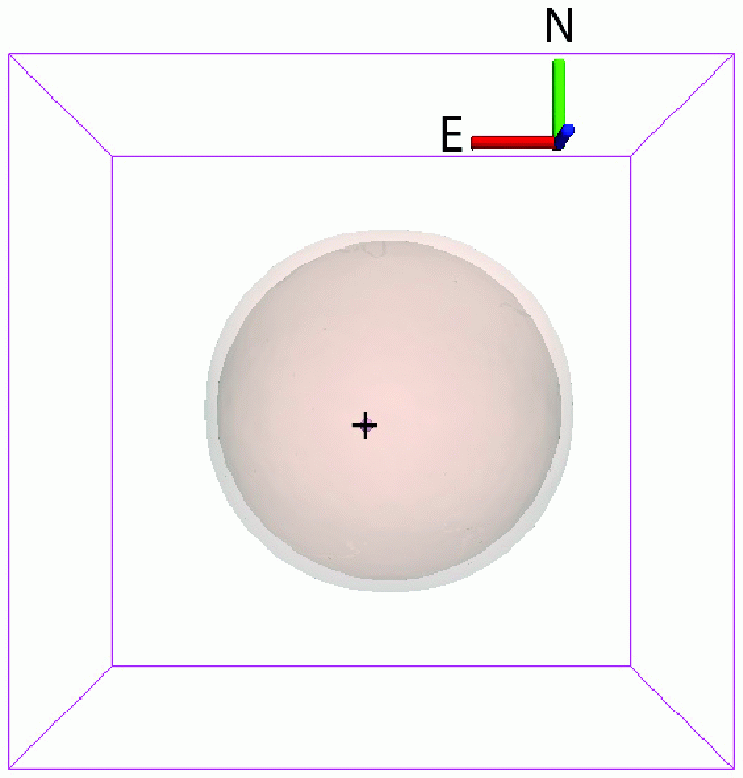}{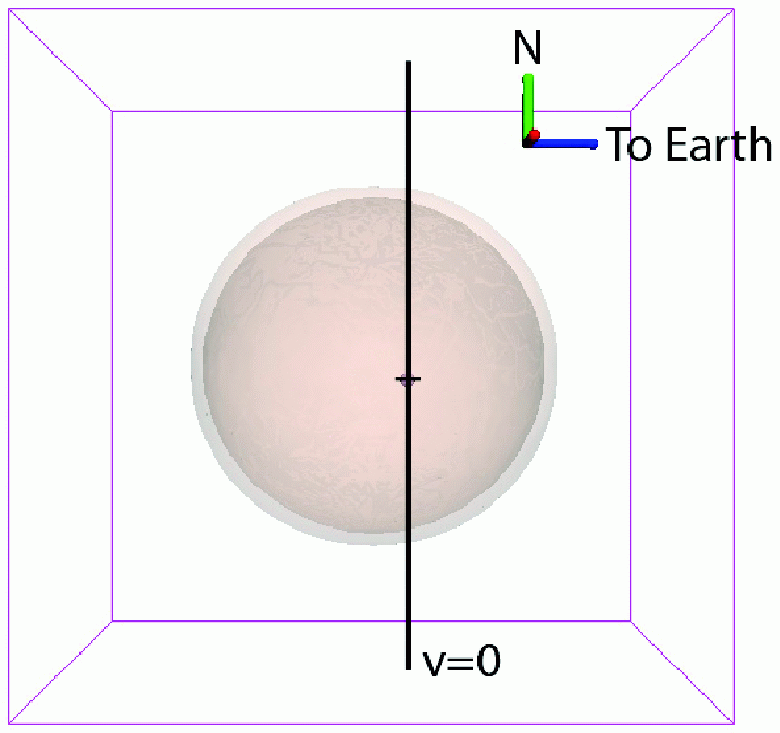}
\caption{Three-dimensional projections showing the fiducial reverse shock
(sphere) and CCO (cross) as seen from Earth (left) and from a 90$\degr$
rotation to the east (right).  The location where ejecta Doppler velocities=0
is identified in the right panel.  Here and in subsequent 3D projections, the
coordinate system is identified at upper right in each panel with red=east,
green=north and blue=to Earth.  The color-coding for this and subsequent 3D
projections is described in Table \ref{colortable}. \label{cartoon1}}
\end{figure}

It is apparent from Figures~\ref{vvsr} and \ref{vvsrxray} that the hot
ejecta are mostly concentrated into a thin shell.  The inside edge of this
``hot'' shell defines the location where ejecta are being heated by passage
through the reverse shock.  In Figure~\ref{cartoon1} we show a 3D model of
the sphere representing the location and shape of the reverse shock in
\casa.  The CCO has been placed in the model at
$\alpha$(J2000)$=23^{\mathrm{h}} 23^{\mathrm{m}} 27\fs94$,
$\delta$(J2000)$=58\degr 48\arcmin 42\farcs5$ \citep{fps06}, and zero
velocity, although its actual line of sight location is unknown, as we
discuss further in \S~\ref{sec:offset}.  The left panel shows the view
from Earth with the CCO identified by the cross and the right panel shows the
model with a 90$\degr$ rotation.  The reverse shock is not centered at $v=0$,
but is shifted into the sky by 859 \kps\ (in the reference frame of the
freely-expanding ejecta) which corresponds to a displacement of 0.31 pc
assuming a distance of 3.4 kpc to \casa.

\subsection{\casa's Rings}
\label{sec:rings}

\begin{figure}
\epsscale{1}
\begin{center}
\plottwo{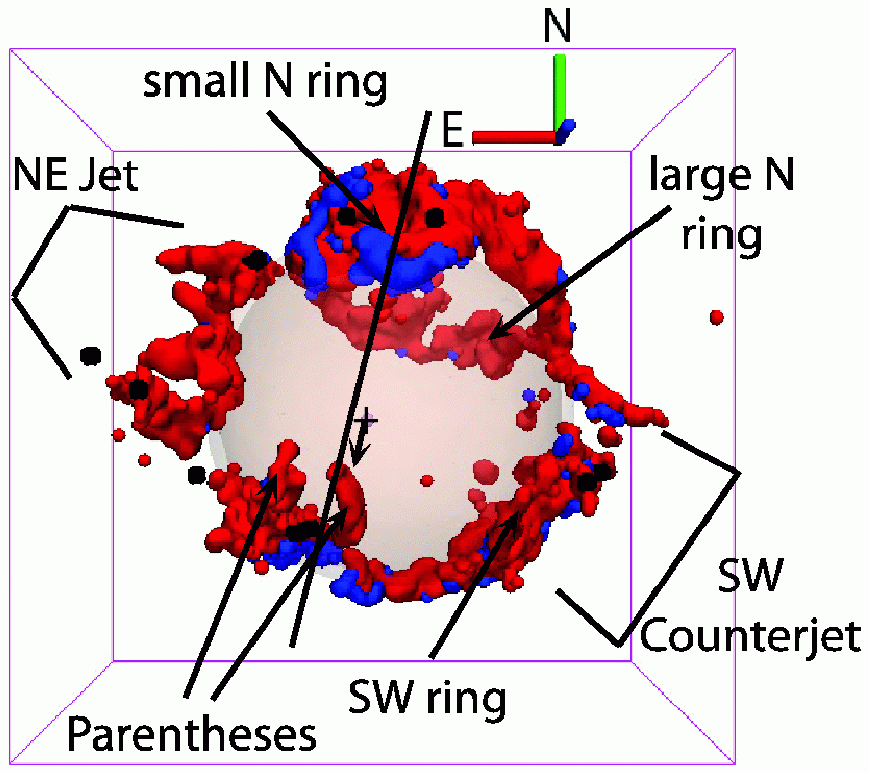}{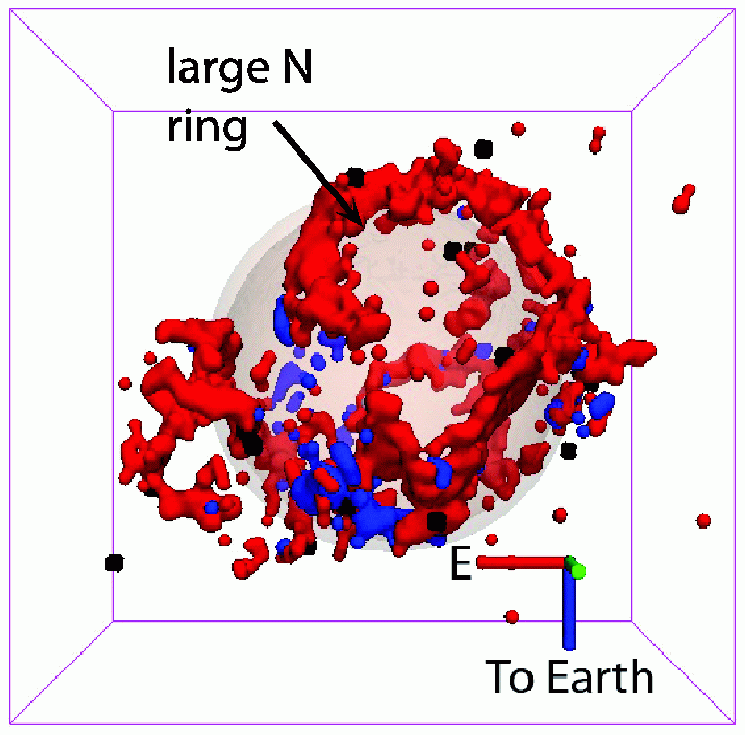}
\plottwo{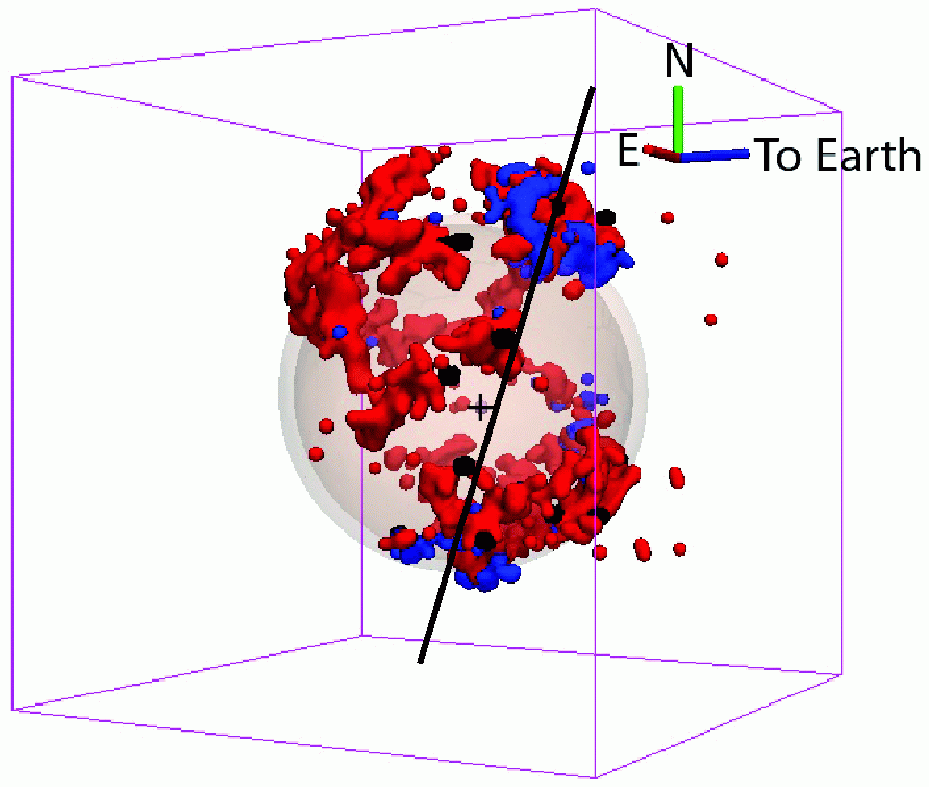}{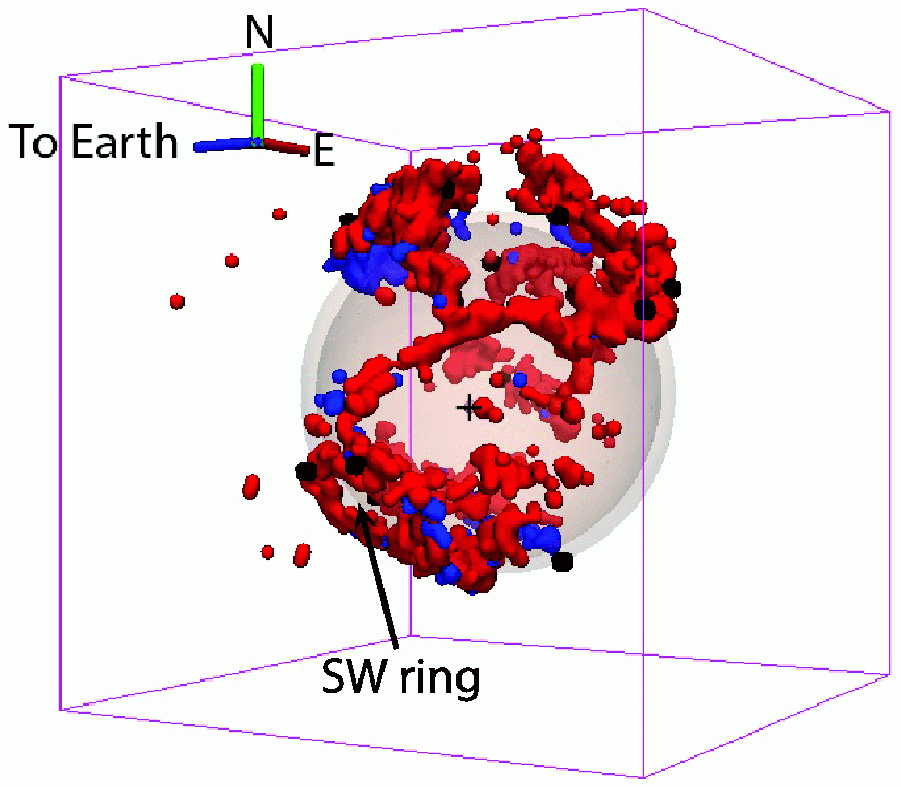}
\end{center}
\caption{Three-dimensional projections of the infrared \ar\ emission (red),
X-ray \xsi\ emission (black), the fiducial reverse shock (sphere), and the
CCO (cross).  Locations where the \neon/\ar\ ratio are high are indicated in
blue.  The lines in the top left and bottom left panels identify an apparent
symmetry axis for the Ne-rich regions.  The inferred CCO proper motion
direction is indicated in the top left panel.  Major structures discussed in
\S\ref{sec:rings} are indicated.  The individual views are: top left - from
Earth, top right - from the north, bottom left - 60$\degr$ rotation to the
east, bottom right - 120$\degr$ rotation to the west.  These same four
projections will be used for all subsequent 3D figures. \label{arviews}}
\end{figure}

To demonstrate how well the shocked ejecta define the reverse shock surface,
we show four projections of the 3D distributions of \ar\ and \xsi\ in
Figure~\ref{arviews} - from Earth, from north, from a 60$\degr$ rotation to
the east, and from a 120$\degr$ rotation to the west.  These same four views
will be used for all subsequent 3D figures.  The blue material in
Figure~\ref{arviews} identifies regions where the \neon/\ar\ ratio is high which will be discussed briefly in Section \S\ref{sec:nemoons}.

The most striking aspect of the \ar, \neon, and \xsi\ emission is that they
do not simply form a limb-brightened sphere - they are organized into a
cellular structure that appears as rings on the surface of a sphere.  When
seen face-on, they define the Bright Ring of \casa.  In the three-dimensional
views, we can identify two complete rings to the north and a complete ring to
the southwest.  Broken or partial rings are found in other locations and we
also identify a ``parentheses'' structure to the southeast of the CCO that
forms an incomplete ring as seen in earlier optical work \citep{lmu95,rhf95}.
While the \ar, \neon, and \xsi\ distributions define a roughly spherical
surface, there are two locations where the ejecta stretch to slightly larger
radii -- to the northeast and the west.  In \S\ref{sec:3dpistons} we show
that these two rings are at the bases of the Jet and the Counterjet.
Presumably the reverse shock is distorted in these two locations accounting
for the conical shape of the heated ejecta found there \citep{fhm06}.

\subsection{The \fe\ Distribution}
\label{sec:fedist}

\begin{figure}
\epsscale{1}
\begin{center}
\plottwo{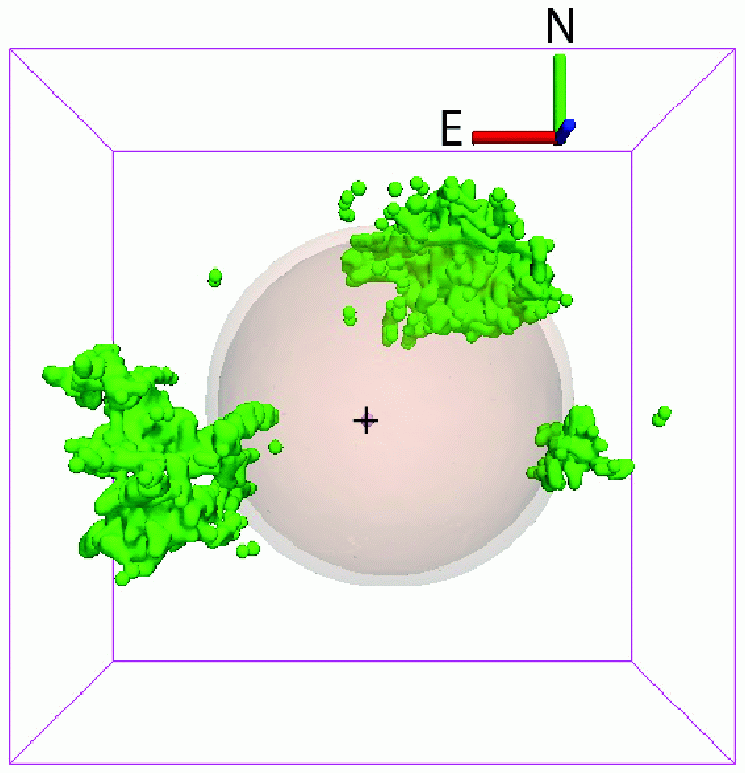}{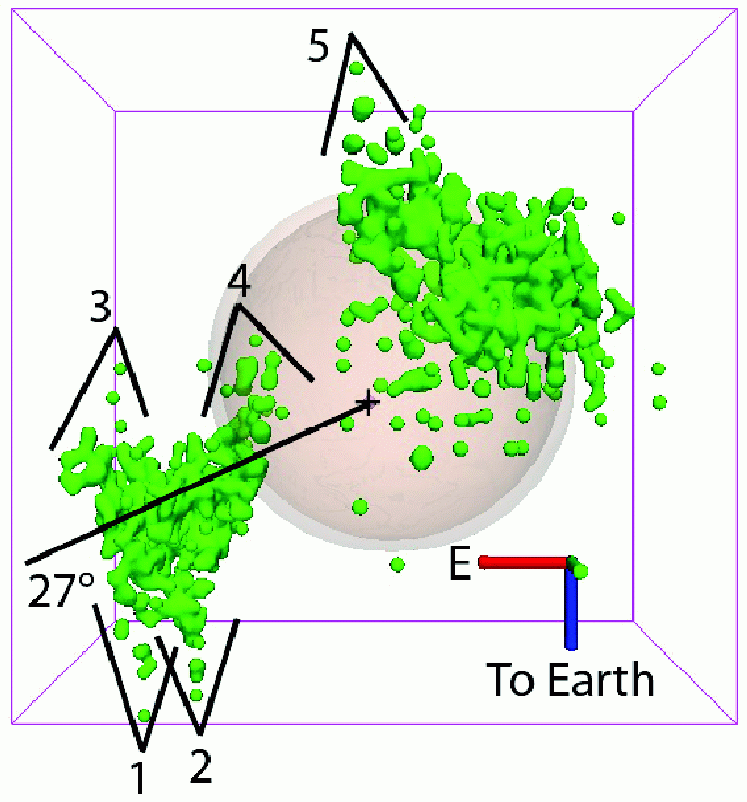}
\plottwo{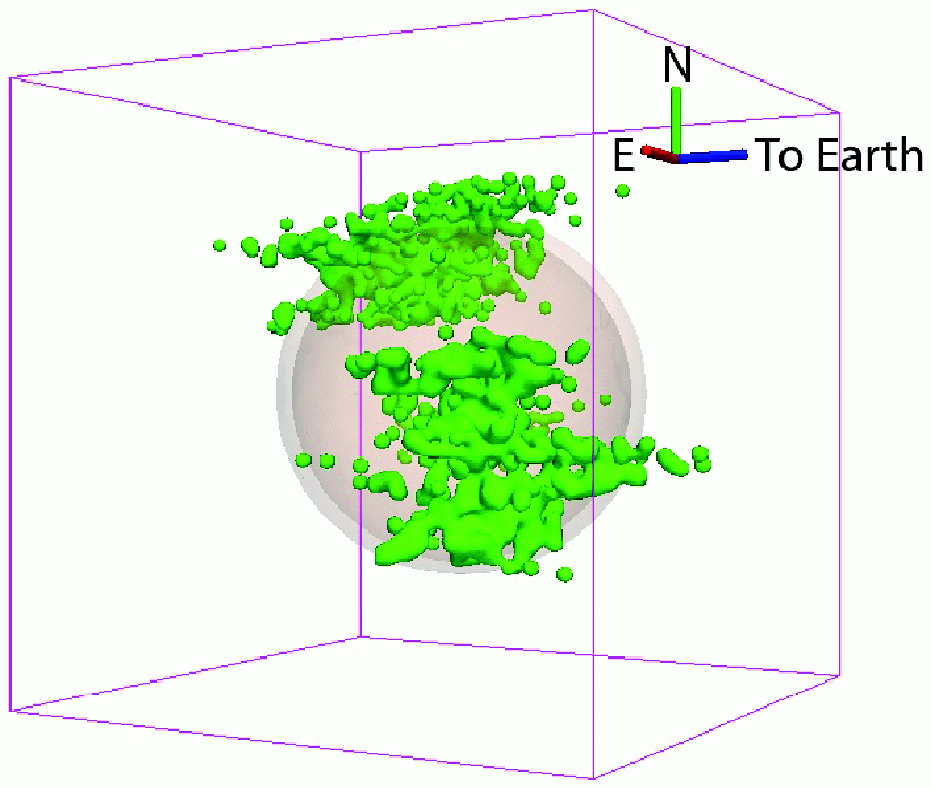}{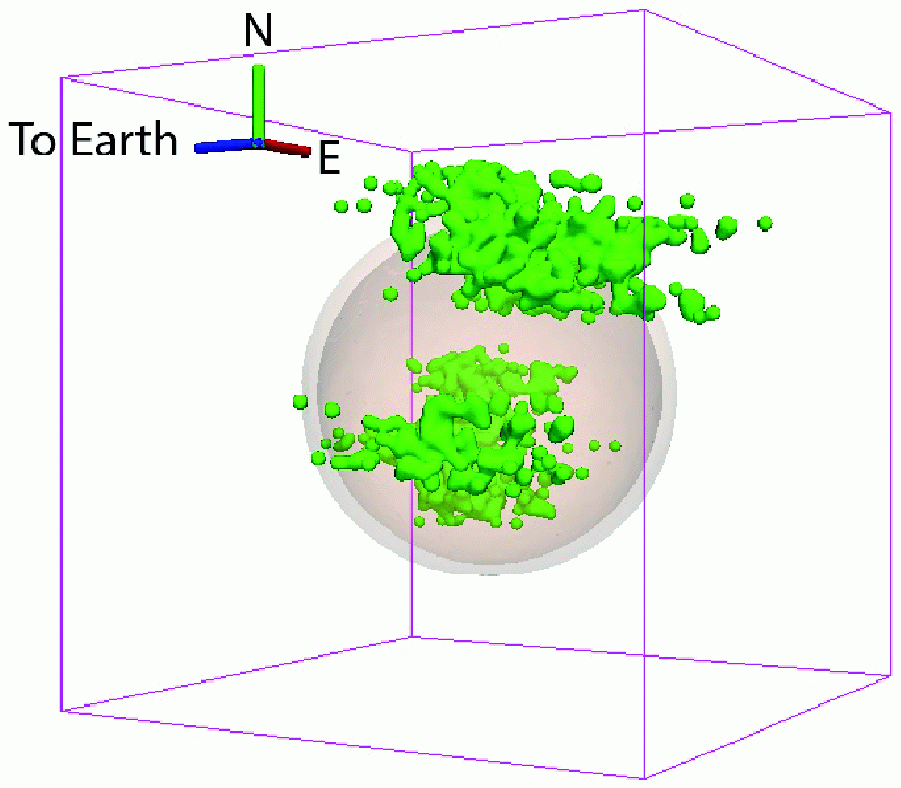}
\end{center}
\caption{Three-dimensional projections of the X-ray \fe\ emission (green) with
the reverse shock (sphere) and CCO (cross).  Regions 1-5 in the
upper right panel identify red- and blue-shifted Doppler fingers that result
from ionization effects.  The Doppler fingers in the southeast \fe\ complex
(1-4) were collimated to lie along the 27$\degr$ line.  \label{feviews}}
\end{figure}

\begin{figure}
\epsscale{1}
\begin{center}
\plottwo{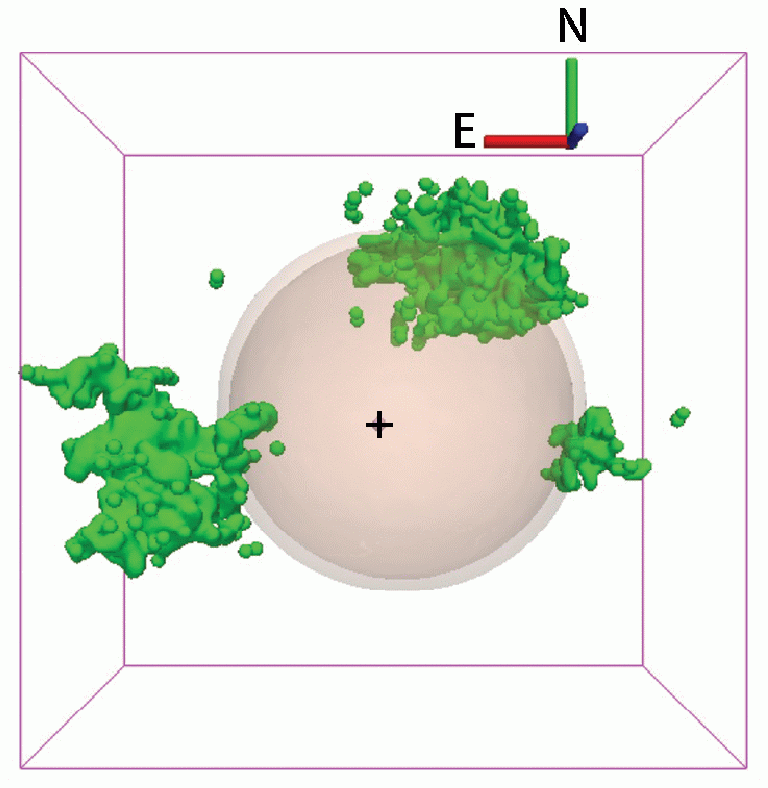}{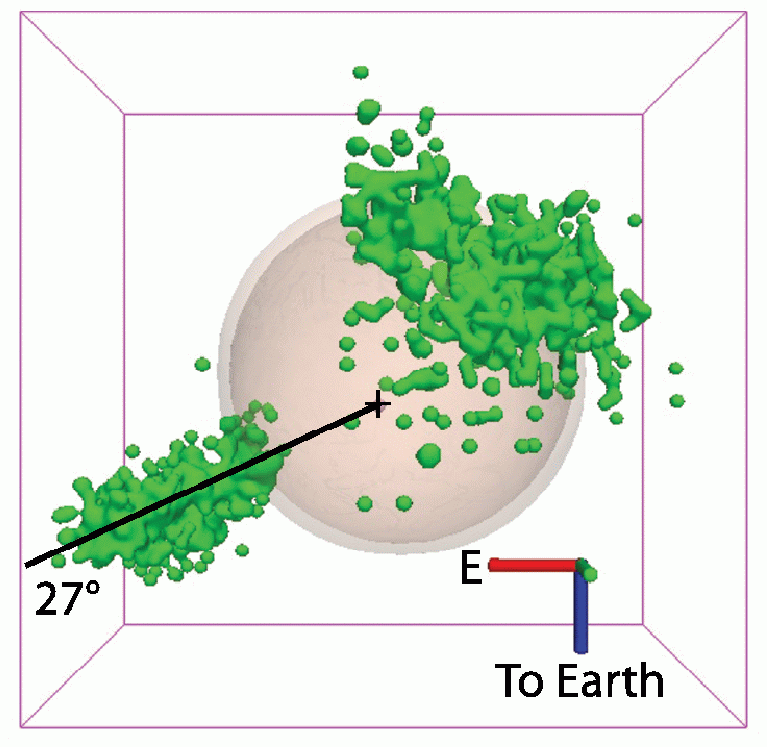}
\plottwo{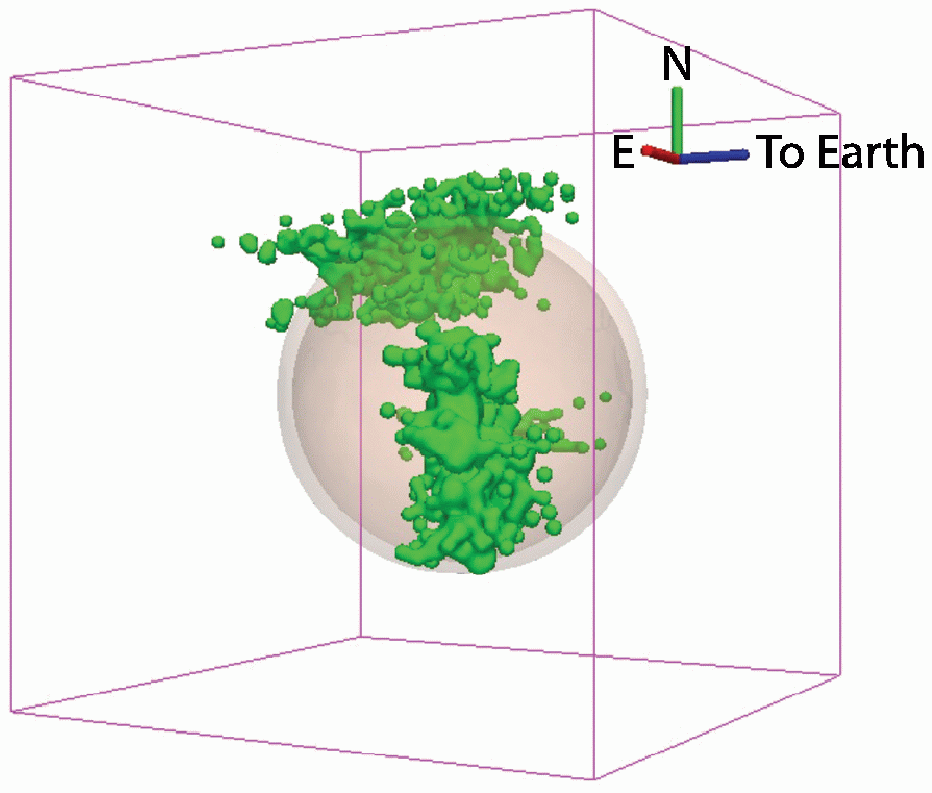}{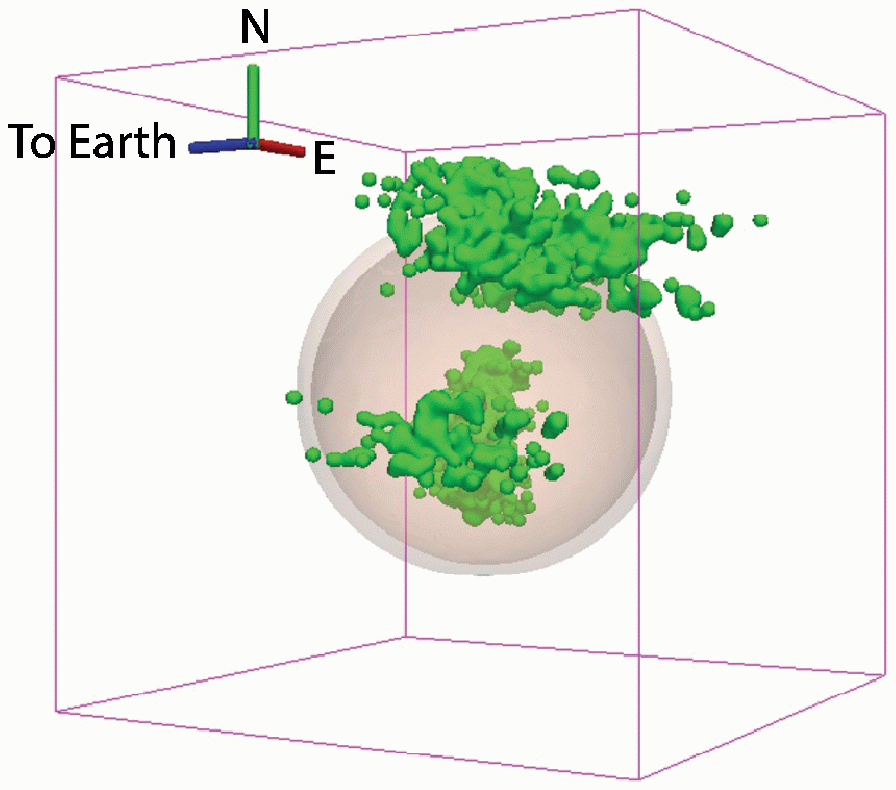}
\end{center}
\caption{Three-dimensional projections of the ionization corrected X-ray \fe\
emission with the reverse shock (sphere) and CCO (cross).  \label{fecorr}}
\end{figure}

In Figure~\ref{feviews} we show four projections of the three-dimensional
\fe\ data plotted with the reverse shock and CCO.  In the upper right panel
of Figure~\ref{feviews}  we can best see the effects of ionization.  Regions
1-5 all form Doppler fingers that extend directly towards or away from Earth.
These fingers are thus likely not velocity structures, but rather regions where the ionization state of the X-ray gas is higher (blue-shifted) or lower
(red-shifted) than the average ionization state of the X-ray Fe ejecta as a
whole.  We corrected for these Doppler fingers in the southeast \fe\ complex
(fingers 1-4) by simply collimating the data - i.e. we shifted the
data composing the Doppler fingers, and only the Doppler fingers, in velocity until they were between the dotted lines in
Figure~\ref{vvsrxray}.  We did not correct any other Doppler fingers in the
\fe\ distribution (i.e. only the southeast \fe\ was collimated).  We show in
Figure~\ref{fecorr} the 3D \fe\ distribution after ionization correction.

As we noted in \S\ref{sec:xrayanal}, the \fe\ emission is localized to three
locations - the west, the north, and the southeast.  In the west and the
north, the \fe\ forms a partial shell on the spherical surface that defines
the fiducial reverse shock, but in the southeast, the \fe\ emission forms a
jet-like structure that extends outward from the reverse shock sphere.  The
three strong \fe\ regions do not fit easily into a bipolar structure - the
north and west regions are 90$\degr$ from each other and the southeast
extension points back to the center of \casa\ and not to either the north or
west regions.

\subsection{Ejecta Pistons and Piston Rings}
\label{sec:3dpistons}

\begin{figure}
\epsscale{1}
\begin{center}
\plottwo{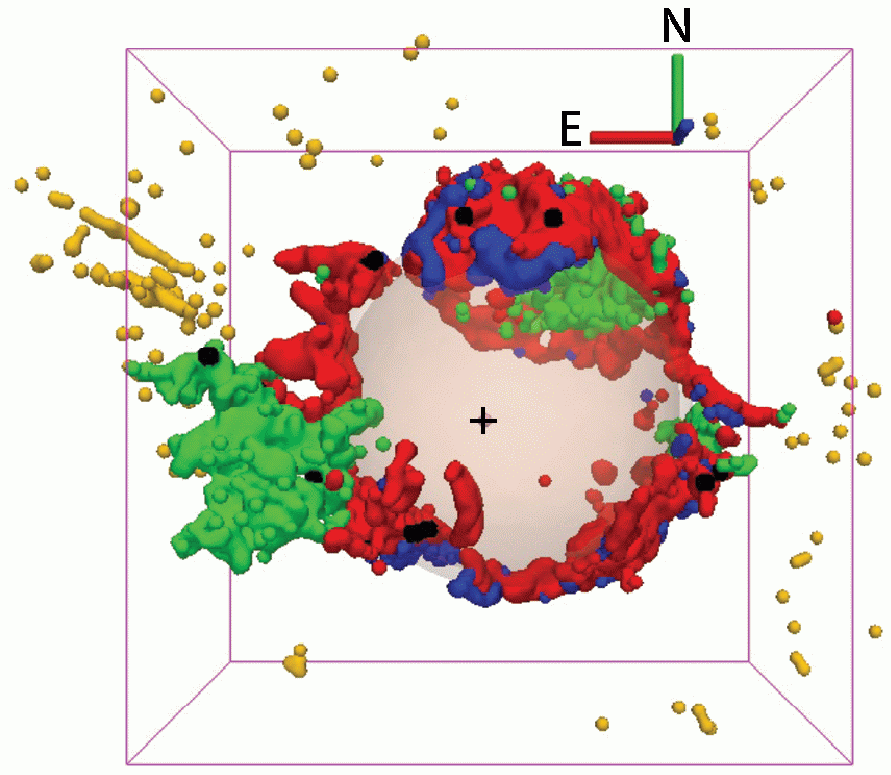}{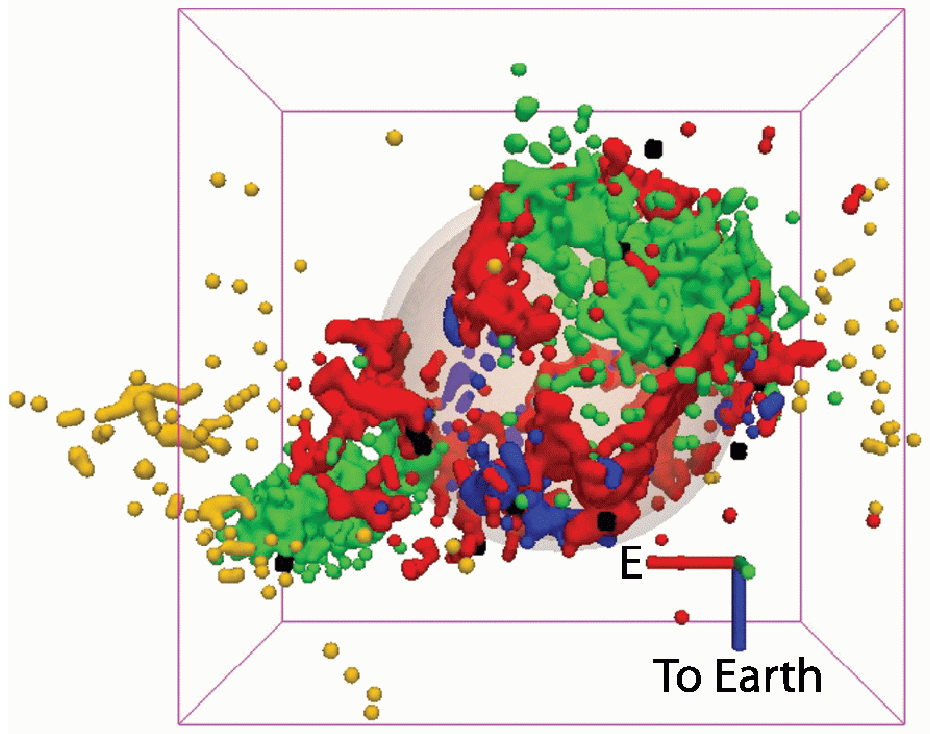}
\plottwo{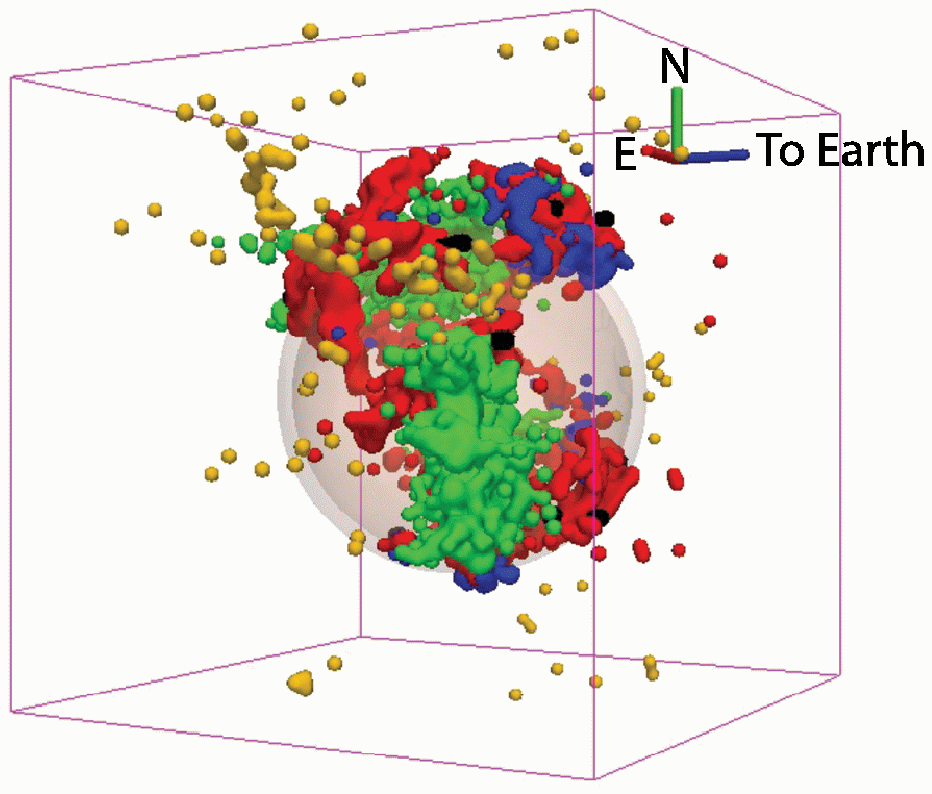}{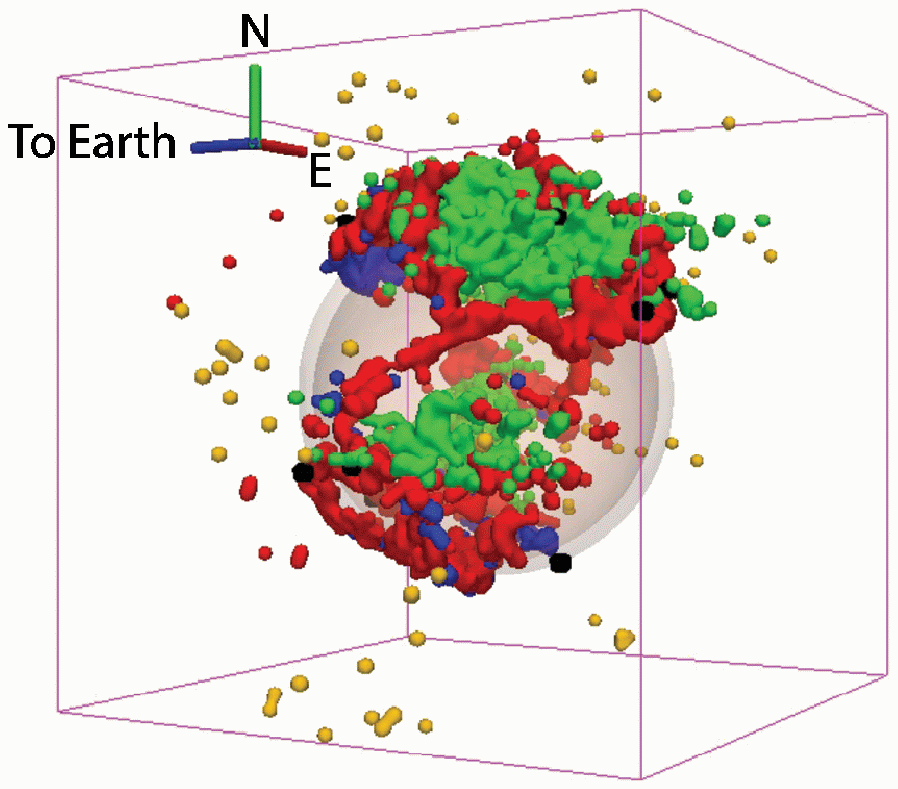}
\end{center}
\caption{Three-dimensional projections of the infrared \ar\ (red), high
infrared \neon/\ar\ ratio (blue), X-ray \xsi\ (black), X-ray \fe\ (green),
outer optical ejecta (yellow), fiducial reverse shock (sphere), and CCO
(cross).  The fastest moving outer optical ejecta defining the northeast Jet
and southwest Counterjet may have moved as much as 12$\arcsec$ in the
intervening years, which is about twice the size of the plotting symbols.
\label{pistons}}
\end{figure}

Figure~\ref{pistons} shows the shock heated Si/Ar/Ne rings plotted with the
\fe\ emission.  We can immediately see that the three \fe\ regions are
circled by rings of \ar\ ejecta.  While the \ar\ ring to the north is
complete, to the west and southeast there are only partial \ar\ rings.
Contrary to reports by \citet{wbv02, wbv03} the \fe\ emission does not appear
behind the higher ejecta layers (Si/Ar) to the north - rather the \fe\
emission is ringed by the \ar\ emission.  Similarly, the southeast \fe\
extension, which has been interpreted as an overturning of ejecta layers
\citep{hrb00}, is not in front of the \ar\ emission, but ringed by the
\ar.  Both the \fe\ and \ar\ emission in the southeast turn on at the reverse
shock, but the \fe-rich ejecta extend farther downstream than the \ar-rich
ejecta and thus the \fe-rich ejecta are at a larger \emph{average} radius
than the adjacent \ar-rich ejecta.

The fact that the \fe\ emission correlates so well with the rings in
the \ar\ emission is a confirmation that the rest wavelength used to
calculate the \fe\ Doppler velocities is correct.  Recall that the rest
wavelength of the \fe\ ``line'' was determined by a fit of the \fe\ data to
the spherical expansion model in the Doppler velocity \emph{vs.} projected
radius plot in Figure~\ref{vvsrxray}.  If a different rest wavelength had
been used, the \emph{entire} \fe\ distribution would have been shifted into
or out of the plane of the sky thus losing alignment with all of the rings.
The alignment of the \fe\ with the \ar\ rings is in no way related to the
ionization correction applied to the \fe\ data.  The ionization correction
was simply a collimation of the southeast \fe\ structure thus no ionization
correction was applied to the \fe\ emission to the north or west, so those
alignments with \ar\ rings are solely due to the choice of rest wavelength
used to calculate Doppler velocity.  To the southeast, collimation was
applied to only the Doppler fingers in the four locations along the
27$\degr$ line in the top right panel of Figure~\ref{feviews}.  The bulk of
the southeast \fe\ data remained untouched.  None of the four locations where
the Doppler fingers appear on the southeast \fe\ structure are on the surface
of the fiducial reverse shock -- locations 1-3 are well outside the fiducial reverse shock and location 4 is inside.  Thus, the collimation of the southeast \fe\ has no effect on the location where the \fe\ crosses the reverse shock or
the alignment of the \fe\ with the \ar\ ring on the surface of the fiducial
reverse shock.

\casa's well-known Jet and Counterjet are also associated with shock heated
Si/Ar/Ne rings.  The prominent Si-rich northeast Jet and the weaker southwest
Counterjet are dominated by a series of linear rays that are visible in
optical \citep[e.g.][]{fg96}, X-ray \citep{hlb04}, and infrared \citep{hin04}
images.  Proper motions and Doppler measurements of outer optical knots show
the high velocity of the jets and their orientation close to the plane of the
sky. \citep{kv76,fg96,fes01,fhm06}.  The number of detected outer optical
ejecta knots has grown considerably since the first detection of the jet
region by \citet{min59}.  The current count stands at 1825 knots of varying
compositions of O, S, and N \citep{fhm06,hf08}.  Of these, only a small subset
($\sim$135) have measured Doppler velocities \citep{fg96,fes01}.  In
Figure~\ref{pistons} we plot the outer optical ejecta in yellow at their 1988
or 1996 locations, depending on the available data, using the
velocity-to-arcseconds scale factor $S=0\farcs022$~per~\kps\ which
is appropriate for freely expanding ejecta.  Strings of optical knots define
the linear features in the Jet while the Counterjet is not easily identified
with this limited sample of outer optical ejecta.  At the bases of the Jet
and Counterjet, where they intersect the spherical reverse shock, we find
broken and distorted rings of \ar\ ejecta.

In \S\ref{sec:pistons} we suggest that the Jet, Counterjet, southeast \fe\
extension, northern \fe\ emission and other less prominent features are all
regions where the ejecta have emerged from the explosion as ``pistons'' of
faster than average ejecta, and that the rings on the spherical surface
represent the intersection points of these pistons with the reverse shock,
similar to the bow-shock structures described by \citet{bgp87}.
Figure \ref{blcartoon} is a cartoon of the geometry involved.  It represents
a simple extension of our earlier cartoon \citep[Figure 10,][]{err06}, where
we now focus on one region of significantly higher velocity -- the piston --
and the ring-like structure at its intersection with the reverse shock that
would appear in 3D.  Emission from the piston itself fades after
reverse-shock passage, which can also leave empty rings on the spherical
reverse shock surface.  The ejecta layering in Figure~\ref{blcartoon} may
remain intact or the piston may ``break through'' the outer layers, as
suggested for the Jet and Counterjet \citep{fhm06}.

\subsection{Bipolarities and other Symmetries}
\label{sec:nemoons}

While the Jet/Counterjet is the most obvious bipolar structure in \casa, the
strong \neon\ emission seen in Figures~\ref{arviews} and \ref{pistons} also
appears to be another example of a bipolar structure, but in a completely
different direction.  These appear quite prominently as the ``Ne-crescents''
in the face-on views of \citet{smith09} (their Figure~8) and \citet{err06}
(their Figure~9).  These regions are also illuminated by optical O emission
\citep{fes01} and distinguished by their distinct lack of X-ray Si emission
\citep{err06}.  \citet{smith09} showed that these two strong Ne regions are
approximately symmetric around the kinematic center of the remnant
\citep{tfv01} and the position of the CCO.  The symmetry axis is quite close
to the direction of the natal ``kick'' inferred by \citet{fhm06}.  In the
bottom panels of Figure~\ref{arviews} we can now also see that these Ne/O
regions are symmetric around the CCO along the line of sight.  While this
might appear to be tantalizing evidence that the Ne-crescents contributed
significantly to the natal kick experienced by the CCO, \citet{smith09}
determined that there is not enough mass/energy in the Ne-crescents to affect
the CCO motion.  In the context of the Figure~\ref{blcartoon} cartoon, the
Ne-crescents would reflect the passage of a slower-moving piston through the
reverse shock, where only the Ne/O layer has been illuminated so far.  While
the Ne-crescents have not caused the CCO motion, they may perhaps arise due
to the same dynamical asymmetries that led to the CCO motion.

All of the ejecta pistons discussed thus far, whether bipolar or not,
all lie in the same broad plane.  Even the outer optical knot distribution,
which forms a giant ring at or beyond the forward shock with a thickness that
is about the same as the distance between the front and back sides of the
\ar\ emission, is roughly oriented along the same plane as the ejecta
pistons.  In the next section we define this broad plane using the infrared
\si\ emission.

\subsection{The Tilted Thick Disk}
\label{sec:disk}

Some of the \si\ data in Figure~\ref{vvsr} map onto the same shell where the
\ar\ is found, but most of the \si\ is found interior to the \ar\ shell.
There are two reasons why the \si\ data might appear interior to the \ar\
data: a) the \si\ emission is spatially interior to the \ar\ emission or b)
the \si\ is spatially coincident with the \ar, but decelerated.  We reject
the latter explanation because we find no evidence that the interior \si\ is
shocked.  The free-free absorption seen in the radio suggests that the
interior \si\ is cool \citep[$\lesssim 1000$K;][]{kpd95}.  The
18.7~\micron/33.48~\micron\ \sthree\ line ratio suggests a low density
\citep[$n_e$ upper limit of 100~cm$^{-3}$;][]{smith09} in the interior and
there is no associated X-ray emission that traces out the same interior
structure as the infrared \si.  These interior conditions are in contrast to
the conditions on the Bright Ring where the temperatures are much higher
\citep[5000-10,000 K;][]{adm99}, the densities are much higher
\citep[$n_e\sim10^4$ cm$^{-3}$;][]{smith09}, and there are spatially
coincident but decelerated X-ray ejecta \citep{drf04}.  Furthermore, the assumption of free expansion for the \si\ ejecta results in a smooth connection between the interior ejecta and the Bright Ring features at the spherical reverse shock shell.  In order to achieve this same smooth connection with a decelerated population of \si, we would have had to adopt a range of distance-to-velocity scale factors with some \si\ even more decelerated than the X-ray ejecta.  Therefore we believe
that there are two populations of \si\ ejecta in \casa\ -- a shocked
population that resides on the Bright Ring and an unshocked, photoionized
\citep{hf88} population that is in free expansion and is physically interior
to the reverse shock.  While we believe that the \si\ ejecta are indeed in
free expansion, the model reconstruction is highly dependent on this
assumption.

In Figure~\ref{siviews} we show four projections of the three-dimensional
\si\ distribution including the CCO and the reverse shock.  The \si\
emission is concentrated onto two concave, wavy sheets - one front and one
rear.  The distribution of the \si\ is not spherically symmetric nor is it
centered on zero velocity.  Most of the \si\ maps interior to the fiducial
reverse shock with portions extending out onto the Bright Ring.  For
comparison, the average 3D radius of the \ar\ ejecta is $\approx110\arcsec$
while the average 3D radius of the \si\ ejecta is $\approx91\arcsec$.  The two
sheets are separated by a series of ``holes'' around the edges.  The plane
containing the holes is not exactly in the plane of the sky, but is oriented
with an $\approx25\degr$ rotation about the N-S axis and an $\approx30\degr$
rotation about the E-W axis.  Our reconstruction shows that the two \si\
sheets are separated by a much lower density region, where only a little
emission is seen (Figure~\ref{siviews}).

We therefore describe the \si\ distribution as a tilted thick disk as shown in
Figure~\ref{cartoon2} where the ``front'' and ``rear'' faces of the thick
disk are shown as thin grey disks.  The area between the faces
contains relatively weaker emission from Si, S, and O ejecta as
demonstrated in Figure~\ref{gausssi} and by \citet{isensee09}.  Presumably
there are unshocked Fe ejecta, but emission from this species has not been
detected.

\subsection{3-D Relationships among Infrared, X-ray and Optical Emission}

In Figure~\ref{allviews} we show the full compilation of data (this figure is
also available as an mpeg animation and a 3D PDF in the electronic edition of
the \emph{Astrophysical Journal}).  It is immediately clear that the rings of
\ar\ match the holes of \si\ supporting the interpretation of the Bright Ring
as the intersection point between the spherical reverse shock and the
flattened ejecta distribution.  \si\ emission is found interior to the strong
\neon\ regions and three of the \si\ holes are filled with \fe\
emission.  To the southeast, specifically, the hole in the \si\ distribution
indicates that there is no unshocked \si\ immediately interior to the \fe\
extension.  Had there been a spatial inversion of ejecta layers to the
southeast, one might expect to see Si-rich ejecta physically interior to the
Fe-rich ejecta.  Thus, the spatial relationship between the \ar\ rings,
the \si\ holes, and the \fe\ emission supports the piston cartoon outlined in
Figure~\ref{blcartoon} in which the whole column of ejecta is displaced
outward along these directions.  While there is a \si\ hole at the base of the
northeast Jet, there is no \fe\ filling that hole.  To the west, there is an
\fe-filled \si\ hole and an empty \si\ hole in the southwest.  Without a
better mapping of the southwest Counterjet, we cannot tell at this time what
the relationship is between the Counterjet and these west/southwest \si\
holes.  Similarly, we cannot tell whether the west \fe\ is related to the
Counterjet, although the distortion of the reverse shock to the west is a
strong indicator of Counterjet interaction.

The complete outer optical knot distribution shown in \citet{fhm06} forms a
giant ring at the location of the forward shock and the limited Doppler
velocity information indicates that this ring is oriented in the same broad
plane defined by the \si\ emission.  The outer optical ejecta show layering 
with S-rich ejecta closer to the reverse shock and N-rich ejecta at larger 
radii.  Previous interpretations of the outer ejecta distribution are that it 
is a limb-brightened shell \citep{lmu95,fes01}.  However, given that the inner 
ejecta distribution is a flattened disk, the outer S- and O-rich optical 
ejecta distribution may simply be an extension of the inner Si-group and O/Ne 
layers out to large radii.  We would not expect to see outer S- and O- rich 
optical ejecta projected across the front and back of \casa\ because those 
layers are still interior to the reverse shock in those directions.  However, 
we might expect to see fast-moving N-rich ejecta projected across the disk of 
\casa\.  In \S~\ref{sec:offset} we will discuss the nature of the ``missing'' 
ejecta towards the front and back of \casa.
There are gaps in the outer optical knot distribution to the north and south 
as shown in \citet{fhm06}.  The northern gap is in the same direction as the 
northern \fe\ cap and \si\ hole while the southern gap is in the same 
direction as a ``closed'' region in the \si\ emission where no large holes 
exist.  The nature of these relationships cannot be determined with the 
limited sample of outer optical knots displayed here.

\section{Discussion}
\label{sec:discussion}

The picture that emerges from the three-dimensional models presented in
Figures \ref{cartoon1} - \ref{allviews} is of a flattened explosion where the
highest velocity ejecta were expelled in a broad plane in the form of jets or
pistons.  As discussed below, the overall flattened ejecta structure is intrinsic to the explosion, with only small effects from later circumstellar medium (CSM) interactions.  In contrast to the flattened ejecta distribution, the Bright Ring, which is formed by the intersection of the ejecta with the reverse shock, maps onto a roughly spherical surface \citep{rhf95}.  Furthermore, the diffuse radio emission is well modeled by a limb-brightened spherical shell \citep{gkr01} and the forward shock appears nearly circular \citep{gkr01}.  

In this discussion, we suggest a coherent dynamical picture of the explosion,
including the various observed symmetries and asymmetries, and the contrasting
structures seen in the shocked vs. un-shocked components.  This picture should
constrain and guide the next generation of simulations, and perhaps even
provide insights into how a star, or at least \casa\ actually explodes.

\subsection{The Flattened Explosion and Thick Disk}

Since the unshocked ejecta have not interacted with the reverse shock, they
are in a pristine state and give direct clues as to the actual
explosion geometry.  Previous indications of the intrinsic nature of the
asymmetries in \casa\ come from the first observations showing the
northeast Jet \citep{min59}, from later X-ray analyses showing the Fe
emission distributed into two lobes oriented nearly 90$\degr$ to the jet axis
\citep{wbv02}, and finally from the presumed natal kick of the CCO
\citep{fhm06}.  The observations presented here of the pre-reverse shock
thick disk show that \casa's explosion led to a flattened ejecta
distribution at least for the inner layers of the star (the possible
geometry of the outer (He/C) layers will be discussed in \S~\ref{sec:offset}).
The origins of this flattening, and resultant disk, are unclear.  It could
simply be the remaining structure interior to the reverse shock from a highly
oblate explosion in which the faster moving jets and pistons in the symmetry
plane have already driven most material beyond the reverse shock.  Since the
breaking of spherical symmetry seems essential in order to produce viable
supernovae \citep{burr07}, the oblate explosion geometry could result from
instabilities such as the Spherical Accretion Shock Instability
\citep[SASI;][]{blon03}, although this instability occurs very early in the
explosion and its geometric signature might be lost at later times.  Other
instabilities also show oblate geometry, such as the neutrino-driven
core-collapse explosion model presented in \citet{burr09} which was calculated
using a 3D adaptive mesh refinement code.

Alternatively, if the explosion were jet-induced, the thick disk could be a
perpendicular toroidal structure, such as proposed by \citet{wmc08}.  Similar
to earlier work by \citet{bwo05} and \citet{jsk05}, they propose that the
prominent northeast and weaker southwest jets \citep{fg96,hlb04,hin04} are
actually secondary to the main energy outflow.  \citet{wmc08} suggest that the
primary energy-carrying jet is associated with the iron-dominated ejecta
in the southeast, approaching us at an angle of $40\degr-50\degr$.  Although
the observations here confirm the Fe southeast jet structure, their model
fails because it predicts an expanding torus roughly perpendicular to the jet
direction, which would contain the ``secondary'' northeast and southwest jets
and other ejecta features.  By contrast, our observations show that all of
these structures, are in a single toroidal-like structure \emph{in the same
broad plane, not perpendicular,} to the primary southeast Fe jet.  At the same
time, the \citet{wmc08} work suggests important possible instabilities in the
explosion that are worth further exploration.  A similar geometry, consisting
of bipolar and equatorial mass ejection, is suggested by \citet{st07} for the
Homunculus Nebula around $\eta$~Carinae, and \citet{SN2006gy} for
SN2006gy\footnote{See illustration at
\url{http://chandra.harvard.edu/photo/2007/sn2006gy/.}}.

In a more speculative vein, the thick disk may be responsible for the
appearance of the light from \casa's explosion. It was underluminous as seen
from Earth \citep{vdbd70,che76,shk79}, which would occur if the optical depth
were high towards the surfaces of the disk (in the direction of Earth),
whereas light could escape more easily in the jet/piston plane.  This curious
idea may be supported by the light echoes which appear to mostly lie in the
plane of the sky \citep{rws08,rfs2010}.  In two locations, echoes are seen out
of the plane of the sky, but this apparent discrepancy may again be related
to the explosion structure.  The position angles of the out-of-plane echoes
are along a) the northeast Jet and b) the north Fe-illuminated piston.  So
the disk surfaces could be swept-up, (optically) thick material, with lower
densities in its jet/piston containing plane.  A calculation of the physical
conditions for this scenario would be of great interest, but is beyond the
scope of the current work.

\subsection{Jets, Pistons and Rings}
\label{sec:pistons}

From the first optical observations describing the Jet
as a ``flare'' \citep{min59} to recent X-ray and infrared images showing a
fainter Counterjet to the southwest \citep{hlb04,hin04}, \casa's striking
jets have been a topic of much discussion.  The jets are not the only ejecta
pistons in \casa, however,  they are simply the Si-group-illuminated ones
(Ar, Si, S), which made them easily visible in X-ray, optical and infrared
emission.  There are Fe-illuminated pistons and Ne/O-illuminated pistons as
well.  The Ne/O-illuminated pistons show a bipolarity similar to the jets,
but the Fe-illuminated pistons show no such symmetry.  The differences in
composition of the various pistons cannot be due to temperature, ionization
state, density, etc., since these differences appear consistently in the
infrared ionic lines \citep{err06} , optical \citep[e.g.,][]{ck79} and X-ray
emission lines \citep[e.g.,][]{hrb00}, and in the dust composition
\citep{rkr08}.  In this paper, the description of the observed structures in
terms of pistons is a simple extension of the description in \citet{err06},
where the ejecta in different directions were inferred to be moving at
different velocities.  All of the pistons currently visible are in the same
broad plane defined by the interior thick disk.

\begin{figure}
\epsscale{1.0}
\plotone{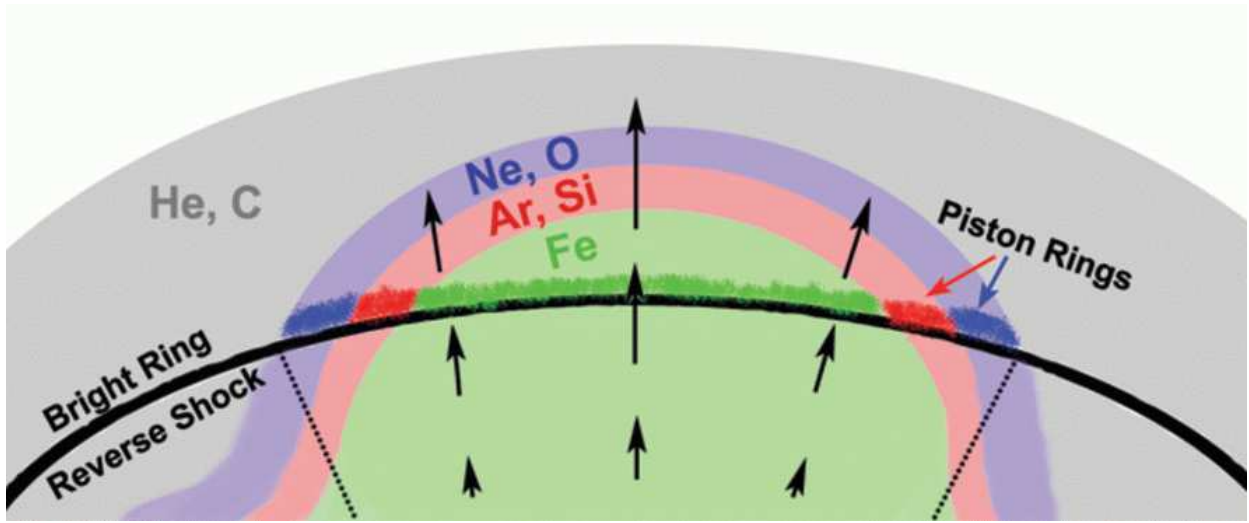}
\caption{Cross-section of a simple ejecta piston model in which the layering
of the progenitor star has been preserved. At the intersection between the
wedge-shaped piston and reverse shock, shock-heated Fe is enclosed by a ring
of shock-heated Ne/O and Ar/Si emission.  Interior to the reverse shock, the
Fe sits within a hole in the unshocked Si distribution.  The center of the
piston is moving faster than the edges of the piston accounting for the Fe
layer having crossed the reverse shock in the center while the Ne/O and Ar/Si
layers are just now reaching the reverse shock at the edges.  All of the
ejecta currently crossing the reverse shock have velocities of
$\approx5000$~\kps. \label{blcartoon}}
\end{figure}

\begin{figure}
\epsscale{1}
\begin{center}
\plottwo{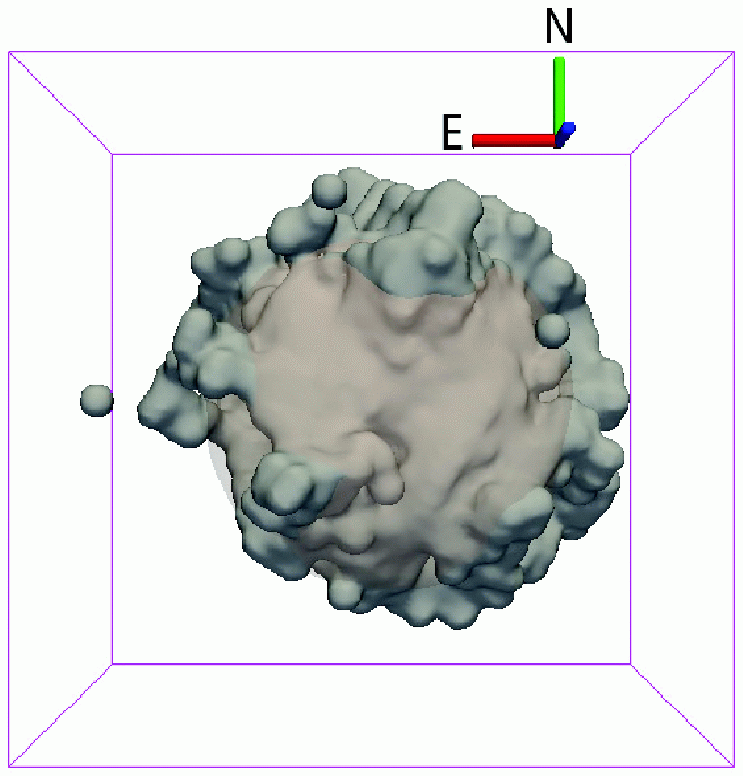}{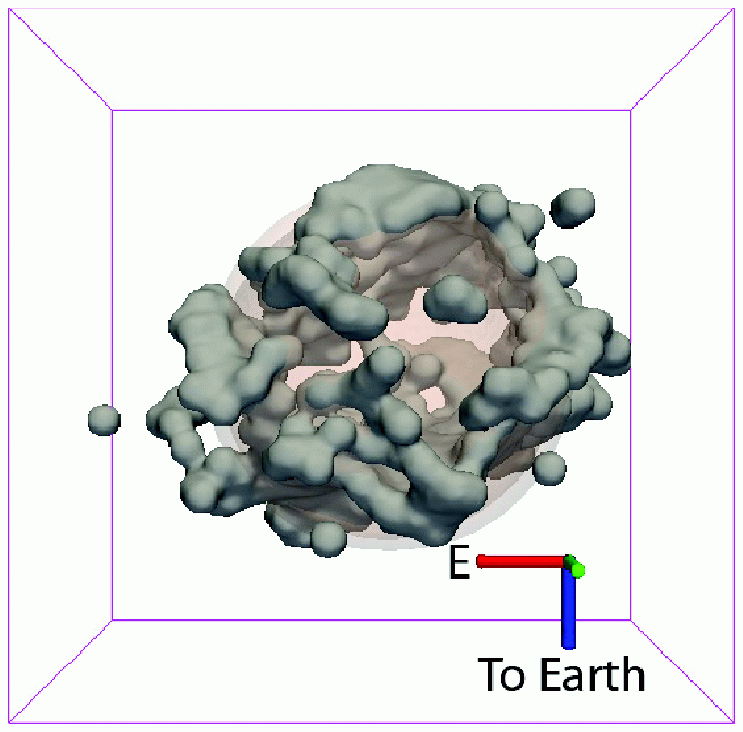}
\plottwo{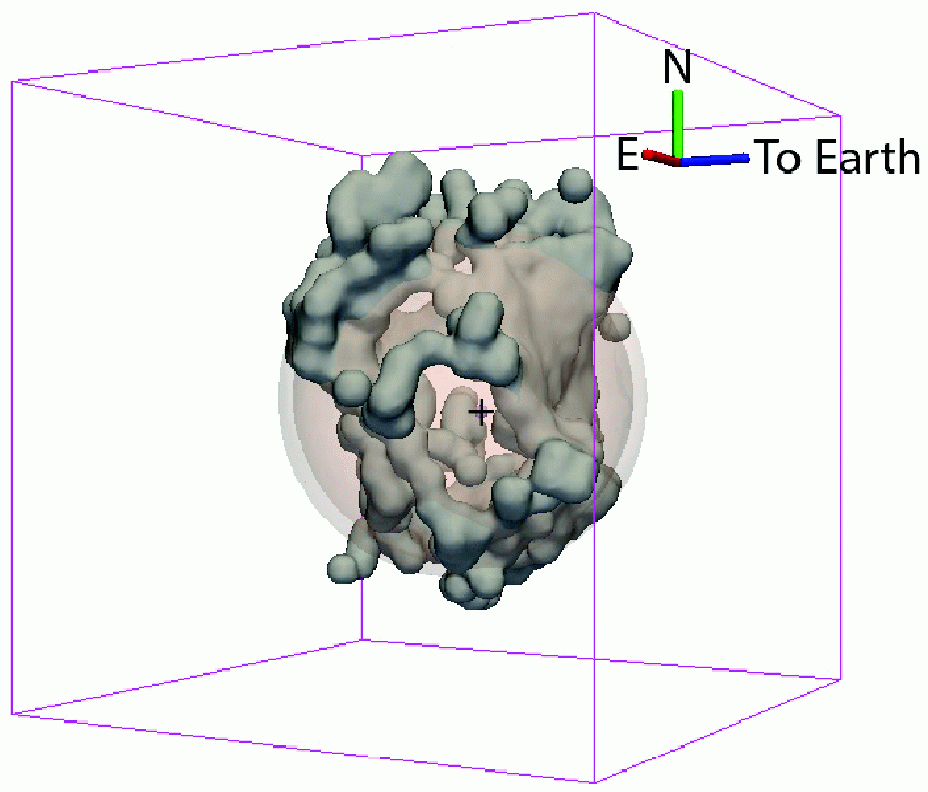}{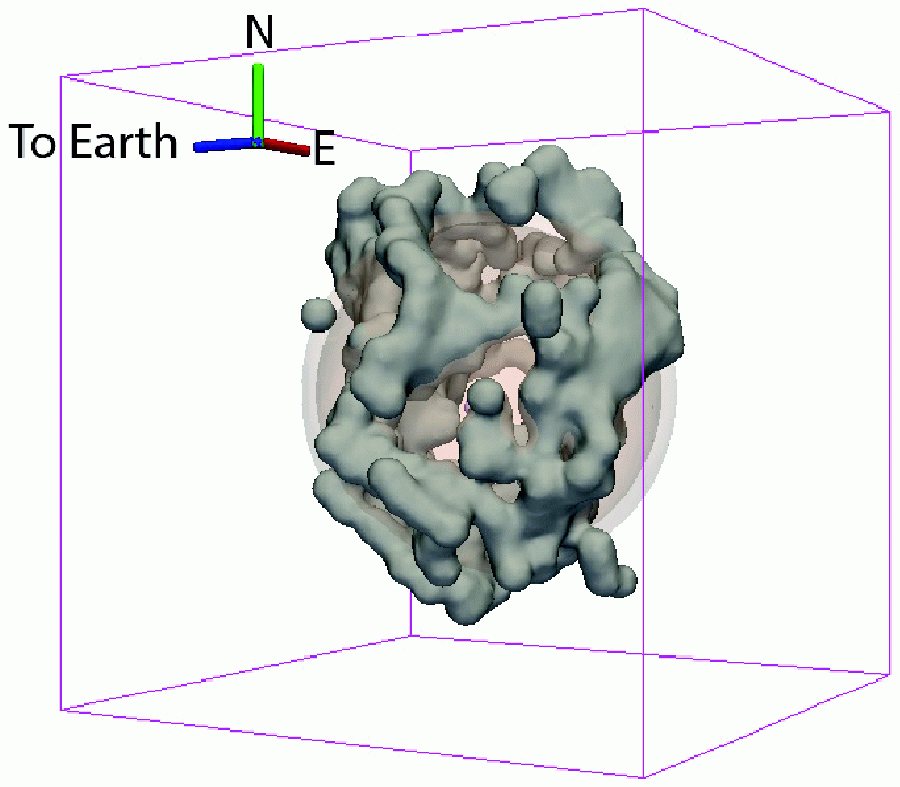}
\end{center}
\caption{Three-dimensional projections of the infrared \si\
emission (grey), fiducial reverse shock (sphere) and CCO (cross).  There are
two populations of \si\ ejecta -- a shocked population that resides on the
Bright Ring and an unshocked population that is physically interior to the
reverse shock.
\label{siviews}}
\end{figure}

\begin{figure}
\epsscale{1}
\begin{center}
\plottwo{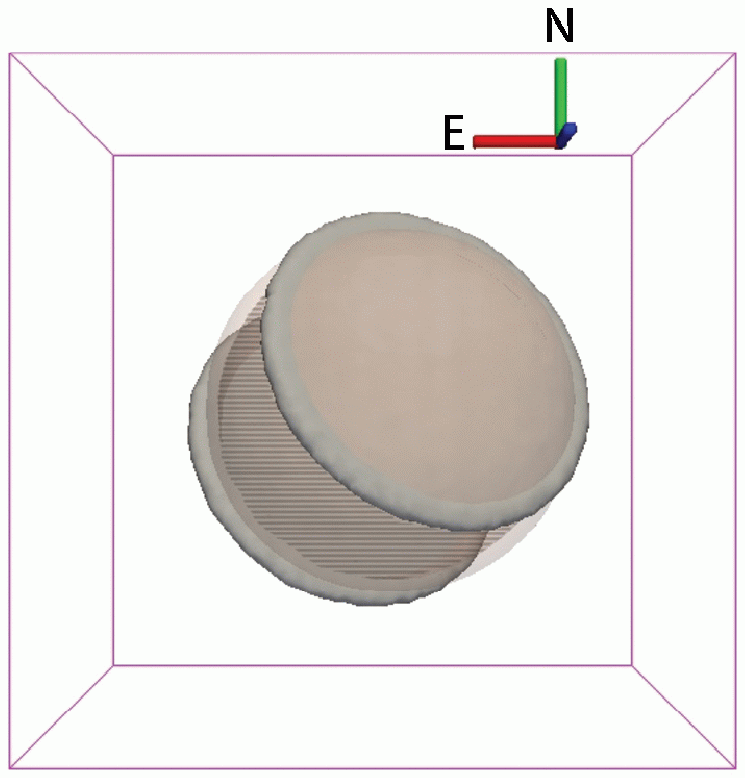}{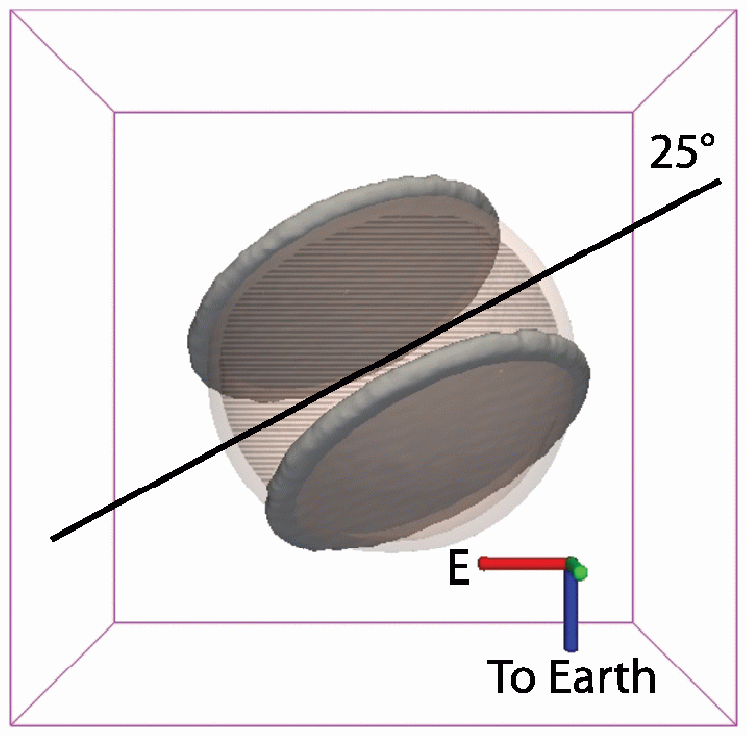}
\plottwo{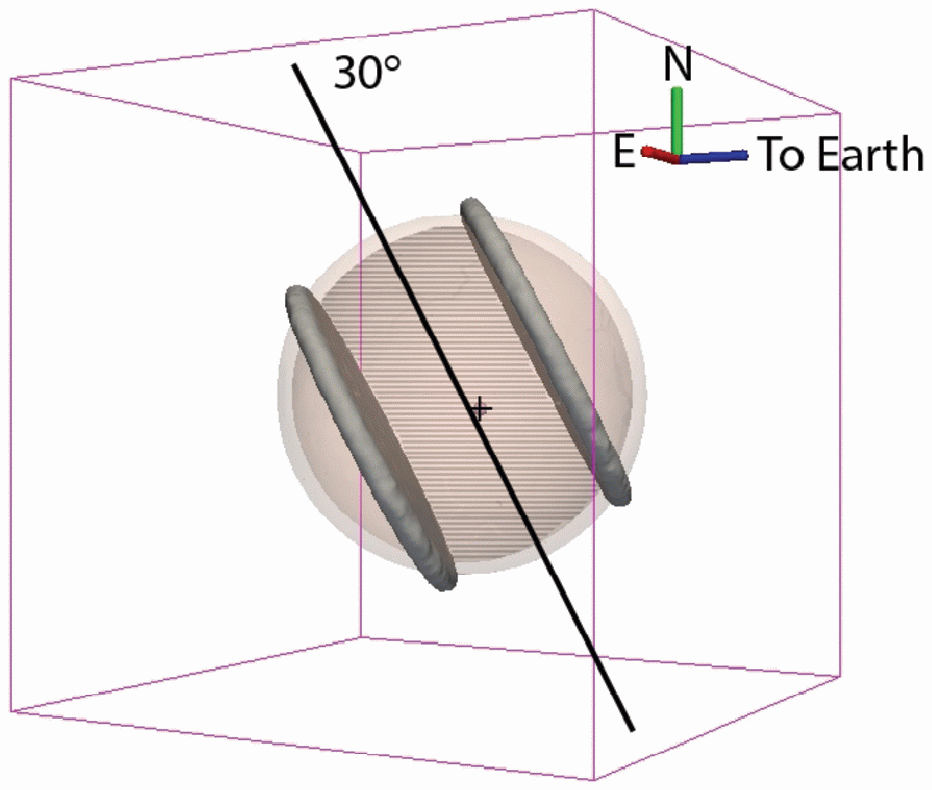}{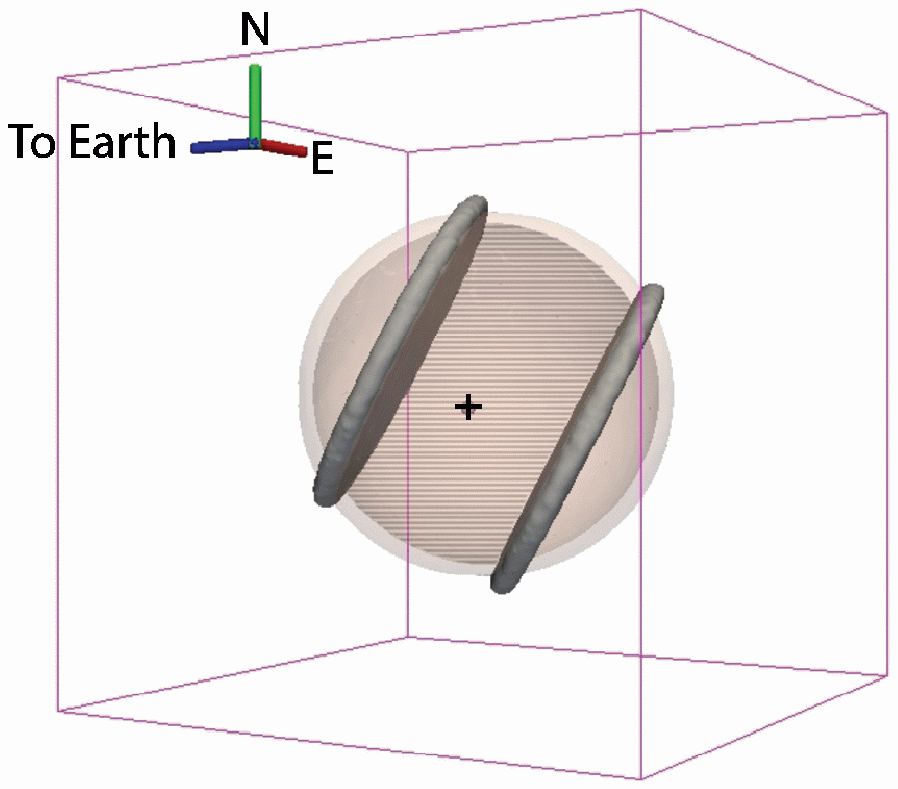}
\end{center}
\caption{Three-dimensional projections showing the tilted thick disk model
(grey and hashed) with the fiducial reverse shock (sphere) and the CCO
(cross).  The hash marks between the faces indicate that the region is not
empty but contains weak emission from unshocked Si, S, and O and probably
undetected unshocked Fe.  Approximate orientations with respect to the
north-south and east-west axes are indicated.  \label{cartoon2}}
\end{figure}

In order to see pistons dominated by different elements, the hydrostatic
nucleosynthetic layers must remain somewhat intact, as illustrated in Figure
\ref{blcartoon}.  In addition, there must be a significant radial velocity
gradient across the layers during the explosion, since otherwise they would
arrive virtually simultaneously, and would never appear segregated, at the
bright ring.  Remembering that all undecelerated ejecta currently
encountering the Bright Ring must be moving at $\sim$5000~\kps, this means
that each piston has a different range of velocities.  For example, if in one
direction, the velocities varied between 1000 \kps\ (Fe) to 5,000 \kps\ (Ne),
then the  Ne layer would be just encountering, and be visible at the Bright
Ring, but Si would not.  If, in another direction, the velocities varied from
1000 \kps\ (Fe) to 10,000 \kps\ (Ne), then the Ne would be beyond the bright
ring, but Si-group emission would be visible.  Thus, we conclude that the
Ne-illuminated pistons have slower average velocities than the
Fe-illuminated pistons, although the currently visible portions are all 
around 5000 km s$^{-1}$.

The pistons represent directions where relatively faster-moving ejecta were 
expelled.  The energy in the pistons can be estimated by the mass in the 
piston and the velocity of the ejecta.  The Fe-illuminated piston to the 
north, for example, contains about 3.2 M$_{\sun}$ of visible material 
(ejecta+CSM) \citep{wbv03}.  Moving at a velocity of 3000 \kps, the total 
kinetic energy would be about 3$\times10^{50}$ ergs.  An estimate for the 
northeast Jet based on hydrodynamic models places the total energy there at 
about 10$^{50}$ ergs \citep{lhr06}.  Thus, the pistons represent a significant 
fraction of the remnant's energy budget.

Any piston pushing out through less dense and chemically distinct layers above would be subject to a series of instabilities such as Rayleigh-Taylor, Richtmeyer-Meshkov, and Kelvin-Helmholtz, which would naturally result in some overturning of the layers \citep[e.g.,][]{kps06}. Certainly small-scale mixing between layers has occurred, and perhaps even some incomplete mixing on large scales, however the instabilities did not lead to a homogenization of the ejecta \citep{hrb00}.  Furthermore, \citet{fhm06}
show, based on their analysis of outer optical ejecta, that the original layering of the star is still intact in some regions.  In addition, the onion-skin nucleosynthetic layers must be remarkably preserved in order to reproduce the observations, as indicated in Figure~\ref{blcartoon}.  The observed size of the rings, which subtend solid angles up to a steradian, require large-scale, low-order mixing modes.  Therefore a large-scale instability or plume mechanism (e.g. the Ni-bubbles of \citet{woo88} as advocated by \citet{hrb00} to explain the southeast Fe structures), that does not result in whole-scale mixing or ejecta layer inversion, is required by the observational data.  Although simulations are beginning to uncover both the larger and the finer scale instabilities \citep{hjm09}, it is not yet clear under what conditions the low-order mixing modes should dominate without producing an overturning of the layers.

The pistons become visible upon their encounter with the reverse shock, where
they are compressed, heated, and ionized.  The physical conditions in the interacting regions are very
inhomogeneous, resulting in X-ray (10$^7$~K), optical (10$^4$~K), infrared
ionic line (10$^3$~K)  and dust (10$^2$~K) emission.  The appearance of the
piston-shock interaction region is illustrated in Figure~\ref{blcartoon},
which shows the formation of rings (e.g., of Ar) and the filling of the rings
by material from deeper (e.g., Fe) layers.  This model is reminiscent of the
geometry described by \citet{bgp87} to explain their radio bow shock
structures and most of their paraboloid extensions (see their Figure~2)
correspond to our ejecta pistons.  The ejecta cool after the reverse
shock passage, so that the emitting material is mostly confined to a shell
above the reverse shock.  The inside edge of the shell thus indicates the
location and shape of the reverse shock and the thickness of the shell is
related to the ejecta lifetime.  Radiative cooling times for the dense
optical knots are quite short (days to months), so their observed
$\sim$30~year lifetimes \citep{vdb83,vdb85} are likely due to continued
heating during the passage of the shock through the knot \citep{mfc04}.  At
typical speeds of $\sim$5000~\kps, this results in a thickness of $\sim$
10\arcsec\ for the illuminated shell.

X-ray emission will also fade after reverse shock passage, but the dominant
physical processes are not yet clear.  The Fe emission in the North, e.g.,
is $<$20\arcsec\ thick, translating to a lifetime of $<$90 years at a typical
space velocity of 3500~\kps\ \citep{drf04}.  However, it should be much
thicker if radiative cooling were dominant, since at the low densities of the
X-ray emitting material (10$^{0.5-2.5}$ cm$^{-3}$) these timescales are of
order 10$^{7-8}$ yr, even for high metallicities \citep{gs07}.  Since the
fading of the X-ray ejecta is clearly not from radiative cooling, this would
suggest that the thin X-ray shell must be thin because the knots are quickly
disrupted, but the models of \citet{hl03} suggest dynamical survival times of
$\sim$200-300~yr.  This discrepancy between the relatively short apparent
lifetimes of the X-ray knots and the long lifetimes predicted by dynamical
models remains to be resolved.

While the ejecta layering appears to be retained in some pistons, in others
there is evidence for a ``breakthrough'' of the outer layers.  For instance,
the rings at the base of the Jet and Counterjet are noticeably distorted and
lift off of the fiducial reverse shock, outlining the surface of a cone-like
structure.  The northeast cone is well-defined while in the west only the top
edge of the cone is evident.  These cones are completely consistent with the
scenario put forth by \citet{fhm06} where the jets represent streams of
high-speed ejecta that were expelled up through the outer layers of the
progenitor star, distorting and perhaps obliterating the reverse shock in the
process.  

The projected forward shock does not show large asymmetries.  If
indeed the forward shock is spherical, the deformities in the reverse shock,
which developed as a reflection of the forward shock, might have occurred after the reverse shock formed.  However, since the forward and reverse shocks are
now well separated and dynamically distinct, any deformations in the early
forward shock shape may have been smoothed out by interaction with the CSM.
In this case, the reverse shock might reveal the original shape of the
forward shock.  Alternatively, the forward shock may indeed still have large
deformities.  For instance, \citet{lhr06} speculate that the forward shock
may indeed extend around the northeast Jet but would be effectively invisible
because the plasma flow is oblique with respect to the shock resulting in
less efficient diffusive shock acceleration.  Given the size of \casa,
\citet{lhr06} further speculate that the forward shock at the Jet tip, where
the plasma flow would be perpendicular to the shock, may actually be outside
the field of view of Chandra's ACIS-S3 chip.

\subsubsection{Southeast Fe Piston and CSM}
\label{sec:overturn}

The X-ray Fe emission in the southeast is of considerable interest for models
of supernova explosions and the subsequent instabilities in the shocked
ejecta.  The southeast was thought to be a region where the Si and Fe ejecta
layers had overturned \citep{hrb00} because the Fe-rich ejecta were at a
larger radius than the Si-rich ejecta.  The subsequent spectral analyses of
\citet{lh03}, \citet{hl03}, and \citet{hl09} showed that the Fe-rich ejecta
to the southeast are at a higher ionization state on average than the Si-rich
ejecta and thus the Fe-rich ejecta crossed the reverse shock at an earlier time and are indeed at a larger radius, on average, than the Si-rich ejecta.  Our 3D reconstruction is completely consistent with the geometry inferred by the earlier spatial and spectral analyses, however our hypothesis is that, rather than a spatial inversion of ejecta layers to the southeast, the whole ejecta column along that direction has been displaced outward.

If there had been an inversion of the ejecta layers, we would expect to see
Si-rich ejecta, either shocked or unshocked, immediately interior to the
Fe-rich ejecta along that explosion direction; this is not observed.  The Fe-K structures all turn on at the reverse shock, implying that there is a
source of unshocked Fe-rich ejecta supplying those regions.  Unfortunately, these unshocked Fe-rich ejecta are not observed here, or anywhere in the interior of the remnant.  Furthermore, the total Fe detected in \casa\ is 
only a few percent of what is predicted by nucleosynthesis models \citep{hl03}, so it is likely that there is a supply of unshocked Fe-rich ejecta interior to the reverse shock.  We note that, if the unshocked Fe were rather low density, as observed for the unshocked ejecta in the Type Ia remnant SN1006 \citep{hfw97}, then it would be well below our detection limit.  A low density Fe layer might also explain why the Fe-rich knots are typically appear more diffuse than the Si-rich knots \citep{hrb00}.  Other scenarios for why we have not detected Fe in the unshocked ejecta are explored by \citet{isensee09}.

There do appear to be N, O, and S optical knots projected \emph{in front of}
the Fe-rich ejecta \citep{fhm06}, but in our 3D picture, the few outer optical knots in the southeast with Doppler velocity measurements are off to the side rather than directly in front of the Fe ejecta.  A Doppler mapping of the complete set of detected outer optical ejecta knots to the southeast is needed to determine their relationship with the Fe-rich ejecta.

The southeast Fe (and the Jet/Counterjet) appear very long-lived compared to
the rest of the Bright Ring, and, in particular, the other bright Fe regions.
Assuming a static reverse shock, the ejecta at the tip of the southeast
column, with total space velocities of $\approx$6000~\kps\ \citep{drf04},
were shocked at least 190 years ago, completely consistent with the long
lifetimes derived by \citet{lh03} for knots in this region of \casa.  The
proper motions indicate that the X-ray Fe has been decelerated; had it
remained in free expansion, the southeast Fe jet would extend out as far
as the northeast Jet.

The long lifetimes of the southeast Fe-rich ejecta are certainly consistent
with the radiative cooling timescale for an X-ray plasma, however it is not
clear why the ejecta should live longer in the southeast than elsewhere on
the Bright Ring.  We note that there is an unusual structure to the CSM in
this direction.  The ionization age of the Fe jet drops rapidly at large
radius, where it may be interacting with the forward shock and the CSM.
Figure~\ref{outersi} shows \si\ observed beyond the forward shock to the
east of \casa.  The grayscale \si\ image has been smoothed to 20$\arcsec$
resolution.  The outer \si\ emission extends as an $\approx30\arcsec$ (0.5~pc)
wide filament from the eastern edge of the remnant out to a distance of
$\approx6\arcmin$ from the center of \casa\ where it meets a broken
north-south ridge or shell.  The observed ridge is $\approx5\arcmin$ long,
but its true extent is unknown because it extends beyond our field of view.
The three linear structures defining the northeast jet and the outermost
optical ejecta observed by \citet{fhm06} are plotted in green in
Figure~\ref{outersi}. We also observe 33.48~\micron\ \sthree\ and
15.56~\micron\ [\ion{Ne}{3}] in these outer structures.  The outer emission
correlates very well with an H$\alpha$ cloud, identified by the blue contours
in Figure~\ref{outersi}, as well as very faint H$\alpha$ filaments farther to
the east of \casa\ \citep{fg96,fes01}.  Doppler velocities for this outer
material are small, only $\pm$ a few 100~\kps\ with errors typically of the
same magnitude, indicating that it is almost certainly CSM.

\subsection{The Offset Spherical Component}
\label{sec:offset}

Despite the large-scale asymmetries of the flattened explosion, and the
prominent jets and pistons, the forward and reverse shocks appear to be
spherical. This issue was addressed many years ago by \citet{bis82}, who
asked, ``[Must] all young remnants whose shape is nearly spherical have been
engendered by a spherically symmetric explosion?  By no means!''  They found
that propagating shocks tend to isotropize with time, although the time to
approach sphericity depends on factors such as density gradients and possibly
shells through which the shocks propagate \citep[e.g.,][]{cs89}.  It would be
very useful to use the observed transition from the interior flattened
explosion to the spherical shock structures in \casa\ to calculate the
physical structure of the pre-SN stellar atmosphere and winds.  From the
standpoint of enabling stars to explode, asymmetries such as we observe
appear to be essential in, e.g., restarting a stalled forward shock
\citep[e.g.,][]{rj00}.

The relationship between the structures observed here and the position of the
CCO can also be helpful in constraining the nature of the explosion.  In the
plane of the sky, the CCO is inferred to be in motion based on its offset of
$7\arcsec$ nearly due south of the optical expansion center \citep{tfv01}.
The derived transverse velocity is $\approx$ 350~\kps\ in the direction of the
southern gap in the outer optical ejecta \citep{fhm06} and is roughly aligned
with the Ne-illuminated pistons, however there is insufficient mass in these
pistons to generate such a kick \citep{smith09}.  The projected motion of the
CCO could be a recoil from the Fe-illuminated ejecta piston to the north,
which has no counter piston, and this scenario should be explored further.
Note that as long as the optical ejecta are undecelerated, this observed
offset cannot be due to a density gradient in the CSM; such a gradient can
shift the apparent center of the shock structures \citep{rhf95,dj96}, but
not the expansion center.

In our analysis, we have positioned the CCO at zero velocity along the line
of sight, although there is no information on its actual velocity in that
direction.  If the CCO were moving very rapidly away from us
($\sim$1000~\kps), then it could be centrally placed between the surfaces of
the thick disk (see Figure \ref{vvsr}), making the flattened explosion
front-back symmetric.  However there is evidence that the observed velocity
asymmetry between the surfaces of the thick disk could well be intrinsic
to the explosion.  In Figures~\ref{gaussfit} and \ref{gausssi} the
blue-shifted \si\ emission is significantly brighter than the red-shifted
emission.  In all of the Gaussian fits to the \si\ emission toward the center,
the near side was found to be brighter than the far side.  In addition,
\citet{isensee09} show that there are major structural differences between
the front and back interior surfaces.  These front and back differences in
brightness and structure could be due to different masses/densities/energies
in the two different directions.

The existence of flattened inner ejecta surrounded by a spherical reverse
shock raises another important issue related to the inward motion (with
respect to the outward moving ejecta) of the reverse shock.  The 
flattened ejecta model requires that the ejecta being encountered by 
the reverse shock on the front and back sides of \casa\ are from an outer 
layer of the star while the inner layers of the star are encountering the 
reverse shock around the edges.  We now know from \casa's light echoes that 
these layers were indeed intact at the time of explosion and that the 
progenitor even had a thin H envelope \citep{kbu08}.  The diffuse radio 
emission is well modeled by a limb-brightened spherical shell indicating that 
the reverse shock is indeed encountering ejecta and particles are being 
accelerated at least to low energies on the near and far sides of \casa\ 
\citep{gkr01}.  Furthermore, if the reverse shock were encountering 
``nothing'' on the near and far sides of \casa, it would propagate much 
faster those directions and lose its spherical shape -- which is not 
observed.  The question is where is the thermal emission from these ejecta 
that the reverse shock must be encountering?

The X-ray data would be ideal for testing the hypothesis of whether the
reverse shock was currently encountering C-rich ejecta on the near and far 
sides of Cas A, as our model requires, rather than ejecta from an inner 
layer of the star.  Unfortunately, there are several technical issues.  
First, the absorption towards \casa\ is quite high, 
$N_H \sim 10^{22}$ cm$^{-2}$ such that the C lines are absorbed away.  Second, 
the grain scattering halo in the direction of \casa\ \citep{ps95}, in 
conjunction with the CCD readout streak \citep{sah06}, essentially insures 
that every location in \casa\ is contaminated by a scaled-down version of the 
global spectrum.  This means that the \emph{diffuse} X-ray emission near the 
center of \casa\ appears relatively abundant in Si because there is no 
appropriate background spectrum that can be used.  This background 
contamination is not a problem for discrete features because nearby diffuse 
regions can be used to model the background and effectively subtract it away.  
When this is done, the only discrete features projected near the center of 
\casa\ are quasi-stationary flocculi (CSM clumps) and non-thermal filaments 
\citep[e.g.][]{drf04}.

Why, then, do we not see discrete ejecta knots, in any waveband, at the
reverse shock location on the near and far sides, especially given that there 
are fast-moving N-rich ejecta seen at large radii in the plane of the sky?  
In standard supernova remnant ejecta profile models the He and C layers 
should have much lower densities than the Ne/O and Si/S layers which are 
visible in X-rays, infrared, and optical emission \citep[see discussion of 
density profiles in, e.g.,][]{tm99}.  Perhaps the densities in the He and C 
layers are simply too low to identify discrete knots.  A low density would 
allow the reverse shock to propagate quickly in the He and C layers, leading 
to a flattened reverse shock, which is not observed.  However, initially 
steep ejecta profiles might be made much shallower or flat by the passage of 
the \emph{first} reverse shock generated while the forward shock is still 
within the atmosphere of the star \citep[see, e.g., Figure 4 of][]{kpj03}.  In 
such a case, the ejecta density profile of the evolved supernova remnant 
would also be shallower and the currently observed \emph{second} reverse 
shock would remain spherical.  However, the absence of discrete ejecta knots 
from the He and C layers then remains a puzzle.

\section{Summary}
\label{sec:summ}

In summary we find a spherical component to the explosion, a tilted thick
disk, and ejecta pistons in the plane of the thick disk. The gross morphology
of \casa\ must be shaped by the explosion rather than a CSM interaction.
The Si- and Ne/O-illuminated pistons are bipolar and the Fe-illuminated
pistons are not.  The southeast Fe jet does not represent an overturned
region; it sits in a hole in the Si emission with no unshocked Si emission
immediately interior to it.  The southeast Fe ejecta knots are unusually 
long-lived compared to the Fe-rich ejecta to the north and west, perhaps the 
result of an interaction with the CSM structures in this direction.

In \citet{isensee09} we explore the detailed structure of the thick disk, and
the relationship between the Si and O layers before they have been influenced
by the reverse shock.  Further optical mapping of the outer knots 
\citep{fes2010} will be useful in seeing if they match up with the Fe 
pistons.  The lack of visible Fe in the interior of \casa\
remains a puzzle, which is important both for the dynamics and our
understanding of nucleosynthesis.  Long wavelength infrared observations
($\sim40-200~\micron$) with \emph{Herschel} and \emph{Sofia} are needed to
search for unshocked Fe emission (\fetwo, [\ion{Fe}{3}]) as well as emission
from the presumed outer C and N layers ([\ion{C}{2}], [\ion{N}{2}],
[\ion{N}{3}]) that should be interacting with the forward shock on the near
and far sides of the remnant.  Finally, whether our piston hypothesis is true
or false, our unprecedented data set and the 3D geometry presented here will
invite theoretical work for a deeper understanding of supernova explosion
mechanisms and nucleosynthesis.

\acknowledgements
This work is based on observations made with the \emph{Spitzer Space
Telescope}, which is operated by the Jet Propulsion Laboratory, California
Institute of Technology under NASA contract 1407.  Partial support for this
work was provided by NASA/JPL through award \#1264030 to the University of 
Minnesota.  This work is also based on observations made with the 
\emph{Chandra X-ray Observatory}.  Partial support for this work was provided 
by NASA through Chandra Award \#AR5-6008X to the University of 
Minnesota.  During the course of this work, T.~D. received partial support 
from NASA through SAO grant GO3-4063A while a post-doc at Harvard.  At MIT, 
T.~D. received support from NASA through SAO contract SV3-73016 to MIT for
support of the \emph{Chandra} X-ray Center and Science Instruments, which is
operated by SAO for and on behalf of NASA under contract NAS8-03060.  While at 
WV Wesleyan College, T.~D. received partial support through a Faculty 
Research Enhancement Grant from the NASA-WV Space Grant Consortium.

The three-dimensional graphics were made using the 3D Slicer program which is
a medical imaging tool being adapted for astronomical use by the Astronomical
Medicine Project which is a part of Harvard's Initiative in Innovative
Computing\footnote{\url{http://am.iic.harvard.edu}}.  T.~D. would like
to thank Megan Watzke for connecting her with the Astronomical Medicine
Project and Michelle Borkin for providing valuable assistance with 3D Slicer.
T.~D. also thanks Mike Noble for help with the 3D PDF graphics and Dan Dewey
for valuable conversations.  T.~D., L.~R., and K.~I. thank Alex Heger for 
guidance and valuable conversations concerning supernova models.  Finally, 
we thank the referee for a thorough review of the manuscript which has 
resulted in significant improvements in the presentation of our results.

\appendix
\section{3D PDF Graphics}

\begin{figure}
\epsscale{1}
\begin{center}
\plottwo{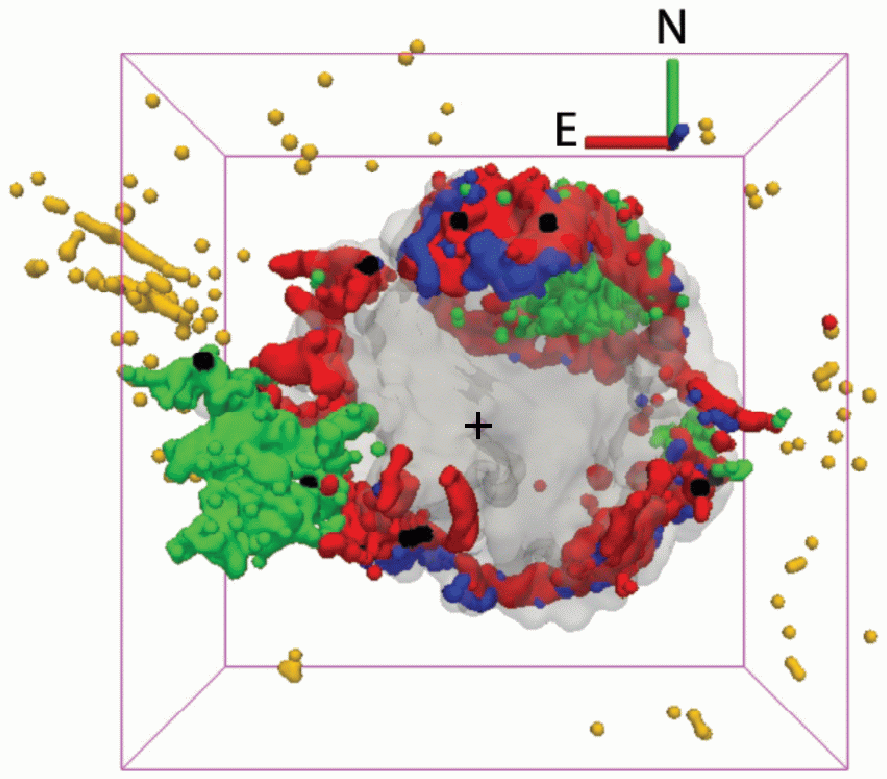}{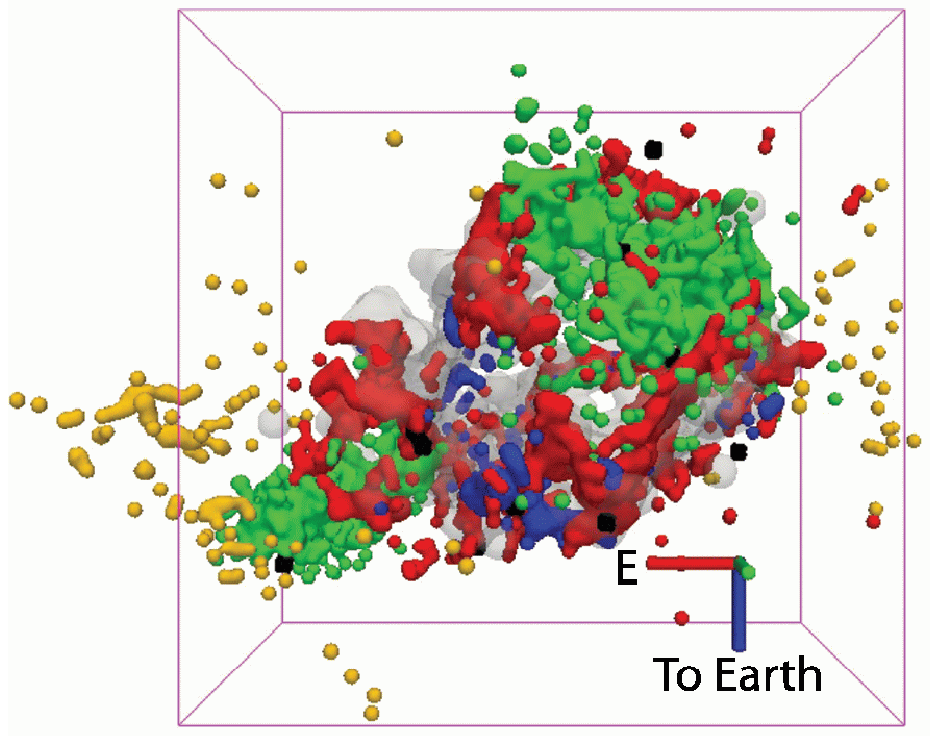}
\plottwo{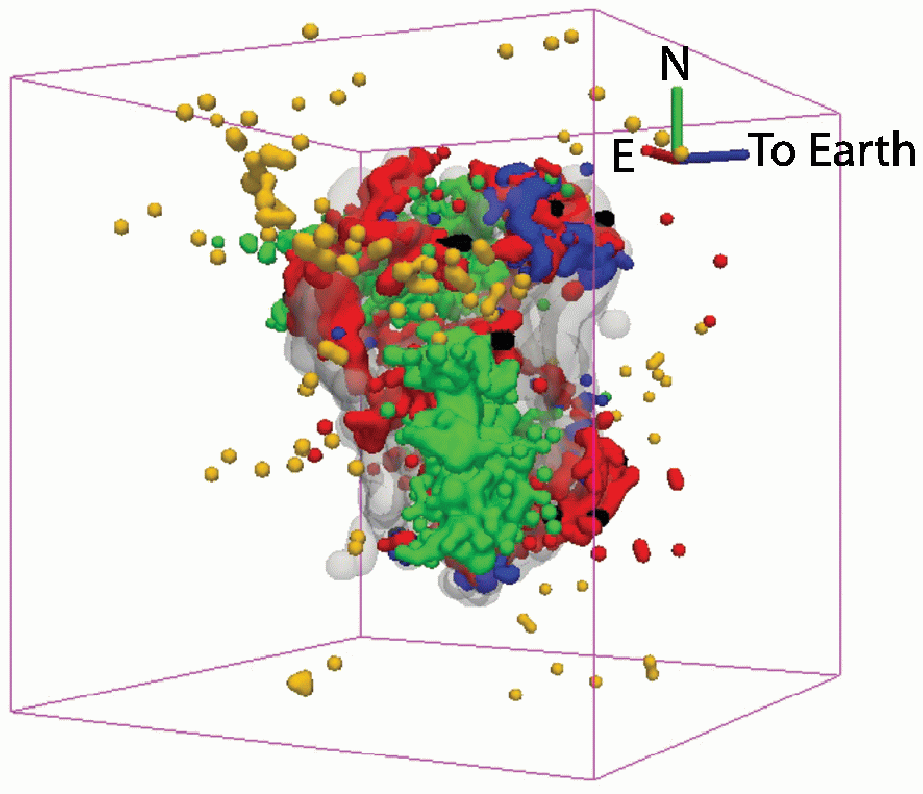}{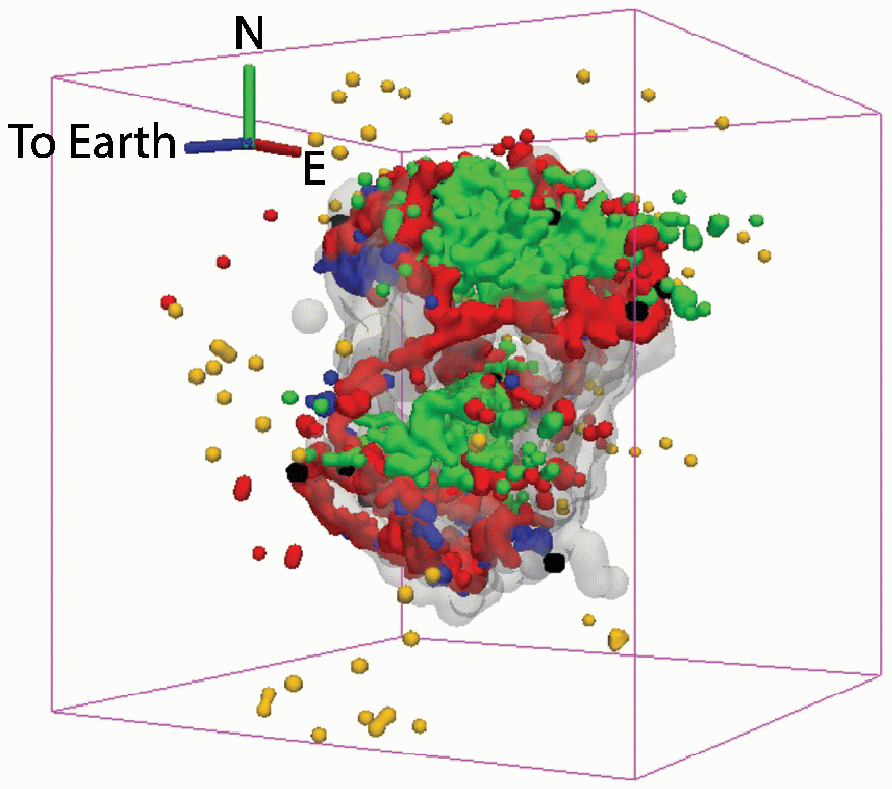}
\end{center}
\caption{Three-dimensional projections of the infrared \ar, high \neon/\ar\
ratio, and \si\ emission, the X-ray \fe\ and \xsi\ emission, and the outer
optical knots.  The color coding is described in Table \ref{colortable}.
This figure is also available as an mpeg animation and a 3D PDF in the
electronic edition of the \emph{Astrophysical Journal}.  The mpeg animation and 3D PDF are also available at http://homepages.spa.umn.edu/$_\textrm{\~{}}$tdelaney/paper.  \label{allviews}}
\end{figure}

\begin{figure}
\epsscale{0.8}
\plotone{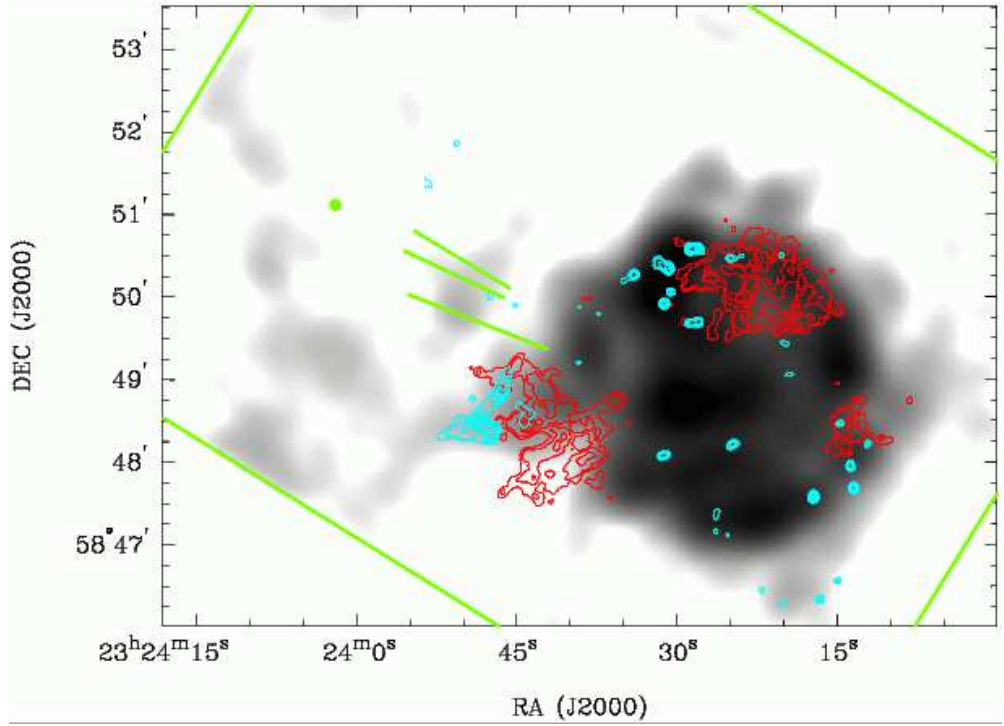}
\caption{Grayscale: 34.48~\micron\ \si\ emission smoothed to 20$\arcsec$
resolution; Red contours: X-ray \fe\ emission; Blue contours: optical
H$\alpha$ and [\ion{N}{2}] emission from diffuse and clumpy CSM.  The green
lines denote the linear northeast jet structures and the green circle
identifies the location of the outermost optical ejecta observed by
\citet{fhm06}.  The edges of the LL1 slit mapping shown in Figure~\ref{slits}
are indicated.  \label{outersi}}
\end{figure}

Figure~\ref{allviews} is available in the electronic edition of the
\emph{Astrophysical Journal} as a 3D PDF file.  The PDF file contains the
full three-dimensional graphics information and allows the user to directly
manipulate the graphics by rotating the figure and zooming in and out.  The
free Adobe\textregistered~ Reader\textregistered~ program version 8 or newer
is required to view the 3D PDF file.  By clicking on the graphic, the 3D
aspect is enabled and a model tree opens allowing the user to turn on and
off individual elements of the figure.  Both the original and ionization
corrected \fe\ emission are provided in the 3D PDF version of
Figure~\ref{allviews} so that the readers can judge for themselves what effect
collimation has had on the southeast \fe\ structure.  The color coding for
the individual components of the model is described in Table
\ref{colortable}.  Coordinate axes are provided to help keep track of
orientation.  The red axis points east, the green axis points north, and the
blue axis points to Earth.  The mpeg animation and 3D PDF are also available 
at http://homepages.spa.umn.edu/$_\textrm{\~{}}$tdelaney/paper.

\newpage
\begin{deluxetable}{llll}
\tablecaption{Color code for the 3D graphics in Figures
\ref{cartoon1}--\ref{pistons}, \ref{siviews}--\ref{allviews}.
\label{colortable}}
\tablehead{\colhead{Component} & \colhead{Telescope/} &
\colhead{Color/} & \colhead{Comments}\\
 & \colhead{Instrument} & \colhead{Symbol} & }
\startdata
\ar\ & \emph{Spitzer}/IRS & red & \nodata \\
high \neon/\ar\ ratio & \emph{Spitzer}/IRS & blue & \nodata \\
\[\si\ & \emph{Spitzer}/IRS & gray & \nodata \\
\fe\ & \emph{Chandra}/ACIS & green & \nodata \\
\xsi\ & \emph{Chandra}/HETG & black & \citet{lds06} \\
outer optical knots & ground based & yellow & \citet{fg96} \\
 & & & \citet{fes01} \\
fiducial CCO & \nodata & cross & placed at $v=0$~\kps \\
 & & & pink in the 3D PDF \\
fiducial reverse shock & \nodata & peach & \nodata \\
\enddata
\end{deluxetable}
\clearpage

\clearpage

\clearpage

\clearpage

\clearpage

\clearpage

\clearpage

\clearpage

\clearpage

\clearpage

\clearpage

\clearpage

\clearpage

\clearpage

\clearpage

\clearpage

\clearpage

\clearpage

\clearpage

\clearpage

\clearpage

\end{document}